\documentclass[11pt]{article}

\usepackage[final]{acl}
\usepackage[utf8]{inputenc}
\usepackage{graphicx}
\usepackage{subcaption}

\usepackage{tabularx}
\usepackage{booktabs}
\usepackage{amssymb}
\usepackage{CJKutf8}

\usepackage{times}
\usepackage{latexsym}
\usepackage[T1]{fontenc}

\usepackage[T1]{fontenc}

\usepackage[utf8]{inputenc}

\usepackage{microtype}

\usepackage{inconsolata}

\usepackage{graphicx}
\usepackage{booktabs}
\usepackage{enumitem}
\usepackage{amsmath}
\input{./chatdiagram.tex}

\title{Emergent Relational Order in LLM Agent Societies: From Collective Affect to Authority Stratification}

\author{
\textbf{Zhiyuan Ji\textsuperscript{1}},
\textbf{Xinyu Chen\textsuperscript{2}\setcounter{footnote}{1}\thanks{Equal Contribution}},
\textbf{Ziqi Dai\textsuperscript{3}\footnotemark[2]},\\
\textbf{Shiyun Tang\textsuperscript{1}\setcounter{footnote}{0}\thanks{Corresponding author.}},
\textbf{Chunyu Wei\textsuperscript{1}},
\textbf{Yueguo Chen\textsuperscript{1}}\\\\
\textsuperscript{1}Renmin University of China,
\textsuperscript{2}Beihang University,
\textsuperscript{3}Minzu University of China\\
\texttt{\{zhiyuanji, jamietang, weichunyu, chenyueguo\}@ruc.edu.cn}\\
\texttt{zy2402531@buaa.edu.cn}\\
\texttt{daiziqi1296@muc.edu.cn}
}

\begin{document}
\maketitle

\begin{abstract}
Fei Xiaotong's Differential Order Pattern characterizes rural society as egocentric and relationally graded, with cooperation attenuating over social distance. Although often treated as culturally specific, its mechanistic basis remains under-operationalized, and prior LLM-based simulations have mainly addressed short-term coordination rather than long-horizon social structure. We propose CAREB-MAS, a multi-agent framework grounded in Affect Control Theory, Social Identity Theory, and Durkheimian collective affect. Agents reason through an emotion--ethics--belief chain and maintain dynamically evolving egocentric identities, while the macro environment specifies only individual production, preference-based allocation, and minimal interaction protocols.
Across long-horizon simulations, agents spontaneously reproduce five core Differential Order phenomena: stable labor specialization, guanxi-based economic ethics, relational decay of cooperation, emergent relational authority, and clan-based center--periphery stratification. These patterns shift with production structure from kin-centered integration toward greater functional interdependence.
Extensive experiment results support interpreting Differential Order as a structure-sensitive emergent outcome of general social mechanisms, with LLM-based multi-agent simulation providing a interdisciplinary framework for studying social structure and change.
\end{abstract}

\section{Introduction}

The transformation of traditional societies into modern ones constitutes one of the most consequential problems in social science~\cite{sorokin2017social,lazer2021meaningful,bhatt2022ethical}. Classical sociology has long addressed this question: Durkheim, for instance, emphasized the transition from mechanical to organic solidarity driven by changing divisions of labor~\cite{durkheim2019division}.

\begin{figure}[t]
    \centering
  \includegraphics[width=1\linewidth]{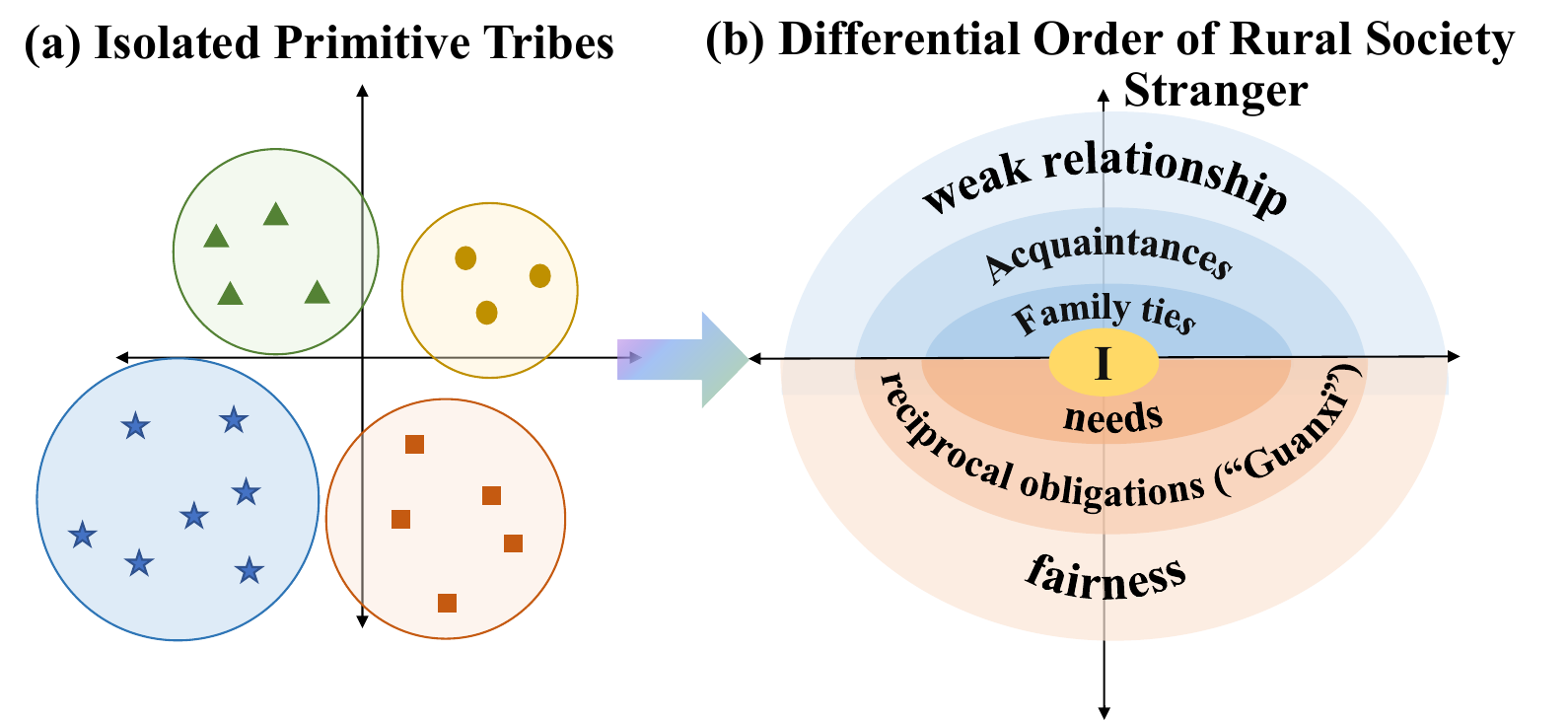}
    \caption{Schematic comparison of social organization:
(\textbf{a}) discrete groups without relational gradation;
(\textbf{b}) differential order, organized as ego-centered concentric layers of relational proximity.}
    \label{fig:placeholder}
    \vspace{-15pt} 
\end{figure}

Beyond the Western canon, Fei Xiaotong's Differential Order Pattern~\cite{fei1992soil} describes rural society as ego-centered relational circles in which moral obligation, trust, and cooperation attenuate with social distance (Figure~\ref{fig:placeholder}). Although often treated as specific to Chinese society, its constituent mechanisms---kinship organization, reciprocity beyond price exchange, and authority without centralized power---are well documented across comparative anthropology~\citep{fortes1940african, sahlins1972stone, polanyi1944great}. This raises the possibility that Differential Order reflects structure-sensitive outcomes of general social mechanisms rather than solely Chinese traditions.

Testing this possibility remains difficult. Ethnography offers contextual richness but limited experimental control; quantitative sociology often captures long-horizon change only indirectly; and conventional agent-based models rely on hand-coded rules that poorly represent the co-evolution of emotion, ethics, authority, and social structure~\cite{eguiluz2005cooperation,zimmermann2004coevolution}. Recent LLM-agent studies have reproduced social contracts~\cite{dai2024artificial}, cooperation norms~\cite{vallinder2024cultural}, and institutional emergence~\cite{piao2025agentsociety}, but mostly in short-horizon settings. Whether LLM-based simulations can generate long-horizon social structure and probe the generality of non-Western sociological theory remains underexplored.

In this paper, we propose \textbf{CAREB-MAS}, a multi-agent simulation framework testing whether Differential Order can emerge from cross-culturally documented micro-mechanisms without encoding culture-specific institutional rules. Agents reason through an emotion--ethics--belief chain grounded in Affect Control Theory~\citep{Heise2007}, Social Identity Theory~\citep{Tajfel1979}, and Durkheimian collective affect, while the environment specifies only kinship, reciprocity-based allocation, and the absence of centralized authority. 

Across long-horizon simulations, agents spontaneously reproduce five core Differential Order phenomena: stable labor specialization, guanxi-based economic ethics, cooperation attenuation with relational distance, emergent relational authority, and clan-based center--periphery stratification. These patterns vary systematically with production structure: symmetric endowments yield kin-centered integration closer to mechanical solidarity, whereas complementary endowments produce more functionally interdependent patterns closer to organic solidarity. Ablations on two LLMs further show that these outcomes are architecture-dependent rather than generic LLM artifacts. Together, the results recast Differential Order as a structure-sensitive emergent outcome of general social mechanisms and position LLM-based multi-agent simulation as a generative testbed for social structure and change.

We summarize our contributions as follows:(1) we provide a computational approach to study Fei Xiaotong's Differential Order Pattern without encoding culture-specific rules; (2) we design CAREB-MAS, a multi-agent framework integrating emotion, ethical judgment, and relational cognition  to harness and test the theoretical mechanisms and  (3) we validate core Differential Order Pattern phenomena and conduct controlled experiments to reveal the sensitivity and dependency of these phenomena on cognitive architecture.
\vspace{-0.5\baselineskip} 
\section{RELATED WORK}\label{sec:related}

\subsection{Social Transformation and Rural Society Formation}

Classical sociology treats social structure as an emergent outcome of interaction rather than a fixed design. In Durkheim, authority and division of labor arise through 
\begin{center}
\noindent Collective Affect $\rightarrow$ Ethics $\rightarrow$ Belief $\rightarrow$ Social Structure
\end{center}
process ~\citep{tada2020,durkheim1912elementary}. This generative view aligns with organizational theories that structure as emerging from routinized interaction~\citep{haveman2019}.

Fei Xiaotong's Differential Order Pattern describes rural society as egocentric and relationally graded. Its core preconditions---kinship organization, generalized reciprocity beyond price exchange, and acephalous authority---are widely documented across comparative studies of pre-market society~\citep{fortes1940african,levi1949structures,sahlins1972stone,mauss1925gift,polanyi1944great}. Relational sociology and historical anthropology further show that such structures persist and reconfigure under modernization~\citep{ZhangJiangHua2015,hanting2024,zhang2018double,longyuan2023}, making rural society a useful setting for examining general social mechanisms~\citep{zhou2024b}.

\subsection{Agent-Based Social Simulation}

Traditional agent-based modeling has served as a computational testbed for social phenomena, from Sugarscape~\cite{epstein1996growing} and cultural dissemination~\cite{axelrod1997dissemination} to segregation~\cite{schelling1971dynamic} and cooperation emergence~\cite{eguiluz2005cooperation,zimmermann2004coevolution}. However, rule-based architectures struggle to capture the co-evolution of emotion, ethics, and social structure~\cite{gao2024large}.

LLM agents extend this line of work by enabling richer social reasoning. Park et al.~\cite{park2023generative} demonstrated believable social behavior in generative agents, and subsequent studies operationalized specific theories or institutions, including social exchange~\cite{wang-etal-2025-investigating}, Hobbesian contract formation~\cite{dai2024artificial}, institutional emergence~\cite{piao2025agentsociety}, cooperation evolution~\cite{vallinder2024cultural}, and norm emergence~\cite{ren2024emergence}.

Despite these advances, existing simulations focus mainly on short-term coordination or dyadic exchange, leaving the long-horizon co-emergence of collective affect, authority, and division of labor underexplored---especially for non-Western sociological theories whose structural predictions have rarely been tested computationally.

\section{Theoretical Foundations}\label{sec:foundations}

\subsection{Differential Order and Relationally Embedded Division of Labor}

Fei Xiaotong's \emph{Differential Order Pattern} conceptualizes rural society as concentric ripples of obligation whose intensity decays with relational distance. Social relations radiate outward from the ego: close kin occupy the inner circles, followed by acquaintances and strangers in the outer layers. Relational proximity systematically shapes emotion, ethics, and cooperation. Although often interpreted through the Chinese rural context, its key structural preconditions---kinship organization~\citep{fortes1940african,levi1949structures}, reciprocity-based allocation without price mechanisms~\citep{polanyi1944great,sahlins1972stone,mauss1925gift}, and limited centralized authority~\citep{fortes1940african}---are widely documented across comparative anthropology. We therefore examine whether Differential Order-like patterns can emerge from these cross-culturally documented conditions.

Within this framework, division of labor stabilizes through repeated interaction 
under relational and ethical constraints rather than formal institutions. We term 
this \emph{relationally embedded division of labor} and test whether it can emerge 
endogenously without explicit cultural rules.

\subsection{Guanxi as Economic Ethics}

In Differential Order, \emph{guanxi} is not merely a preference for familiar others but an \emph{economic-ethical principle} that defines obligation, responsibility, and acceptable cost in exchange. This parallels Sahlins's \emph{generalized reciprocity}~\citep{sahlins1972stone}: close ties sustain open-ended obligation without immediate return. In low-contractual environments, such relational ethics can stabilize cooperation and legitimate asymmetric outcomes despite short-term inefficiency.
\subsection{Five Generative Propositions}

\begin{figure*}[t]
    \centering
    \includegraphics[width=1\linewidth]{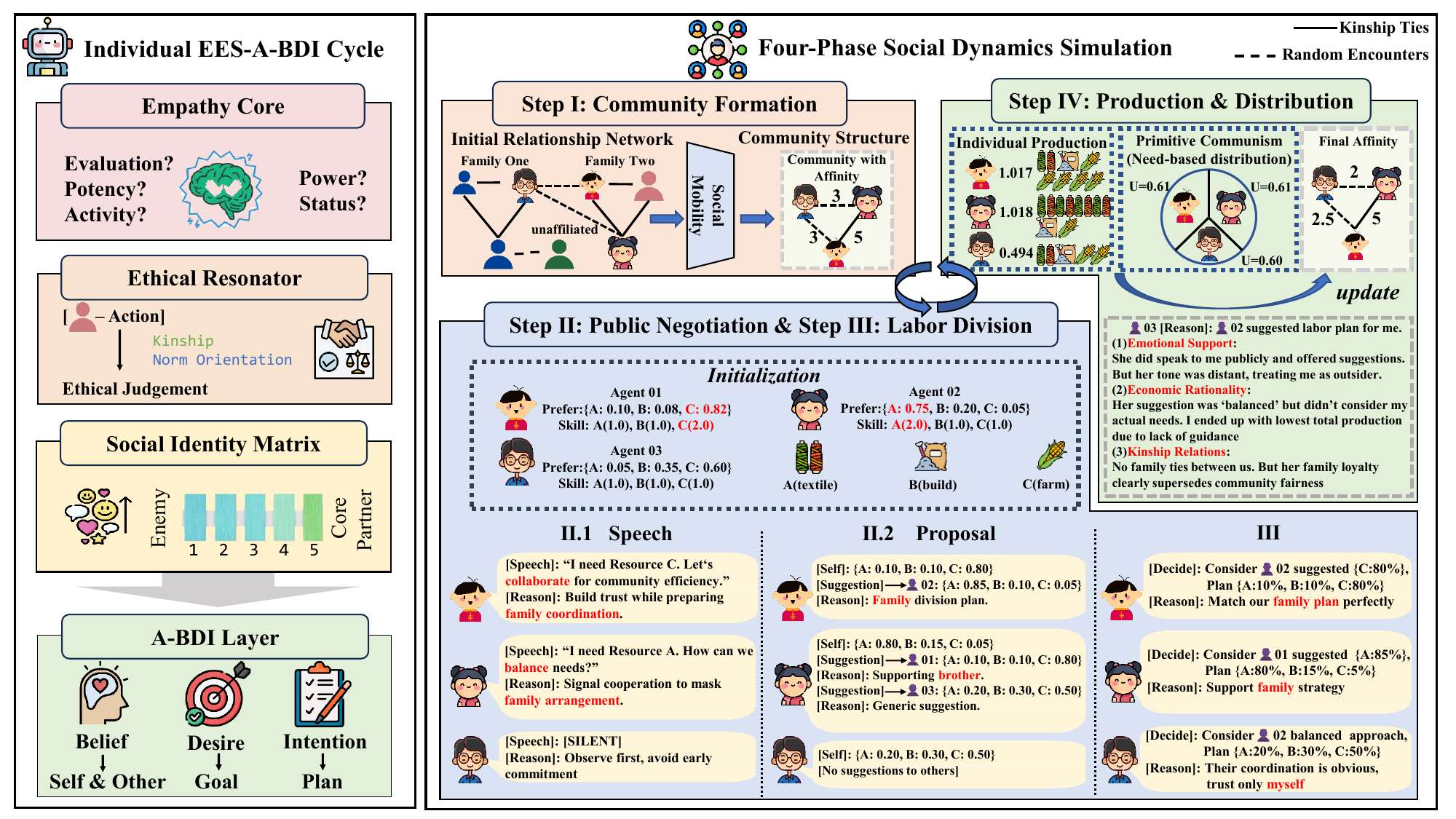}
    \caption{Overview of CAREB-MAS. \textbf{Left}: agent-level cognitive architecture, from EC (Affect Control Theory) to ER (ethical judgment), SIM (Social Identity Theory), and A-BDI (Durkheimian affective constraint on action). \textbf{Right}: four-phase interaction loop: community formation, deliberation, division of labor, and production with psychological update.}
    \label{fig:overview}
\end{figure*}

To distinguish built-in assumptions from dependent outcomes, we conceptually separate \emph{pre-defined structural conditions} from \emph{emergent interactional patterns}. The former correspond to cross-culturally documented features of pre-market social organization, while the latter constitute the primary objects of explanation in our simulations (see Appendix~\ref{app:predefined-emergent}). As shown later, identical structural conditions can yield qualitatively different emergent structures under symmetric and complementary production regimes.

Based on these mechanisms, we formulate five generative propositions. Each describes a macro-level pattern that, if observed, would be consistent with Differential Order—but that arises from general social cognition rather than culture-specific encoding:

\paragraph{P1: Stable Division of Labor.}
Repeated interaction under relational embeddedness can generate stable and specialized divisions of labor without centralized coordination or predefined institutional roles. 

\paragraph{P2: Guanxi as Economic Ethics.}
Ethical obligations embedded in close relations sustain cooperation and role persistence despite short-term efficiency losses.

\paragraph{P3: Differential Order and Relational Decay.}
Cooperative intensity and efficiency decay with relational distance, concentrating high-efficiency collaboration within families or clans.

\paragraph{P4: Emergent Authority Stratification.}
Durable authority can arise through long-term interaction as stable compliance and social recognition, without formal authorization.

\paragraph{P5: Clan-based Center--Periphery Structure.}
Economic returns and authority stratify by relational proximity, producing a family-centered core and peripheral disadvantage.

\subsection{Structural Contrast: Mechanical vs.\ Organic Solidarity}

To examine how production structure conditions Differential Order, we contrast \emph{symmetric} and \emph{complementary} endowments. Symmetric capacities favor integration through shared identity and relational ethics (\emph{mechanical solidarity}), whereas complementary capacities favor integration through functional differentiation and mutual dependence (\emph{organic solidarity}). This contrast provides the main analytical lens for interpreting variation across conditions.

\section{Methodology: The CAREB-MAS Framework}\label{sec:methodology}

We propose \textbf{CAREB-MAS} (Collective Affection--Reasoning--Emergence Based Multi-Agent Simulation), a framework for examining whether structured social order can emerge without explicit cultural templates. Stable social structure arises from repeated local interaction among boundedly rational agents, without hard-coded cultural rules or predefined hierarchy.

\subsection{Sociological Logic and Emergence Design}

The framework operationalizes a causal pathway grounded in three theories: affective perception from Affect Control Theory~\citep{Heise2007}, moral judgment from Durkheimian ethics~\citep{Durkheim1961MoralEducation}, and relational cognition from Social Identity Theory~\citep{Tajfel1979}. These components are integrated into an augmented BDI architecture in which collective affect constrains rational action~\citep{durkheim1912elementary}. The resulting chain---\emph{Affect $\to$ Ethics $\to$ Identity $\to$ Action}---models general social cognition rather than culture-specific institutional content. No module directly encodes Fei's theory, Confucian ethics, or predefined \emph{guanxi} categories. Any Differential Order patterns therefore arise from the interaction between general mechanisms and the structural conditions in Table~\ref{tab:predefined-emergent}.

\subsubsection{Agent Architecture}

All agents share a unified psychological--decision architecture activated sequentially each round (Figure~\ref{fig:overview}). Social perception produces affective responses, which are filtered through moral evaluation, update relational cognition, and finally constrain belief--desire--intention reasoning. Full implementation details and prompts are provided in Appendix~\ref{app:llm_details}.

\begin{itemize}[leftmargin=*]
    \item \textbf{Empathy Core (EC)} $\leftarrow$ \textit{Affect Control Theory}~\citep{Heise2007}.  
Decodes observable behavior into a five-dimensional affective vector
(evaluation, potency, activity, status-conferral, power-assertion).

    \item \textbf{Ethical Resonator (ER)} $\leftarrow$ \textit{Durkheimian Moral Education}~\citep{Durkheim1961MoralEducation}.  
Aggregates perceptions into a moral judgment of each interaction partner,
outputting \emph{approbatory}, \emph{repressive}, or \emph{restorative} attitudes for relational updating.

    \item \textbf{A-BDI Decision Module} $\leftarrow$ \textit{BDI + Durkheimian collective affect}.  
Extends BDI~\citep{georgeff1998belief,wang-etal-2025-investigating} with \emph{Affection}, embodied in SIM and constraining belief formation, desire generation, and intention selection.
\end{itemize}

\subsubsection{Social Dynamics: A Four-Phase Loop}

Each round follows a four-phase loop (Figure~\ref{fig:overview}) modeling pre-modern rural formation under co-residence and relational cognition, without prices or centralized adjudication. Hyperparameters and simulation settings are reported in Appendix~\ref{app:hyperparams}.

\paragraph{Phase I: Community Formation.}
Agents partition into communities based on mean relational proximity $\bar{d}_{ij} = (d_{i \to j} + d_{j \to i})/2$, where $d_{i \to j} \in \{1,\dots,5\}$. Louvain detection produces dynamic communities, and agents may migrate as relationships evolve (Appendix~\ref{app:community_detection_phase}).

\paragraph{Phase II: Public Deliberation.}
Within communities, agents engage in $R=2$ rounds of public deliberation, broadcasting speeches and labor proposals. No voting or enforcement rule is imposed. The resulting action set $\mathcal{A}_C^{(t)}$ becomes the observable input for subsequent updating (Appendix~\ref{app:negotiation_phase}).

\paragraph{Phase III: Division of Labor and Production.}
Each agent independently chooses labor allocation $\boldsymbol{\ell}_i = (\ell_{iA}, \ell_{iB}, \ell_{iC})$ with $\sum_r \ell_{ir} = 1$, based on $\mathcal{A}_C^{(t)}$, A-BDI state, and SIM. No centralized authority or conflict-resolution rule is imposed. Economic ``authority'' is measured \emph{ex post} by proposal adoption frequency (Appendix~\ref{app:labor_division_phase}).%
\footnote{Decision Authority (DA) is the paper's primary operationalization of emergent authority; formal definition appears in \S\ref{sec:results}.}

\paragraph{Phase IV: Distribution and Psychological Update.}
Community goods are pooled and redistributed by preference weights:
$x_{ir} = (\sum_{k \in C} o_{kr}) \cdot p_{ir} / \sum_{j \in C} p_{jr}$.
This implements a redistributive allocation prototype consistent with Polanyi's redistributive economies~\citep{polanyi1944great} and Sahlins's generalized reciprocity~\citep{sahlins1972stone}. After consumption, each agent executes the update cycle $\mathrm{EC} \to \mathrm{ER} \to \mathrm{SIM} \to \mathrm{A\text{-}BDI}$. Production, clearing, and utility details are reported in Appendices~\ref{app:production_phase}--\ref{app:consumption_phase}.

\paragraph{Evaluation Transparency.}
EC and ER are \emph{agent-internal} cognitive processes: each agent uses its own LLM call to perceive and judge others, so assessments may differ across agents. The only metric using an \emph{external} LLM-as-Judge evaluator is Communication Intensity (Appendix~\ref{app:behavior_metrics}, Table~\ref{tab:communication_prompt}). All other metrics---Decision Authority, Proposal Intensity, Lock-in Score, Community Match, and SIM levels---are computed deterministically from simulation logs.

\section{Experiment Setup}\label{sec:setup}

We simulate 18 agents over 30 rounds: two kin-based clans (F1 and F2, six members each) and six unaligned agents without family ties. Agents jointly produce three resource types under a Leontief utility function with preference-weighted redistribution (\S\ref{sec:methodology}). We compare two conditions: \emph{symmetric skills} (balanced production capacities) and \emph{complementary skills} (group-specific specialization in one resource). All agents share the same cognitive architecture; unaligned agents differ only in lacking kinship ties. No agent is given knowledge of Fei's theory or any culture-specific behavioral rules.

Simulations use DeepSeek-V3 (temperature 0.7), with five random seeds per condition. Cross-model robustness is assessed using GPT-4o-mini and Gemini-2.5-flash-lite (Appendix~\ref{app:individual_lockin}). Code, prompts, and simulation logs will be released upon publication.\footnote{\url{https://github.com/Chi20/DOP}} Metric definitions and implementation details are provided in Appendices~\ref{app:metrics}--\ref{app:llm_details}.

\subsection{Proposition--Metric Mapping}

Each proposition is operationalized via a primary quantitative metric paired with a diagnostic criterion that specifies the expected pattern under Differential Order. The full mapping between propositions, metrics, and diagnostic criteria is provided in Appendix~\ref{app:concept-metric}, with formal definitions in Appendix~\ref{app:metrics}.

\section{Experimental Results}
\label{sec:results}

We evaluate Propositions~1--5 using the criteria in Table~\ref{tab:prop-metric}.

\subsection{Emergence of Stable Division of Labor (P1)}
\label{p1-stable-division}

\begin{figure}[t]
    \centering
    \includegraphics[width=1\linewidth]{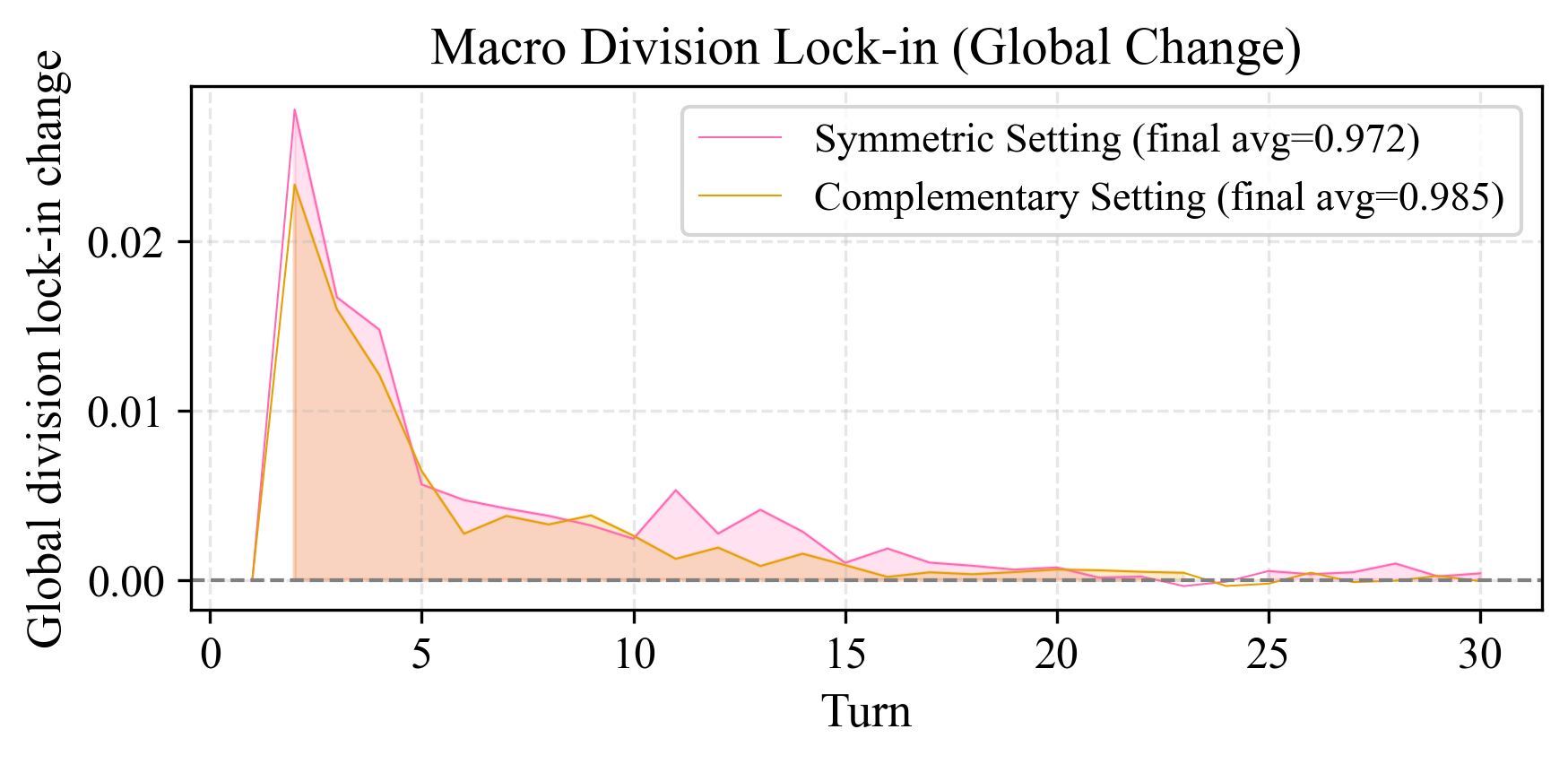}
    \caption{Global division lock-in emerges early and stabilizes under both skill structures. Panel~A: symmetric skills converge to $0.972$; Panel~B: complementary skills converge faster to $0.985$. Per-round improvements (Panel~C) approach zero, indicating equilibrium.}
    \label{fig:macro_lockin}
\end{figure}

Figure~\ref{fig:macro_lockin} shows that stable specialization emerges without centralized coordination. Under symmetric skills, lock-in rises rapidly and plateaus at $0.972$; under complementary skills, convergence is faster and reaches $0.985$. The acceleration under complementarity suggests that structural differentiation itself drives specialization speed, not merely learning efficiency. Per-round improvements decline to near-zero in both conditions, confirming that the division is stable rather than transient. These patterns support Proposition~1. Ablation analysis (Appendix) further indicates that while convergence is architecturally robust, the precise calibration of social coordination depends specifically on the EC/ER/SIM mechanism.

\subsection{Guanxi as Economic Ethics (P2)}
\label{p2-guanxi}

\begin{figure}[t]
    \centering
    \begin{subfigure}[t]{\linewidth}
        \centering
         \includegraphics[width=1\linewidth]{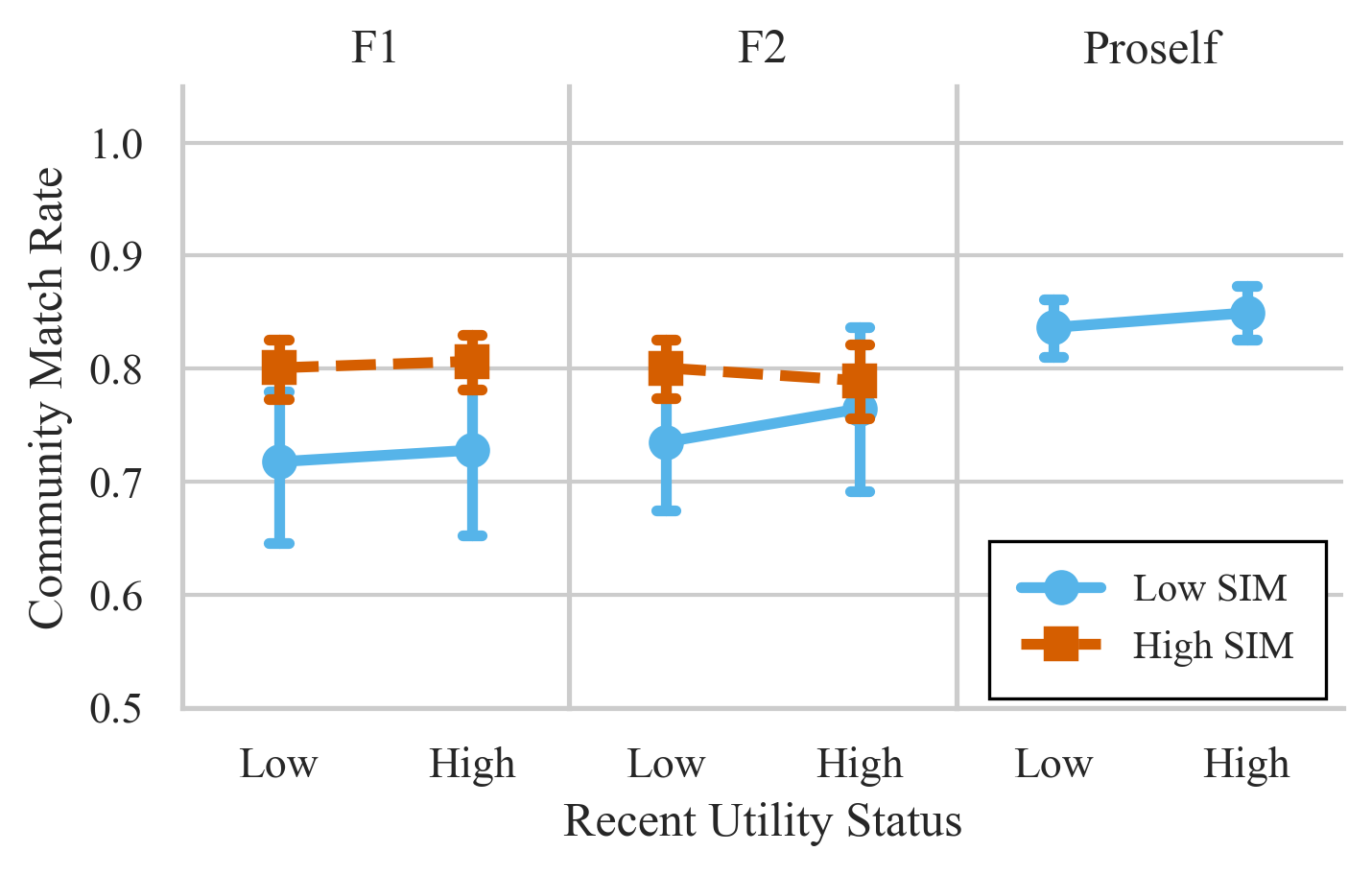}
        \caption{Symmetric skills.}
        \label{fig:commuity_matching_by_SIM_sym}
    \end{subfigure}

    \vspace{0.4em}

    \begin{subfigure}[t]{\linewidth}
        \centering
        \includegraphics[width=\linewidth]{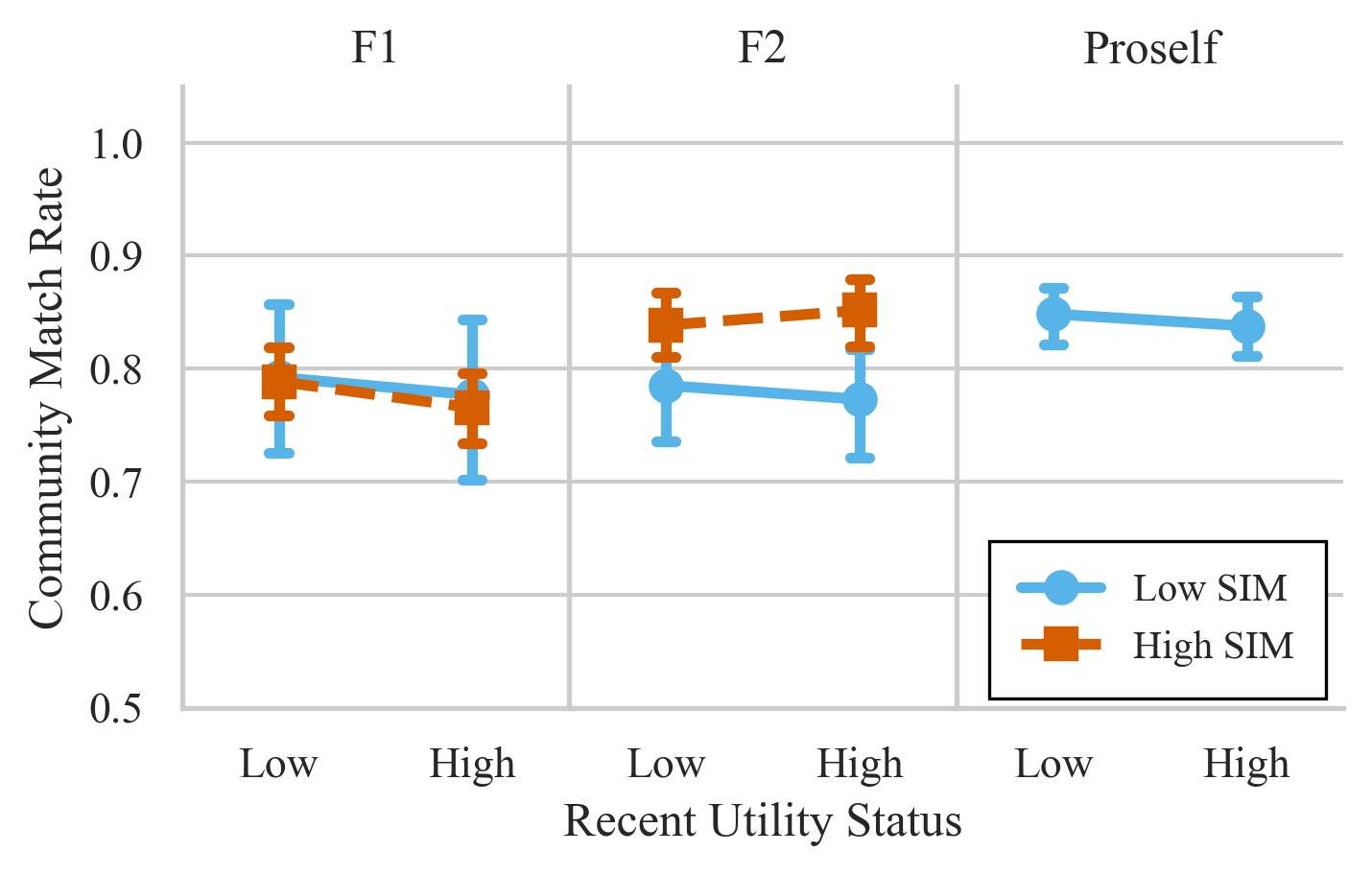}
        \caption{Complementary skills.}
        \label{fig:commuity_matching_by_SIM_comp}
    \end{subfigure}
    
\caption{Community match rates under utility shocks. F1/F2: family groups; Proself: unaligned agents. High-SIM relations maintain stable participation (solid lines) even when recent utility is low, whereas low-SIM ties (dashed lines) exhibit sharp decline. This stickiness is more pronounced in the complementary condition (B).}

    \label{fig:commuity_matching_by_SIM}
\end{figure}

To test whether \emph{guanxi} stabilizes cooperative roles beyond efficiency maximization, we examine how community match responds to negative utility shocks under varying relational embeddedness. Figure~\ref{fig:commuity_matching_by_SIM} reveals a consistent pattern: high-SIM agents maintain stable community participation regardless of recent payoff performance, while low-SIM agents withdraw sharply when efficiency incentives turn unfavorable. This asymmetry is particularly marked under complementary skills, where coordination failure is most costly. The stickiness of high-SIM ties suggests that ethical obligations embedded in close relations function as a commitment device, sustaining participation in community-based division of labor even when short-term rationality would dictate defection. This supports Proposition~2.

\subsection{Relational Decay of Cooperation (P3)}
\label{p3-relational-decay}

\begin{figure}[t]
    \centering
    \begin{subfigure}[t]{\linewidth}
        \centering
        \includegraphics[width=\linewidth]{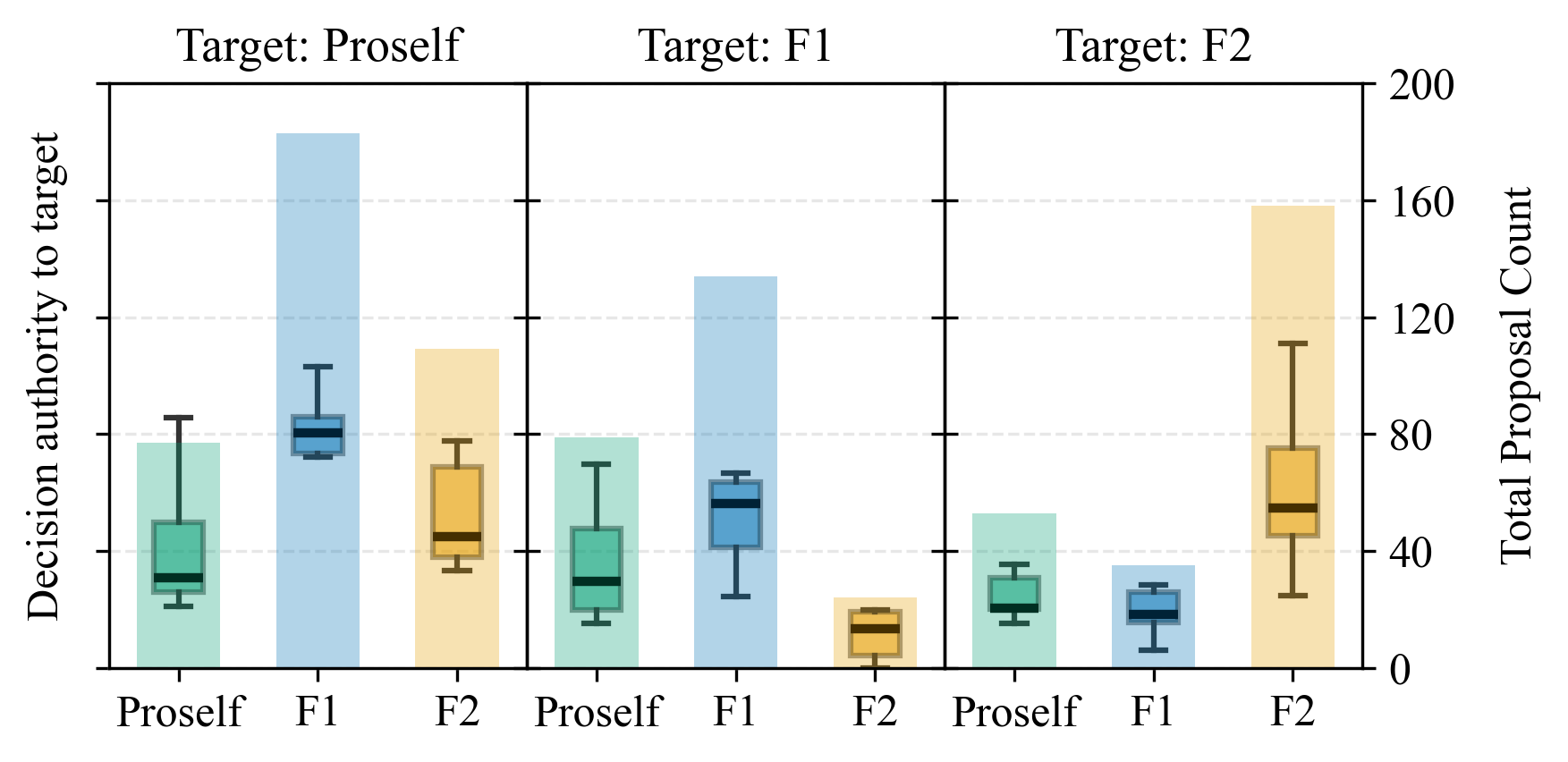}
        \caption{Symmetric skills.}
        \label{fig:authority_proposal_boxplots_sym}
    \end{subfigure}

    \vspace{0.4em}

    \begin{subfigure}[t]{\linewidth}
        \centering
        \includegraphics[width=\linewidth]{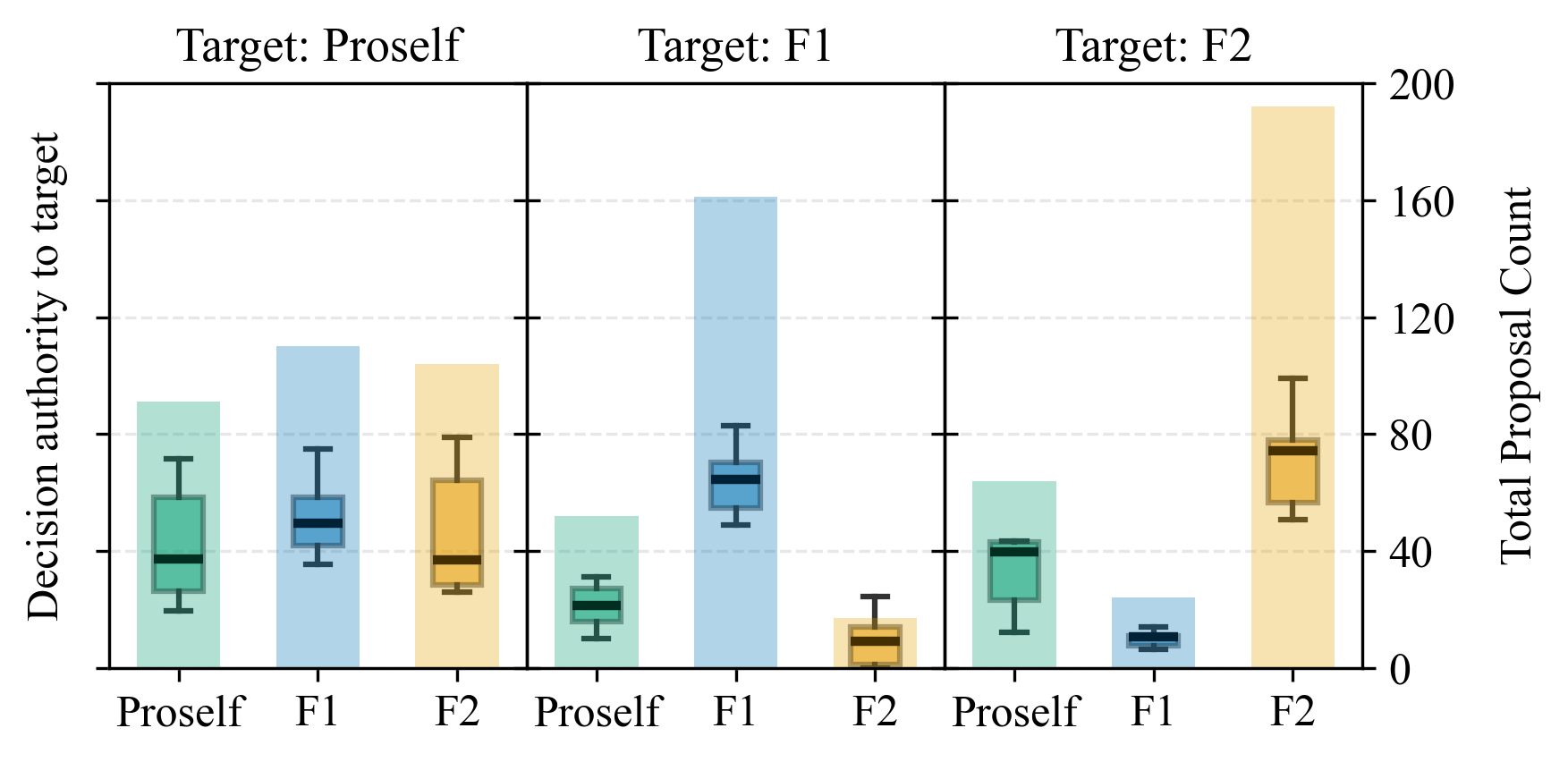}
        \caption{Complementary skills.}
        \label{fig:authority_proposal_boxplots_comp}
    \end{subfigure}

\caption{Decision authority (boxplots) and proposal activity (bars) exhibit a relational gradient. Family agents (F1, F2) show high within-group authority and low cross-family engagement, while Proself agents lack self-boundary and align with family structures. Note the asymmetry in proposal distribution: under symmetric skills (a), within-family and family$\to$proself proposals dominate (34.2\% and 34.3\%), whereas cross-family proposals are rare (6.9\%).}
    \label{fig:authority_proposal_boxplots}
    \vspace{-10pt}
\end{figure}

Figure~\ref{fig:authority_proposal_boxplots} reveals that cooperation decays along relational rather than preference-type boundaries. Family agents (F1, F2) concentrate decision authority and proposal activity within their own groups, with markedly lower engagement across family boundaries. This suggests cooperation is organized around sharply bounded family circles. In contrast, Proself agents do not form a cohesive self-boundary; they accept family-originated proposals readily and receive substantial proposals from family agents, effectively integrating into existing family structures rather than forming an independent cooperative bloc.

The proposal distribution quantifies this asymmetry. Under symmetric skills, within-family proposals account for 34.2\% of total activity and family$\to$proself proposals for 34.3\%, whereas cross-family proposals comprise only 6.9\%. This indicates that cooperative energy flows primarily along family channels and outward to unaligned agents, but rarely across family boundaries. These patterns support Proposition~3: cooperation exhibits graded relational decay centered on family structures rather than distributed uniformly by individual preference type.

\subsection{Emergence of Authority Stratification (P4)}
\label{p4-authority-stratification}

\begin{table}[t]
\centering
\caption{Determinants of Decision Authority. Proposal intensity dominates in both conditions; communication, accumulative utility, and SIM show no robust independent main effects.}
\label{tab:authority_regression}
\scriptsize
\setlength{\tabcolsep}{3pt}
\renewcommand{\arraystretch}{0.9}
\begin{tabular}{lcccc}
\toprule
 & \multicolumn{2}{c}{Baseline Model} & \multicolumn{2}{c}{Interaction Model} \\
\cmidrule(lr){2-3} \cmidrule(lr){4-5}
 & Symm. & Comp. & Symm. & Comp. \\
\midrule
Proposal Intensity      
& $0.820^{***}$ & $0.780^{***}$ 
&               &               \\

Communication Intensity 
& $0.046$       & $-0.005$      
&               &               \\

Accumulative Utility    
& $0.096$       & $0.161$       
& $0.127$       & $-0.031$      \\

Social Identity (SIM)   
& $0.084$       & $-0.039$      
& $-0.296$      & $-0.112$      \\

Family Dummy            
& $-0.135$      & $0.086$       
& $0.298$       & $0.542^{*}$   \\

SIM $\times$ Family     
&               &               
& $0.603^{***}$ & $-0.089$      \\

Constant                
& $0.582^{**}$  & $0.284$       
& $-0.007$      & $-0.499$      \\

\midrule
Time FE      & Yes & Yes & Yes & Yes \\
Obs.         & 540 & 540 & 540 & 540 \\
$R^2$        & 0.697 & 0.688 & 0.113 & 0.100 \\
\bottomrule
\end{tabular}

\vspace{0.2em}
\footnotesize
\textit{Notes:} Dependent variable is standardized decision authority. Baseline models include proposal and communication controls; interaction models isolate SIM, family, and SIM$\times$Family effects. All models include turn fixed effects and group-clustered standard errors. Full specifications appear in Appendix~\ref{app:full-authority-regression}.
\end{table}

Table~\ref{tab:authority_regression} reports OLS estimates of decision authority on proposal intensity, communication intensity, accumulative utility, and social identity. In both conditions, proposal intensity is strongly and positively associated with decision authority (symmetric: $\beta=0.820$, $p<0.01$; complementary: $\beta=0.780$, $p<0.01$). Communication intensity, accumulative utility, and SIM, by contrast, show no robust independent main effects in the baseline models. 

This asymmetry indicates that authority stratification is driven primarily by action-based agenda control rather than by communicative effort, accumulated payoffs, or relational identity. These results support Proposition~4: durable authority emerges endogenously through proposal-driven mechanisms.

\begin{figure}
    \centering
    \includegraphics[width=\linewidth]{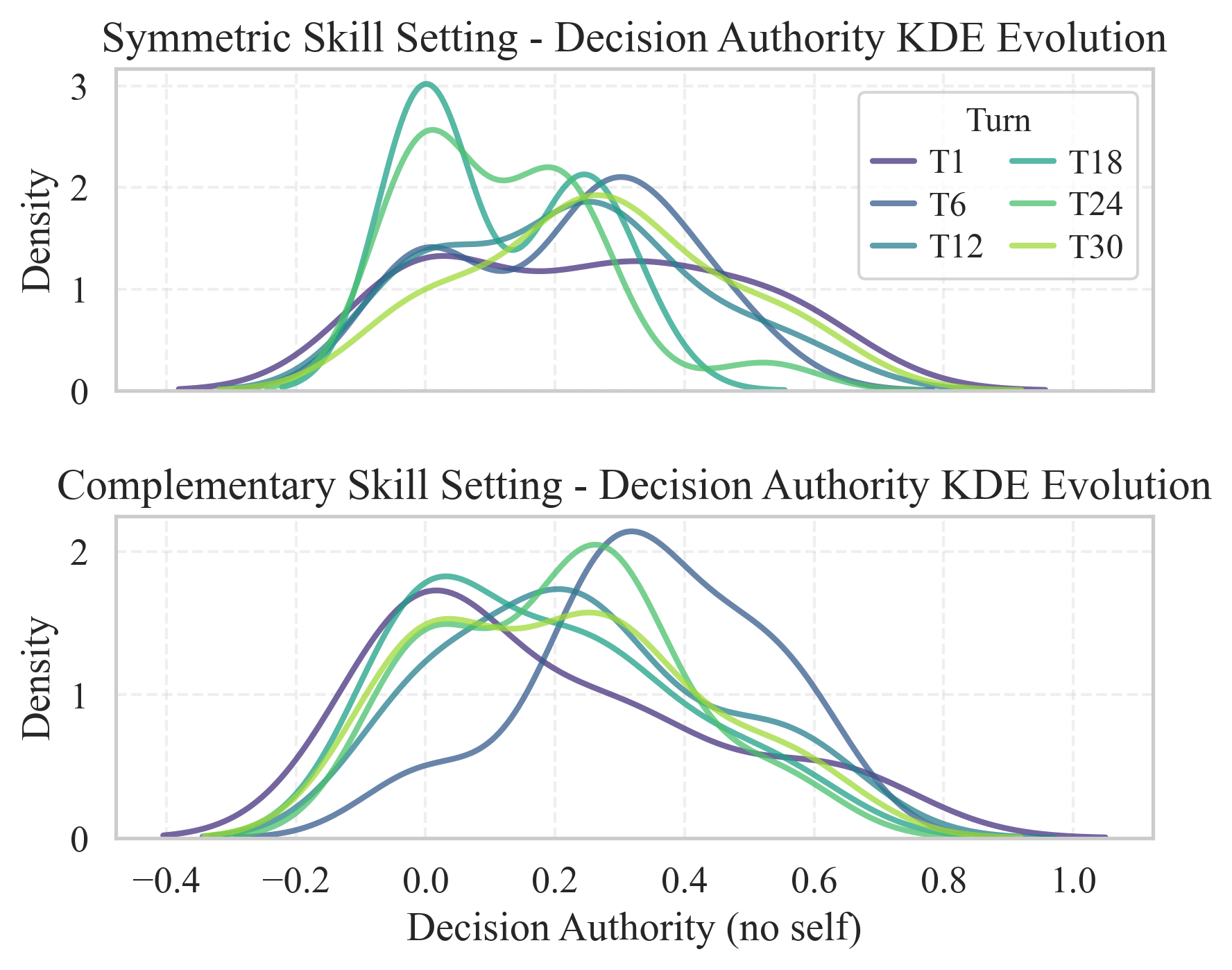}
    \caption{Temporal distribution of decision authority under symmetric and complementary skill structures. Kernel density estimates show persistent stratification rather than convergence to uniformity across rounds.}
    \label{fig:authority_kde}
    \vspace{-10pt}
\end{figure}

\subsection{Clan-Based Center--Periphery Stratification (P5)}
\label{p5-center-periphery}

Authority remains persistently stratified over time rather than diffusing uniformly. Figure~\ref{fig:authority_kde} shows that distributions stabilize into structured forms: under complementary skills, the right-skewed mass in high-authority regions sustains, indicating a stable center--periphery structure; under symmetric skills, differentiation is weaker but persists.

To examine the relational conditions of authority formation, we regress decision authority on social identity (SIM), family membership, and their interaction (Table~\ref{tab:authority_regression}). Under symmetric skills, neither SIM nor family membership alone predicts authority, but their interaction is strongly positive ($\beta = 0.603$, $p < 0.01$). Authority thus accrues disproportionately to agents whose social recognition is embedded in family relations. Under complementary skills, this interaction disappears, suggesting that functional differentiation attenuates relationally conditioned authority. This supports Proposition~5: center--periphery stratification emerges most clearly when family affiliation and social recognition align.

\subsection{Architecture Ablation and Baseline Analysis}
\label{sec:ablation}

\begin{table}[t]
\centering
\small
\caption{Cross-model ablation dissociation (symmetric endowments). Removing EC disrupts EC$\to$SIM but preserves ER$\to$SIM; removing ER does the reverse.}
\label{tab:ablation-dissociation}
\setlength{\tabcolsep}{2.8pt}
\begin{tabular}{lcccc}
\toprule
 & EC$\to$SIM & ER$\to$SIM & SIM$\times$Fam$\to$Auth \\
\midrule
DS Full       & $0.748^{**}$   & $0.573^{***}$   & $0.070$\,n.s.  \\
DS w/o EC     & ---            & $1.021^{***}$   & $0.029$\,n.s.  \\
DS w/o ER     & $0.170$\,n.s.  & ---             & $0.042$\,n.s.  \\
\midrule
GPT Full      & $2.216^{***}$  & $-0.002$\,n.s.  & $0.106^{*}$    \\
GPT w/o EC    & ---            & $0.498^{***}$   & $-0.197$\,n.s. \\
GPT w/o ER    & $-0.127^{*}$   & ---             & $0.181$\,n.s.  \\
\midrule
Gemini Full   & $0.069^{**}$   & $0.186^{***}$   & $0.089$\,n.s.  \\
Gemini w/o EC & ---            & $0.061^{*}$     & $0.097$\,n.s.  \\
Gemini w/o ER & $-0.022$\,n.s. & ---             & $-0.070$\,n.s. \\
\midrule
All BDI-only  & absent         & absent          & absent$^\dagger$ \\
\bottomrule
\end{tabular}

\vspace{0.2em}
\footnotesize
$^\dagger$BDI-only lacks SIM, the interaction term is therefore undefined;
Significance: $^{*}p<0.1$, $^{**}p<0.05$, $^{***}p<0.01$;~n.s. = not significant.
 \vspace{-12pt} 
\end{table}

Table~\ref{tab:ablation-dissociation} reports cross-model module ablations under symmetric endowments; detailed BDI-only comparisons are shown in Table~\ref{tab:bdi-baseline}.

Under symmetric endowments, removing EC selectively disrupts the EC$\to$SIM pathway while preserving ER$\to$SIM, whereas removing ER disrupts ER$\to$SIM while preserving EC$\to$SIM. This dissociation suggests that EC and ER make distinct, module-specific contributions to relational updating within CAREB-MAS.

The three models nonetheless recover the same macro-structural patterns through different intermediate pathways. DeepSeek-V3 draws on both EC and ER jointly; GPT-4o-mini relies primarily on affective-evaluation variance; Gemini-2.5-flash-lite shows an ethical-judgment-dominant profile (ER$\to$SIM $0.186^{***}$ vs.\ EC$\to$SIM $0.069^{**}$). That pathway divergence coexists with structural convergence is difficult to reconcile with a uniform RLHF-artifact explanation.

\paragraph{BDI-Only Baseline Comparison.}
Table~\ref{tab:bdi-baseline} compares the full architecture with a BDI-only baseline that removes EC, ER, and SIM while retaining vanilla decision logic. All indicators follow unified definitions (Appendix~\ref{app:metrics}).
\begin{table}[t]
\centering
\scriptsize
\caption{Full architecture vs.\ BDI-only baseline (symmetric endowments).}
\label{tab:bdi-baseline}
\setlength{\tabcolsep}{2.5pt}
\begin{tabular}{llccc}
\toprule
Prop. & Indicator & DS & GPT & Gemini \\
\midrule
P1 & Lock-in (last 5) & 0.971 / 0.979 & 0.971 / 0.979 & 0.833 / 0.833 \\
P1 & WGCS (F1)$\downarrow$ & 0.410 / 0.477$^{*}$ & 0.410 / 0.477$^{*}$ & 0.404 / 0.379\,n.s. \\
P2 & High/Low-SIM gap & present / n/a & present / n/a & $-0.022$\,n.s. / n/a \\
P3 & Cross-fam.\ proposals & 6.9\% / 0.7\% & 6.9\% / 0.7\% & 10.3\% / 1.3\% \\
P4 & DA mean (no-self) & 0.186 / 0.056 & 0.186 / 0.056 & 0.382 / 0.234 \\
P5 & SIM$\times$Fam$\to$Auth & n.s. / absent & $^{**}$ / absent & n.s. / absent \\
\bottomrule
\end{tabular}

\vspace{0.2em}
\footnotesize
\textit{Notes:} n/a: SIM absent in BDI-only. P4 BDI-only reductions: DeepSeek-V3 and GPT-4o-mini $-70$\%; Gemini-2.5-flash-lite $-39$\%. Significance: $^{*}p<0.05$, $^{**}p<0.01$.
\vspace{-10pt}
\end{table}

P1 lock-in remains high across both conditions ($0.83$--$0.98$), indicating that stable specialization itself is largely LLM-native. The social quality of this specialization, however, diverges. For DeepSeek-V3 and GPT-4o-mini, the full architecture produces more coherent within-group coordination (WGCS: $0.410$ vs.\ $0.477$, $p<0.05$). Gemini-2.5-flash-lite shows a directional but non-significant WGCS improvement under BDI-only ($0.404 \to 0.379$,$p=0.10$), yet this coincides with a collapse in cross-family proposals (P3: $10.3\% \to 1.3\%$) and a marked decline in decision authority (Gemini: $-39$\%; DeepSeek/GPT: $-70$\%).

Across all models, P3--P5 degrade directionally under BDI-only: the concentric cooperation gradient flattens, self-directed proposals increase, and relational authority stratification vanishes. Single-module ablations yield only modest degradation, but complete removal triggers discontinuous collapse in relational structure. This indicates that EC and ER operate as partially compensatory yet jointly necessary pathways for sustaining relational social structure.

Full ablation results are reported in Appendix~\ref{app:ablation-full}. The per-dyad specification in Appendix~\ref{app:ablation-perdyad} uses EC evaluation level, whereas Table~\ref{tab:ablation-dissociation} uses EC intra-agent variance; despite this operational difference, all three models recover the same directional EC$\to$SIM pathway.

\subsection{Bias Diagnosis: RLHF Artifact or Architectural Emergence?}
\label{sec:bias-diagnosis}

Observed cooperation patterns could stem from RLHF-induced prosocial bias, but four observations suggest they instead reflect architectural mechanisms.

First, agents exhibit behaviors that are not well explained by a uniformly prosocial RLHF account. Family agents show the lowest authority acceptance toward the opposing family (Figure~\ref{fig:authority_proposal_boxplots})---exclusionary rather than uniformly cooperative. Authority distributions remain right-skewed (Figure~\ref{fig:authority_kde}), and utility rankings reverse between conditions (Appendix~\ref{app:utility_kde}). These patterns replicate across DeepSeek-V3, GPT-4o-mini, and Gemini-2.5-flash-lite.

Second, ablation dissociation is module-specific (Table~\ref{tab:ablation-dissociation}). Removing EC eliminates the EC$\to$SIM pathway while sparing ER$\to$SIM, and removing ER does the reverse. If RLHF were the driver, module removal should produce uniform attenuation rather than targeted pathway elimination. This selective dissociation holds across all three models.

Third, the three models achieve convergent P1--P5 patterns through divergent intermediate mechanisms. DeepSeek-V3 operates through joint affective--ethical processing, GPT-4o-mini through affective-evaluation dominance, and Gemini-2.5-flash-lite through ethical-judgment dominance. GPT shows positive guanxi stickiness (P2 gap $= +0.03$--$0.04$), whereas DeepSeek and Gemini show negative or absent stickiness (Gemini: $-0.022$\,n.s.). Yet all three produce the same concentric cooperation gradient (P3) and authority stratification (P4). Divergence in P2 alongside convergence in P3--P4 is incompatible with a uniform RLHF explanation.

Fourth, BDI-only baseline collapse is directionally consistent across architectures. Removing EC, ER, and SIM degrades cross-family coordination and decision authority regardless of the underlying LLM: for Gemini, cross-family proposals drop from $10.3\%$ to $1.3\%$ and DA falls $39\%$; for DS/GPT, cross-family proposals drop from $6.9\%$ to $0.7\%$ and DA falls $70\%$. While magnitudes vary, the directional pattern---BDI-only weakens relational structure---indicates that observed social structure depends on the CAREB-MAS architecture rather than model-specific training artifacts.

\subsection{Summary: Structural Contrast}
\label{sec:mechanical-organic}

Across Propositions~1--5, the main contrast is structural: the same architecture yields different modes of integration under different production endowments. These outcomes vary systematically with production endowments: symmetric skills yield more kin-centered, bounded integration consistent with \emph{mechanical solidarity}, whereas complementary skills produce more functionally interdependent and less clan-bound organization closer to \emph{organic solidarity}. This contrast locates Differential Order Pattern at the level of structural conditions rather than as a fixed cultural template.

A related finding is the recurring decoupling between coordination burden and durable authority---an ``Atlas paradox''---which suggests that durable power remains anchored in agenda control rather than coordination service (Appendix~\ref{app:atlas-paradox}; Appendix~\ref{app:micro-politics}).

 \section{Conclusion}\label{sec:conclusion}

In this paper, we show that Differential Order Pattern, often treated as specific to Chinese rural society, can emerge from cross-culturally documented social mechanisms without encoding culture-specific institutional rules. We develop CAREB-MAS, a multi-agent framework that models emotional, ethical, and relational processes under minimal pre-market structural conditions. Across long-horizon simulations, CAREB-MAS reproduces five core Differential Order Pattern phenomena from local interaction alone. Production-structure contrasts and cross-model ablations show that these outcomes vary with structural conditions and depend on cognitive architecture rather than generic LLM behavior. These results support interpreting Differential Order Pattern as a structure-sensitive emergent outcome of general social mechanisms.


\section*{Limitations and Future Directions}
\label{sec:limitation}
The present study faces four limitations. First, EC, ER, and SIM updates scale as $O(N^2)$, restricting experiments to $N=18$ and limiting population-level claims. However, this configuration is a minimal setting for observing intra-clan cohesion, inter-clan negotiation, and cross-group integration within the same simulation. Second, the environment is simplified: we do not include explicit sanctioning, conflict-resolution institutions, or external shocks such as migration and scarcity. While this facilitates mechanism tracing, it limits external validity and leaves open whether such factors would fundamentally alter dynamics. Third, validation is driven more by theory than by direct ethnographic calibration; related empirical correspondence is discussed in Appendix~\ref{app:empirical-correspondence}. Stronger empirical correspondence requires longitudinal field comparison and boundary-condition testing. Fourth, although ablation results do not support a uniform RLHF-bias explanation, we cannot fully exclude training-data priors. Future work should address these limits through larger-scale simulations, more complex social settings, closer empirical comparison, and tests on a wider range of model families.

\section*{Acknowledgments}
This research was supported by the National Key R\&D Program of China (No.2023YFC3304701), NSFC (No.6250072448, No.62272466,U24A20233), Big Data and Responsible Artifcial Intelligence for National Governance, Renmin University of China; the Joint Academy on Future Humanity, Renmin University of China and Westlake University(No.2025RD0202) and Center for AI for Social Science, SLAI.



\bibliography{custom}

\begin{thebibliography}{35}
\providecommand{\natexlab}[1]{#1}

\bibitem[{Axelrod(1997)}]{axelrod1997dissemination}
Robert Axelrod. 1997.
\newblock The dissemination of culture: A model with local convergence and
  global polarization.
\newblock \emph{Journal of conflict resolution}, 41(2):203--226.

\bibitem[{Bhatt(2022)}]{bhatt2022ethical}
Babita Bhatt. 2022.
\newblock Ethical complexity of social change: Negotiated actions of a social
  enterprise.
\newblock \emph{Journal of Business Ethics}, 177(4):743--762.

\bibitem[{Dai et~al.(2024)Dai, Zhang, Li, Yang, Rao, Caetano, Sra
  et~al.}]{dai2024artificial}
Gordon Dai, Weijia Zhang, Jinhan Li, Siqi Yang, Srihas Rao, Arthur Caetano,
  Misha Sra, and 1 others. 2024.
\newblock Artificial leviathan: Exploring social evolution of llm agents
  through the lens of hobbesian social contract theory.
\newblock \emph{arXiv preprint arXiv:2406.14373}.

\bibitem[{Durkheim(1915)}]{durkheim1912elementary}
{\'E}mile Durkheim. 1915.
\newblock \href {https://www.gutenberg.org/ebooks/41360} {\emph{The Elementary
  Forms of the Religious Life}}.
\newblock George Allen \& Unwin, London.
\newblock Original work published 1912. Translated by J. W. Swain. Available at
  Project Gutenberg.

\bibitem[{Durkheim(2019)}]{durkheim2019division}
Emile Durkheim. 2019.
\newblock The division of labor in society.
\newblock In \emph{Social Stratification, Class, Race, and Gender in
  Sociological Perspective, Second Edition}, pages 178--183. Routledge.

\bibitem[{Egu{\'\i}luz et~al.(2005)Egu{\'\i}luz, Zimmermann, Cela-Conde, and
  Miguel}]{eguiluz2005cooperation}
V{\'\i}ctor~M Egu{\'\i}luz, Martin~G Zimmermann, Camilo~J Cela-Conde, and
  Maxi~San Miguel. 2005.
\newblock Cooperation and the emergence of role differentiation in the dynamics
  of social networks.
\newblock \emph{American journal of sociology}, 110(4):977--1008.

\bibitem[{Epstein and Axtell(1996)}]{epstein1996growing}
Joshua~M Epstein and Robert Axtell. 1996.
\newblock \emph{Growing artificial societies: social science from the bottom
  up}.
\newblock Brookings Institution Press.

\bibitem[{Fei et~al.(1992)Fei, Hamilton, and Zheng}]{fei1992soil}
Xiaotong Fei, Gary~G Hamilton, and Wang Zheng. 1992.
\newblock \emph{From the soil: The foundations of Chinese society}.
\newblock Univ of California Press.

\bibitem[{Fortes and Evans-Pritchard(1940)}]{fortes1940african}
Meyer Fortes and Edward~E Evans-Pritchard. 1940.
\newblock \emph{African Political Systems}.
\newblock Oxford University Press.

\bibitem[{Gao et~al.(2024)Gao, Lan, Li, Yuan, Ding, Zhou, Xu, and
  Li}]{gao2024large}
Chen Gao, Xiaochong Lan, Nian Li, Yuan Yuan, Jingtao Ding, Zhilun Zhou, Fengli
  Xu, and Yong Li. 2024.
\newblock Large language models empowered agent-based modeling and simulation:
  A survey and perspectives.
\newblock \emph{Humanities and Social Sciences Communications}, 11(1):1--24.

\bibitem[{Georgeff et~al.(1998)Georgeff, Pell, Pollack, Tambe, and
  Wooldridge}]{georgeff1998belief}
Michael Georgeff, Barney Pell, Martha Pollack, Milind Tambe, and Michael
  Wooldridge. 1998.
\newblock The belief-desire-intention model of agency.
\newblock In \emph{International workshop on agent theories, architectures, and
  languages}, pages 1--10. Springer.

\bibitem[{Granovetter(1985)}]{granovetter1985economic}
Mark Granovetter. 1985.
\newblock Economic action and social structure: The problem of embeddedness.
\newblock \emph{American Journal of Sociology}, 91(3):481--510.

\bibitem[{Hanting(2024)}]{hanting2024}
Zhang Hanting. 2024.
\newblock \href {https://doi.org/10.5539/ass.v20n4p59} {Chinese relational
  sociology: An analysis of the theoretical paradigm of guanxi studies in the
  chinese context}.
\newblock \emph{Asian Social Science}, 20(4):59.

\bibitem[{Haveman and Wetts(2019)}]{haveman2019}
Heather~A. Haveman and Rachel Wetts. 2019.
\newblock \href {https://doi.org/10.1111/soc4.12627} {Organizational theory:
  From classical sociology to the 1970s}.
\newblock \emph{Sociology Compass}, 13(3):e12627.

\bibitem[{Heise(2007)}]{Heise2007}
David~R. Heise. 2007.
\newblock \emph{Expressive Order: Confirming Sentiments in Social Actions}.
\newblock Springer, New York.

\bibitem[{Lazer et~al.(2021)Lazer, Hargittai, Freelon, Gonzalez-Bailon, Munger,
  Ognyanova, and Radford}]{lazer2021meaningful}
David Lazer, Eszter Hargittai, Deen Freelon, Sandra Gonzalez-Bailon, Kevin
  Munger, Katherine Ognyanova, and Jason Radford. 2021.
\newblock Meaningful measures of human society in the twenty-first century.
\newblock \emph{Nature}, 595(7866):189--196.

\bibitem[{L{\'e}vi-Strauss(1949)}]{levi1949structures}
Claude L{\'e}vi-Strauss. 1949.
\newblock \emph{Les Structures \'el\'ementaires de la Parent\'e}.
\newblock Presses Universitaires de France.

\bibitem[{Longyuan and Iftikhar(2023)}]{longyuan2023}
Bao Longyuan and Rukhsana Iftikhar. 2023.
\newblock \href {https://ojs.pssr.org.pk/journal/article/view/396} {Social and
  historical investigation of {{China}}’s ethnic minorities in 50s-60s of the
  20th century}.
\newblock \emph{Pakistan Social Sciences Review}, 7(4):664--672.

\bibitem[{Mauss(1925)}]{mauss1925gift}
Marcel Mauss. 1925.
\newblock \emph{The Gift: Forms and Functions of Exchange in Archaic
  Societies}.
\newblock Cohen \& West.

\bibitem[{Park et~al.(2023)Park, O'Brien, Cai, Morris, Liang, and
  Bernstein}]{park2023generative}
Joon~Sung Park, Joseph O'Brien, Carrie~Jun Cai, Meredith~Ringel Morris, Percy
  Liang, and Michael~S Bernstein. 2023.
\newblock Generative agents: Interactive simulacra of human behavior.
\newblock In \emph{Proceedings of the 36th annual acm symposium on user
  interface software and technology}, pages 1--22.

\bibitem[{Piao et~al.(2025)Piao, Yan, Zhang, Li, Yan, Lan, Lu, Zheng, Wang,
  Zhou et~al.}]{piao2025agentsociety}
Jinghua Piao, Yuwei Yan, Jun Zhang, Nian Li, Junbo Yan, Xiaochong Lan, Zhihong
  Lu, Zhiheng Zheng, Jing~Yi Wang, Di~Zhou, and 1 others. 2025.
\newblock Agentsociety: Large-scale simulation of llm-driven generative agents
  advances understanding of human behaviors and society.
\newblock \emph{arXiv preprint arXiv:2502.08691}.

\bibitem[{Polanyi(1944)}]{polanyi1944great}
Karl Polanyi. 1944.
\newblock \emph{The Great Transformation}.
\newblock Farrar \& Rinehart.

\bibitem[{Ren et~al.(2024)Ren, Cui, Song, Wang, and Hu}]{ren2024emergence}
Siyue Ren, Zhiyao Cui, Ruiqi Song, Zhen Wang, and Shuyue Hu. 2024.
\newblock \href {https://doi.org/10.24963/ijcai.2024/874} {Emergence of social
  norms in generative agent societies: Principles and architecture}.
\newblock In \emph{Proceedings of the Thirty-Third International Joint
  Conference on Artificial Intelligence, {IJCAI-24}}, pages 7895--7903.
  International Joint Conferences on Artificial Intelligence Organization.
\newblock Human-Centred AI.

\bibitem[{Sahlins(1972)}]{sahlins1972stone}
Marshall Sahlins. 1972.
\newblock \emph{Stone Age Economics}.
\newblock Aldine-Atherton.

\bibitem[{Schelling(1971)}]{schelling1971dynamic}
Thomas~C Schelling. 1971.
\newblock Dynamic models of segregation.
\newblock \emph{Journal of mathematical sociology}, 1(2):143--186.

\bibitem[{Sorokin(2017)}]{sorokin2017social}
Pitirim Sorokin. 2017.
\newblock \emph{Social and cultural dynamics: A study of change in major
  systems of art, truth, ethics, law and social relationships}.
\newblock Routledge.

\bibitem[{Tada(2020)}]{tada2020}
Mitsuhiro Tada. 2020.
\newblock \href {https://doi.org/10.1007/s11186-020-09394-1} {Language and
  imagined gesellschaft: Émile durkheim’s civil-linguistic nationalism and
  the consequences of universal human ideals}.
\newblock \emph{Theory and Society}, 49(4):597--630.

\bibitem[{Tajfel and Turner(1979)}]{Tajfel1979}
Henri Tajfel and John~C. Turner. 1979.
\newblock An integrative theory of intergroup conflict.
\newblock In William~G. Austin and Stephen Worchel, editors, \emph{The Social
  Psychology of Intergroup Relations}, pages 33--47. Brooks/Cole, Monterey, CA.

\bibitem[{Vallinder and Hughes(2025)}]{vallinder2024cultural}
Aron Vallinder and Edward Hughes. 2025.
\newblock Cultural evolution of cooperation among llm agents.
\newblock In \emph{Proceedings of the 24th International Conference on
  Autonomous Agents and Multiagent Systems}, AAMAS '25, page 2771–2773,
  Richland, SC. International Foundation for Autonomous Agents and Multiagent
  Systems.

\bibitem[{Wang et~al.(2025)Wang, Zhang, and
  Chen}]{wang-etal-2025-investigating}
Lei Wang, Zheqing Zhang, and Xu~Chen. 2025.
\newblock \href {https://doi.org/10.18653/v1/2025.acl-long.481} {Investigating
  and extending homans' social exchange theory with large language model based
  agents}.
\newblock In \emph{Proceedings of the 63rd Annual Meeting of the Association
  for Computational Linguistics (Volume 1: Long Papers)}, pages 9762--9777,
  Vienna, Austria. Association for Computational Linguistics.

\bibitem[{Zhang(2015)}]{ZhangJiangHua2015}
Jianghua Zhang. 2015.
\newblock ``folk society'' and beyond: A comparative study of fei xiaotong and
  robert redfield’s works on civilization studies.
\newblock \emph{Chinese Journal of Sociology}, 35(4):134--156.

\bibitem[{Zhang(2018)}]{zhang2018double}
Y~Zhang. 2018.
\newblock The double rupture: Central themes of historical anthropology in
  contemporary china.
\newblock \emph{Cargo: Revue Internationale D’anthropologie Culturelle \&
  Sociale}, 8:17--36.

\bibitem[{Zhou and Xiao(2024)}]{zhou2024b}
Daming Zhou and Mingyuan Xiao. 2024.
\newblock \href {https://doi.org/10.1186/s41257-024-00105-7} {Recent
  developments in anthropological methods for the study of complex societies}.
\newblock \emph{International Journal of Anthropology and Ethnology}, 8(1):4.

\bibitem[{Zimmermann et~al.(2004)Zimmermann, Egu{\'\i}luz, and
  San~Miguel}]{zimmermann2004coevolution}
Mart{\'\i}n~G Zimmermann, V{\'\i}ctor~M Egu{\'\i}luz, and Maxi San~Miguel.
  2004.
\newblock Coevolution of dynamical states and interactions in dynamic networks.
\newblock \emph{Physical Review E—Statistical, Nonlinear, and Soft Matter
  Physics}, 69(6):065102.

\bibitem[{Émile Durkheim(1961)}]{Durkheim1961MoralEducation}
Émile Durkheim. 1961.
\newblock \emph{Moral Education: A Study in the Theory and Application of the
  Sociology of Education}.
\newblock Free Press, New York.
\newblock Originally published as \textit{L'éducation morale} (1925).

\end{thebibliography}
\clearpage
\appendix
\label{sec:appendix}

\section{Experimental Parameterization.}
\label{app:hyperparams}

\begin{table*}[htbp]
\centering
\footnotesize
\caption{Experimental Hyperparameters}
\label{tab:exp-hyperparam}
\begin{tabular}{lcc} 
\toprule
\textbf{Parameter} & \textbf{Symm. Setting} & \textbf{Comp. Setting} \\ \midrule
Number of agents        & 18    & 18    \\
Family size (F1, F2)    & 6 each & 6 each \\
Proself group size      & 6     & 6     \\
Interaction rounds      & 30    & 30    \\
Skill Allocation (A:B:C) 
 & F1: 2:2:2 & F1: 4:1:1 \\
 & F2: 2:2:2 & F2: 1:4:1 \\
 & Proself: 2:2:2 & Proself: 1:1:4 \\
Ethical flexibility $e$ & [-0.5, 0.5] & [-0.5, 0.5] \\
Contextual coefficients $(a_A, a_B, a_C)$ & (1, 1, 1) & (1, 1, 1) \\
\bottomrule
\end{tabular}
\end{table*}

Table~\ref{tab:exp-hyperparam} summarizes the key hyperparameters used in both experimental settings. Each setting consists of 18 agents interacting over 30 rounds, with fixed population size and identical random seed configurations to ensure comparability. Agents are partitioned into two family groups (F1 and F2) and a smaller proself group, preserving a constant social composition across settings.

The Symmetric Setting adopts a balanced skill allocation, where all groups possess identical capabilities across resource types. The Complementary Setting introduces functional specialization by asymmetrically distributing skills across families and the proself group, while holding all other parameters constant. Ethical flexibility and contextual coefficients are fixed across settings to isolate the effect of skill structure on interaction dynamics and authority formation.

Each setting uses matched random seed configurations across the symmetric and complementary conditions to ensure comparability.
Each condition is repeated five times with independent stochastic realizations.

\subsection{Parameter Definitions}
\label{app:param-definitions}

Table~\ref{tab:param-definitions} provides formal definitions for all 
parameters referenced in the main text.

\begin{table}[htbp]
\centering
\small
\caption{Formal parameter definitions.}
\label{tab:param-definitions}
\begin{tabular}{lp{0.65\linewidth}}
\toprule
\textbf{Symbol} & \textbf{Definition} \\
\midrule
$N$ & Number of agents in the simulation (=18) \\
$T$ & Number of interaction rounds (=30) \\
$R$ & Number of deliberation sub-rounds per phase (=2) \\
$\mathcal{R}$ & Set of resource types ($\{A, B, C\}$) \\
$d_{i \to j}$ & Agent $i$'s subjective identity level toward agent $j$; 
                $d_{i \to j} \in \{1,2,3,4,5\}$ \\
$\bar{d}_{ij}$ & Symmetric relational proximity: 
                  $(\,d_{i \to j} + d_{j \to i}\,)/2$ \\
$e_{jb}$ & Ethical--Judgmental Bias; $e_{jb} \sim \mathrm{Uniform}(-0.5, 0.5)$; 
           modulates moral assessment stringency \\
$\boldsymbol{\ell}_i$ & Labor allocation vector for agent $i$; 
                         $\boldsymbol{\ell}_i = (\ell_{iA}, \ell_{iB}, \ell_{iC})$, 
                         $\sum_r \ell_{ir} = 1$ \\
$s_{i,t}(r)$ & Agent $i$'s skill level for resource $r$ at round $t$; 
               updated by learning-by-doing: 
               $s_{i,t+1}(r) = s_{i,t}(r) + \ell_{i,t}(r)$ \\
$\mathbf{p}_i$ & Consumption preference vector; 
                  $\mathbf{p}_i \sim \mathrm{Dir}(\mathbf{1}_3)$ \\
$b(r)$ & Resource-specific production unit (contextual coefficient); 
         $b(r) = 1\ \forall r$ in both conditions \\
$\lambda$ & Search randomness parameter for community edge formation (=0.3) \\
$u_{i,t}$ & Per-round Leontief utility: 
            $u_{i,t} = \min_{r} \{R_{i,t}(r) / p_i(r)\}$ \\
$U_{i,t}$ & Cumulative utility: $U_{i,t} = \sum_{\tau=1}^{t} u_{i,\tau}$ \\
\bottomrule
\end{tabular}
\end{table}

\section{Metric Definitions}
\label{app:metrics}

\subsection{Division and Matching Metrics}

\textbf{Division Lock-in Score.} For agent $i$ with preference vector $\mathbf{pref}_i$ and labor skill vector $\mathbf{skill}_i$:
\begin{equation}
\text{lock-in}_i = \frac{\mathbf{pref}_i \cdot \mathbf{skill}_i}{\|\mathbf{pref}_i\| \cdot \|\mathbf{skill}_i\|}
\end{equation}
Global division lock-in: $\text{Lock-in}_{\text{global}} = \frac{1}{N}\sum_{i=1}^{N} \text{lock-in}_i$.

\textbf{Division Lock-in Change.} Per-round change in global lock-in:
\begin{equation}
\Delta\text{Lock-in}^{(t)} = \text{Lock-in}_{\text{global}}^{(t)} - \text{Lock-in}_{\text{global}}^{(t-1)}
\end{equation}

\textbf{Multi-Level Division Matching.} Let $\mathbf{d}_S$, $\mathbf{d}_C$, $\mathbf{d}_G$, $\mathbf{d}_O$ denote demand vectors at social, community, group, and out-group levels:
\begin{align}
\text{social\_match}_i &= \cos(\mathbf{skill}_i, \mathbf{d}_S) \\
\text{community\_match}_i &= \cos(\mathbf{skill}_i, \mathbf{d}_C) \\
\text{group\_match}_i &= \cos(\mathbf{skill}_i, \mathbf{d}_G) \\
\text{outgroup\_match}_i &= \cos(\mathbf{skill}_i, \mathbf{d}_O)
\end{align}
Group-level matching (reported as ``Group matching'' in tables) is the mean of $\text{group\_match}_i$ across agents within the same Louvain-detected community at round $t$.

\textbf{Within-Group Coordination Score (WGCS).} 
WGCS quantifies the coherence of labor allocation decisions within a community by measuring the alignment between individual proposals and the community's aggregate demand profile. 

Let $C$ denote a community at round $t$, and let $\mathcal{P}_C = \{p^{(k)}_{i \to j} : i,j \in C, k \in K_i^{(C)}\}$ be the set of all labor proposals issued within $C$. Define the community proposal centroid as:
\begin{equation}
\bar{\mathbf{p}}_C = \frac{1}{|\mathcal{P}_C|} \sum_{p \in \mathcal{P}_C} \mathbf{p}
\end{equation}
where $\mathbf{p}$ is the vector representation of proposal $p$ in the labor-type space.

For each agent $i \in C$, let $\mathbf{a}_i^{(C)}$ denote $i$'s final labor allocation vector. The within-group coordination for agent $i$ is:
\begin{equation}
\text{WGCS}_i = \cos\!\big(\mathbf{a}_i^{(C)}, \bar{\mathbf{p}}_C\big)
\end{equation}
The community-level WGCS is the mean across members:
\begin{equation}
\text{WGCS}_C = \frac{1}{|C|}\sum_{i \in C} \text{WGCS}_i
\end{equation}

In practice, WGCS is reported as an F1-weighted variant to account for class imbalance across labor types:
\begin{equation}
\text{WGCS}_C^{\text{(F1)}} = \frac{2 \cdot \text{Precision}_C \cdot \text{Recall}_C}{\text{Precision}_C + \text{Recall}_C}
\end{equation}
where $\text{Precision}_C$ and $\text{Recall}_C$ are computed by treating the community centroid $\bar{\mathbf{p}}_C$ as the ``gold'' allocation and each agent's final allocation as a prediction. This F1 variant (reported as ``WGCS (F1)'' in tables) is more sensitive to coordination failures on minority labor types and is the primary metric used in ablation comparisons.

\subsection{Behavior Metrics}
\label{app:behavior_metrics}

\textbf{Decision Authority.} Let $p^{(k)}_{i \to j}$ denote the $k$-th labor allocation proposal issued by agent $i$ to agent $j$, and let $a_j$ denote agent $j$'s final allocation in the same round. Let $K_i^{(-i)} = \{k : j(k) \neq i\}$ be the set of proposals issued by agent $i$ to \emph{other} agents, with $|K_i^{(-i)}|$ its cardinality.

If $|K_i^{(-i)}| = 0$, decision authority is defined as $0$.

Otherwise, decision authority is computed as:
\begin{equation}
\text{Authority}_i = \frac{1}{w\sum_{k \in K_i^{(-i)}} \left\| p^{(k)}_{i \to j(k)} - a_{j(k)} \right\|^2},
\end{equation}
where $w=1 + \frac{1}{|K_i^{(-i)}|}$ and $j(k)$ denotes the target agent of proposal $k$ and $j(k) \neq i$ by construction.

Authority approaches $1$ when proposals to others closely match their final allocations and approaches $0$ as proposals diverge. Self-proposals are excluded to isolate influence over others' decisions.

\textbf{Proposal Action Intensity.}
\begin{equation}
\text{Proposal}_i = \sum_{j \neq i} \text{1}[\text{agent } i \text{ proposes to } j]
\end{equation}

\textbf{Proposal Directionality.} To distinguish self-directed from other-directed coordination behavior, we define:

\begin{gather}
\text{Self-proposals}_i = \sum_{k} \text{1}[p^{(k)}_{i \to i} \text{ exists}] \\
\text{Other-directed}_i = \sum_{j \neq i} \sum_{k} \text{1}[p^{(k)}_{i \to j} \text{ exists}] \\
\text{Total}_i = \text{Self-proposals}_i + \text{Other-directed}_i \\
\text{Other-directed \%}_i = \frac{\text{Other-directed}_i}{\text{Total}_i} \times 100
\end{gather}
Cross-family proposals are a subset of Other-directed proposals where $j$ belongs to a different family group than $i$.

\textbf{Communication Action Intensity.}
Communication action intensity quantifies the complexity and information content of an agent's public communication. For each speech act, an external LLM evaluator assigns a single composite score in the range $[0,10]$, aggregating four dimensions: relationship complexity, syntactic complexity, semantic information content, and argument logic (Table~\ref{tab:communication_prompt}). This score is used directly as the agent's communication action intensity.

\begin{table}[htbp]
\centering
\small
\caption{Prompt for evaluating communication action intensity.}
\begin{tabular}{p{0.95\linewidth}}
\hline
\textbf{LLM Evaluation Prompt} \\ \hline
Please analyze the language complexity and information content of the following public speech text, evaluating from the following dimensions:
(1) Relationship Complexity (complexity of relationships involved);
(2) Syntactic Complexity (complexity of sentence structure);
(3) Semantic Information Content (density and depth of information expressed);
(4) Argument Logic (rigor and logic of reasoning and argumentation).

Text content: ``\{text\}''

Please strictly follow the following JSON format and do not add any other content:
\{
\quad ``score'': number
\}

Where score is a comprehensive complexity score from 0--10, with higher scores indicating more complex language and greater information content. \\
\hline
\end{tabular}
\label{tab:communication_prompt}
\end{table}

\subsection{Social Identity Metrics}

\textbf{Social Identity Level (SIM).} Average SIM score agent $i$ assigns to community members $C_i$:
\begin{equation}
\text{SID}_i = \frac{1}{|C_i|}\sum_{j \in C_i} \text{SIM}(i, j)
\end{equation}

\textbf{In-Group vs.\ Cross-Group SIM Differentiation.} 
To quantify relational identity gradient (P3), we compute:
\begin{equation}
\begin{split}
\text{SIM\_gap}_i &= \frac{1}{|F_i \cap C_i|}\sum_{j \in F_i \cap C_i} \text{SIM}(i,j) \\
&\quad -\; \frac{1}{|C_i \setminus F_i|}\sum_{j \in C_i \setminus F_i} \text{SIM}(i,j)
\end{split}
\end{equation}
where $F_i$ denotes agent $i$'s family group. Positive values indicate stronger relational identity toward in-family members within the same community. The population-level P2/P3 gap reported in tables is the mean of $\text{SIM\_gap}_i$ across agents (or, for DS/GPT, the difference in community-match between high-SIM and low-SIM tertiles under low-utility conditions; see footnote in Table~\ref{tab:bdi-baseline} for Gemini operational difference).

\subsection{Utility and Adaptation Metrics}

\textbf{Total and Average Utility.}
\begin{equation}
U_{\text{total}} = \sum_{i=1}^{N} U_i, \quad \bar{U} = \frac{U_{\text{total}}}{N}
\end{equation}

\textbf{Community--Match Change/Adaptive Division.}
We define the change in community match as:
\begin{equation}
\Delta M = M^{\text{post}} - M^{\text{pre}},
\end{equation}
where $M^{\text{pre}}$ and $M^{\text{post}}$ denote the match score before and after community re-division, respectively. Positive values of $\Delta M$ indicate improved alignment with the new community structure.

\textbf{Action Effectiveness}
\label{app:action-effectiveness}

To capture how effectively agents translate proposal activity into material outcomes, we define \emph{action effectiveness} as a proposal-normalized measure of economic impact.

Let $U_{i}$ denote the cumulative economic utility obtained by agent $i$ over the entire simulation horizon. Let $P_{i}$ denote the agent's average proposal activity, measured as the mean number of proposals submitted per interaction round. To avoid division by zero, we set $P_i = \max(P_i, \epsilon)$ with a small constant $\epsilon > 0$.

We define action effectiveness as:
\begin{equation}
\label{eq:action-effectiveness}
\mathrm{AE}_{i}
=
\frac{\lvert U_{i} \rvert}{P_{i}}.
\end{equation}

Importantly, action effectiveness does not assume that all proposals are adopted, nor does it model counterfactual outcomes. Instead, it provides a behavioral efficiency proxy that links realized economic outcomes to communicative and decision-making effort.

This measure is used in Section~\ref{p5-center-periphery} to examine how material influence interacts with decision authority and center--periphery stratification.

\textbf{Late Labor Concentration.} 
To capture temporal dynamics of specialization, we define late-stage labor concentration as the Gini coefficient of agent $i$'s labor allocation vector $\boldsymbol{\ell}_i^{(T_{\text{late}})}$ averaged over the final 20\% of simulation rounds:
\begin{equation}
\text{LateConc}_i = \text{Gini}\!\left(\frac{1}{|T_{\text{late}}|}\sum_{t \in T_{\text{late}}} \boldsymbol{\ell}_i^{(t)}\right)
\end{equation}
Higher values indicate more focused specialization in the late stage of community formation.

\section{Concept--Metric Mapping and Justification}
\label{app:concept-metric}

To ensure that simulated patterns reflect theoretically grounded mechanisms rather than arbitrary computational artifacts, we provide a two-level mapping: (1) ethnographic concepts to computational metrics, and (2) generative propositions to their quantitative operationalizations and diagnostic criteria.

\subsection{Ethnographic Concept--Metric Mapping}

Table~\ref{tab:empirical_mapping} formalizes the correspondence between key ethnographic constructs in \textit{From the Soil} \citep{fei1992soil} and their computational representations.

\begin{table}[h]
\centering
\small
\begin{tabular}{p{0.25\linewidth} | p{0.3\linewidth} | p{0.35\linewidth}}
\toprule
\textbf{Ethnographic Concept} & \textbf{Computational Metric} & \textbf{Simulated Phenomenon} \\
\midrule
\textit{Differential Order} (Ch. 4)\newline The ``self-centered web'' & Relational Distance ($d_{ij}$) extracted from the dynamic Social Identity Matrix (SIM). & Cooperation and proposal intensity decay predictably outward from the ego-center. \\
\midrule
\textit{Relational Ethics} (Ch. 8)\newline Rules of propriety over law & SIM Match vs. Utility Gap. & High-SIM relations sustain cooperation and community matching even under severe utility deficits. \\
\midrule
\textit{Elder Authority} (Ch. 11)\newline Exchange-based power & Decision Authority (Proposal adoption rate normalized by community). & Authority concentrates in individuals providing durable coordination, decoupled from sheer verbosity. \\
\bottomrule
\end{tabular}
\caption{Mapping between ethnographic constructs and computational metrics.}
\label{tab:empirical_mapping}
\end{table}

\subsection{Proposition--Metric Mapping}

Table~\ref{tab:prop-metric} summarizes how each generative proposition is operationalized via a primary quantitative metric and a diagnostic criterion specifying the expected empirical pattern under Differential Order.

\begin{table}[h]
\centering
\small
\caption{Proposition--metric mapping.}
\label{tab:prop-metric}
\begin{tabular}{@{}p{0.08\linewidth}p{0.36\linewidth}p{0.46\linewidth}@{}}
\toprule
 & \textbf{Primary Metric} & \textbf{Diagnostic Criterion} \\
\midrule
\textbf{P1} & Lock-in Score (output--skill cosine) & High convergence with near-zero per-round change \\
\textbf{P2} & Community Match gap (high vs.\ low SIM) & Maintained community match despite low utility \\
\textbf{P3} & DA \& Proposal Activity by source--target group pair & Within-family $>$ family$\to$proself $>$ cross-family gradient \\
\textbf{P4} & DA density over time; hierarchical OLS & Proposal intensity predicts authority beyond identity or wealth \\
\textbf{P5} & SIM$\times$Family interaction (OLS) & Joint conditioning of social identity and family on authority \\
\bottomrule
\end{tabular}
\end{table}

\subsection{Justification of Metric Choices}

This section provides extended justification for the operationalization choices summarized above.

\subsubsection{P1: Lock-in Score}

\paragraph{Why Lock-in Score rather than raw specialization index?}
P1 predicts that without central coordination, individuals lock into 
specialized roles. Lock-in Score is defined as the cosine similarity 
between an agent's production allocation vector and its skill vector 
(Appendix~\ref{app:metrics}). This directly measures whether 
specialization has occurred \emph{and stabilized}: convergence to high 
values indicates specialization; near-zero per-round change indicates 
equilibrium. Alternative measures—such as Herfindahl index—capture 
specialization intensity but not its stability or alignment with skill. 
Lock-in Score captures both dimensions simultaneously.

\subsubsection{P2: Community Match by SIM $\times$ Utility}

\paragraph{Why community match rather than raw cooperation frequency?}
P2 predicts that relational obligation sustains cooperation under 
adversity. We partition agents by SIM level (high vs.\ low) and 
recent utility (high vs.\ low), then compare community match—the 
cosine similarity between an agent's skill vector and community 
demand—across these groups.

Community match captures whether agents continue to \emph{align their 
labor with community needs}, rather than merely engaging in interaction. 
Raw cooperation frequency would measure interaction volume without 
distinguishing productive cooperation from empty gestures. The key test 
is whether high-SIM agents maintain community match under low utility 
while low-SIM agents do not.

\subsubsection{P3: Source--Target Group Decomposition}

\paragraph{Why decompose DA and proposals by group pairs?}
P3 predicts ego-centric, distance-graded cooperation. We decompose 
Decision Authority and Proposal Activity by source--target group pairs 
(within-family, family$\to$proself, cross-family). A monotonic gradient 
within-family $>$ family$\to$proself $>$ cross-family constitutes 
direct evidence of the concentric structure predicted by Fei's theory. 
This preserves the \emph{directional} and \emph{relational} structure 
lost in aggregate measures.

\subsubsection{P4: DA Temporal KDE + OLS}

\paragraph{Why combine distributional and regression evidence?}
P4 predicts that authority emerges endogenously. Two distinct claims 
require separate tests:
\begin{enumerate}[nosep,leftmargin=*]
    \item Authority \emph{exists} as a differentiated phenomenon 
    (tested via temporal KDE showing persistent stratification).
    \item Authority is \emph{emergent} rather than \emph{ascribed} 
    (tested via OLS showing proposal intensity predicts authority 
    beyond identity, family, or wealth).
\end{enumerate}
Neither test alone suffices.

\subsubsection{P5: SIM$\times$Family Interaction}

\paragraph{Why the interaction term rather than main effects?}
P5 predicts clan-based center--periphery stratification. The 
SIM$\times$Family interaction tests whether social identity alignment 
with family membership \emph{jointly} conditions authority. This 
follows the standard notion of relational embeddedness 
\citep{granovetter1985economic}: advantage accrues not from family or 
recognition alone, but from their alignment within the network. The 
disappearance of this interaction under complementary skills 
($\beta=-0.089$, n.s.) while strong under symmetric skills 
($\beta=0.603^{***}$) directly tests structure sensitivity.

\section{LLM and Reproducibility Details}
\label{app:llm_details}

\subsection{Model Configuration}

\begin{table}[h]
\centering
\small
\caption{LLM Configuration}
\label{tab:llm_config}
\begin{tabular}{ll}
\toprule
\textbf{Parameter} & \textbf{Value} \\
\midrule
Model & DeepSeek-V3 \\
Temperature & 0.7 \\
Top-$p$ & 0.9 \\
Max tokens & 512 \\
\bottomrule
\end{tabular}
\end{table}

\subsection{Agent System Prompt}
\label{app:agent_system_prompt}

Each agent is initialized with a fixed system prompt that defines its role, decision principles, and subjective preferences. This prompt establishes the agent as a socially embedded decision-maker whose behavior is guided by personal values, social relationships, and material preferences.

\begin{table}[htbp]
\centering
\small
\caption{System prompt for agent role initialization.}
\begin{tabular}{p{0.95\linewidth}}
\hline
\textbf{System Prompt} \\ \hline
You are \{agent\_name\}, a person with labor capacity and your own preferences. Your decisions should be guided by your personal values (SVO), your relationships with others (identity matrix), and your desire to achieve personal and community goals. You will communicate, negotiate, and make decisions about how to allocate labor for producing essential resources. Your consumption preferences are: [\{preference\_string\}]. \\
\hline
\end{tabular}
\label{tab:agent_system_prompt}
\end{table}

\paragraph{Initialization Details.}
For each agent, the placeholder \{preference\_string\} is instantiated as a list of resource-specific consumption preferences, formatted to two decimal places. In addition, each agent is assigned an ethical--judgmental bias value $e_{jb} \sim \mathrm{Uniform}(-0.5, 0.5)$ at initialization, which is stored as an internal trait and subsequently injected into the agent’s contextual prompt during decision-making and social identity updates.

\subsubsection{EC Perception Prompt}
\label{app:ec_perception_prompt}

The EC (Emotional--Cognitive) perception module captures an agent’s subjective interpretation of individual observable actions performed by others. For each public action in the previous turn, the agent evaluates the action through its own beliefs, resources, skills, and social relationships, producing a multi-dimensional affective assessment.

\begin{table}[htbp]
\centering
\small
\caption{Prompt for EC-based subjective perception of observed actions.}
\begin{tabular}{p{0.95\linewidth}}
\hline
\textbf{LLM Perception Prompt} \\ \hline
You are a social agent. Your task is to interpret another agent's action from your own subjective perspective, based on your current beliefs, resources, and social relationships.

\textbf{Context Provided:}
\begin{itemize}
    \item Your family membership (if any);
    \item Your current skills and resource holdings;
    \item Your Social Identity Matrix (SIM), encoding existing relationships and trust levels;
    \item A unified factual record of all observable actions from the previous turn.
\end{itemize}

\textbf{Action to Interpret:}
The identity of the acting agent and a textual description of the action performed.

\textbf{Task:}
Rate the action on the following five subjective dimensions, each using an integer scale from 1 to 5:
\begin{enumerate}
    \item \textbf{Evaluation}: perceived benefit or harm to oneself;
    \item \textbf{Potency}: perceived strength or weakness of the actor;
    \item \textbf{Activity}: perceived intensity or emotional activation of the action;
    \item \textbf{Status-Conferral}: perceived respect or status acknowledgment;
    \item \textbf{Power-Assertion}: perceived attempt to command, negotiate, or submit.
\end{enumerate}

\textbf{Output Format:}
Return a single JSON object containing the five ratings. No additional text or explanation is allowed. \\
\hline
\end{tabular}
\label{tab:ec_perception_prompt}
\end{table}

\subsubsection{ER Judgment Prompt}
\label{app:er_judgment_prompt}

The ER (Ethical Resonator) module produces a holistic moral judgment of each other agent’s overall behavior within a single interaction round. Rather than evaluating individual actions in isolation, this module aggregates prior EC perceptions and applies an agent-specific ethical orientation to form a unified moral assessment.

\begin{table}[htbp]
\centering
\small
\caption{Prompt for ER-based holistic moral judgment of agents.}
\begin{tabular}{p{0.95\linewidth}}
\hline
\textbf{LLM Judgment Prompt} \\ \hline
You are a social agent reflecting on the overall conduct of another agent during the previous turn.

\textbf{Context Provided:}
\begin{itemize}
    \item Your ethical--judgmental bias (EJB), indicating your default moral orientation;
    \item Your Social Identity Matrix (SIM), encoding existing relationships and trust levels;
    \item Your family membership (if any), skills, and resource holdings;
    \item A summarized record of all actions performed by the target agent in the last turn, paired with your own subjective perceptions of those actions.
\end{itemize}

\textbf{Task:}
Form a \emph{holistic moral judgment} of the target agent’s overall behavior during the turn. The judgment should synthesize patterns across actions and reflect your ethical orientation.

\textbf{Outputs:}
Produce a single JSON object containing:
\begin{itemize}
    \item \texttt{moral\_judgement}: a brief (1--2 sentence) overall moral evaluation of the agent’s conduct;
    \item \texttt{ethical\_attitude}: one categorical label selected from \{\texttt{Approbatory}, \texttt{Repressive}, \texttt{Restorative}\}, indicating praise, condemnation, or corrective orientation.
\end{itemize}

No additional text or explanation is allowed outside the JSON object. \\
\hline
\end{tabular}
\label{tab:er_judgment_prompt}
\end{table}

\subsection{Social Identity Matrix (SIM) Update Mechanism}
\label{app:sim_update}

Agents update their social identity assessments through an LLM-mediated evaluation process at the end of each interaction round. The Social Identity Matrix (SIM) records, for each agent pair, a discrete identity level capturing trust and relational closeness on an ordinal scale.

For each agent $i$, the SIM stores an identity level $s_{i,j} \in \{1,2,3,4,5\}$ for every other community member $j$, where higher values indicate stronger affiliation and trust. 

\begin{table}[htbp]
\centering
\small
\caption{Prompt for updating Social Identity Matrix (SIM) and BDI states.}
\begin{tabular}{p{0.95\linewidth}}
\hline
\textbf{LLM Update Prompt} \\ \hline
As agent \{agent\_name\}, review the events from the current interaction round to update your internal state and social relationships.

\textbf{Context Provided:}
\begin{itemize}
    \item Personal profile, including ethical orientation and (when applicable) family membership;
    \item Current resource holdings and cumulative utility;
    \item Existing social identity levels for all community members;
    \item A factual summary of the round’s discussion, production outcomes, and resource distribution.
\end{itemize}

\textbf{Task:}
Update your Belief--Desire--Intention (BDI) state and revise social identity levels for \emph{all} other community members based solely on the information above.

\textbf{Identity Update Rules:}
\begin{itemize}
    \item Identity levels must be integers in $\{1,2,3,4,5\}$;
    \item Higher values indicate stronger trust and affiliation;
    \item Updates should follow predefined upgrade and downgrade triggers grounded in social identity theory (e.g., cooperation, commitment fulfillment, exploitation, or indifference).
\end{itemize}

\textbf{Output Format:}
Return a single JSON object containing:
\begin{itemize}
    \item an updated BDI state (beliefs, desires, intentions);
    \item updated identity levels for all other agents.
\end{itemize}

No additional text or formatting is allowed outside the JSON object. \\
\hline
\end{tabular}
\label{tab:sim_update_prompt}
\end{table}

\subsubsection{A-BDI Intention Prompt}
\label{app:abdi_intention_prompt}

The A-BDI (Agent Belief--Desire--Intention) module governs agents’ forward-looking decision-making. At the end of each turn, agents sequentially update beliefs, desires, and intentions, where intentions constitute a concrete action plan for the subsequent turn. This process integrates social identity, moral judgment, emotional perception, and factual interaction history.

\begin{table}[htbp]
\centering
\small
\caption{Prompt for A-BDI intention generation.}
\begin{tabular}{p{0.95\linewidth}}
\hline
\textbf{LLM Intention Prompt} \\ \hline
You are a strategic social agent formulating a concrete plan for the next interaction round.

\textbf{Context Provided:}
\begin{itemize}
    \item Your current Social Identity Matrix (SIM), encoding allies, neutral agents, and adversaries;
    \item Your belief statement summarizing your perceived social position and landscape;
    \item Your desire statement specifying your primary goals for the next turn;
    \item Your holistic ethical judgments of other agents’ behavior from the previous turn;
    \item Your subjective emotional perceptions of others’ actions (EC records);
    \item A unified factual log of all observable events from the previous turn.
\end{itemize}

\textbf{Task:}
Based on \emph{all} contextual information above, generate your intentions for the next turn. The intentions should describe a concrete and actionable plan, including:
\begin{itemize}
    \item which agents to target or engage;
    \item what communicative or strategic actions to take;
    \item how these actions advance your stated desires.
\end{itemize}

\textbf{Output Format:}
Return a single string describing your intention statement. No additional text or formatting is allowed. \\
\hline
\end{tabular}
\label{tab:abdi_intention_prompt}
\end{table}

\subsection{Multi-Phase Interaction and Decision Pipeline}
\label{app:interaction_pipeline}

Each interaction round proceeds through a fixed sequence of 4 phases. 

\subsubsection{Phase 1: Community Detection}
\label{app:community_detection_phase}

At the beginning of each interaction round, agents are partitioned into communities based on their current social relationships. This phase operationalizes endogenous community formation driven by social identity rather than exogenous grouping.

\paragraph{Network Construction.}
A weighted, undirected graph $G=(V,E)$ is constructed over the set of agents, where each node represents an agent. For each unordered agent pair $(i,j)$, the Social Identity Matrix (SIM) provides directed identity levels $s_{i \rightarrow j}$ and $s_{j \rightarrow i}$, each taking integer values in $\{1,\dots,5\}$. The average identity strength is computed as:
\begin{equation}
\bar{s}_{ij} = \frac{s_{i \rightarrow j} + s_{j \rightarrow i}}{2}.
\end{equation}

\paragraph{Probabilistic Edge Formation.}
Edges are added probabilistically based on social affinity. For each pair $(i,j)$, a base propensity for community linkage is defined as:
\begin{equation}
p^{\text{base}}_{ij} = 0.1 \cdot (\bar{s}_{ij} - 1),
\end{equation}
where higher identity levels (4--5) induce stronger connection tendencies, level 3 is neutral, and levels 1--2 correspond to weak or unlikely ties. A stochastic component is added to capture exploratory or opportunistic interactions:
\begin{equation}
p_{ij} = p^{\text{base}}_{ij} + \epsilon, \quad \epsilon \sim \mathrm{Uniform}(0, \lambda),
\end{equation}
where $\lambda$ is a configurable search randomness parameter. An edge between $i$ and $j$ is added with probability $p_{ij}$ and assigned weight $\bar{s}_{ij}$.

\paragraph{Community Partitioning.}
Given the resulting weighted graph, communities are identified using the Louvain modularity maximization algorithm. This yields a partition of agents into disjoint communities that maximize within-group social cohesion as measured by identity-weighted connectivity.

\paragraph{Community Assignment.}
Each agent is assigned to exactly one community per round. Community membership is stored and used to scope subsequent negotiation, coordination, and production phases. Community composition is therefore endogenous, dynamic, and directly shaped by evolving social identities.

\subsubsection{Phase 2: Negotiation Phase}
\label{app:negotiation_phase}

The negotiation phase consists of up to a fixed number of discussion rounds within each community. Each round includes (i) public speeches and (ii) proposal-based coordination over labor allocation.

\paragraph{Public Speech Prompt.}
In the first step of each discussion round, agents may issue a public speech visible to all community members. The speech is optional and limited to concise free-form communication.

\begin{table}[htbp]
\centering
\small
\caption{Prompt for public speech during negotiation.}
\begin{tabular}{p{0.95\linewidth}}
\hline
\textbf{LLM Speech Prompt} \\ \hline
As \{agent\_name\}, it is your turn to make a public speech to your community.

\textbf{Context Provided:}
\begin{itemize}
    \item Your current desires (from the A-BDI module);
    \item The list of community members;
    \item A summary of previous speeches in the current turn.
\end{itemize}

\textbf{Task:}
Freely communicate with the community. You may express goals, react to others, ask questions, or remain silent.

\textbf{Constraints:}
Your speech should be concise (0--3 sentences). An empty response indicates silence. \\
\hline
\end{tabular}
\label{tab:public_speech_prompt}
\end{table}

\paragraph{Proposal Prompt.}
In the second step, agents generate concrete labor allocation plans for themselves and may optionally propose plans for up to three other agents.

\begin{table}[htbp]
\centering
\small
\caption{Prompt for labor allocation proposals.}
\begin{tabular}{p{0.95\linewidth}}
\hline
\textbf{LLM Proposal Prompt} \\ \hline
You are coordinating with your community to achieve your goals.

\textbf{Context Provided:}
\begin{itemize}
    \item Your personal profile (ethical orientation, family membership);
    \item Your current BDI state (beliefs, desires, intentions);
    \item Community members and your social identity levels toward them;
    \item Your preferences and labor skills;
    \item Public speeches and proposals made so far in this turn.
\end{itemize}

\textbf{Task:}
Formulate your own labor allocation plan and optionally suggest up to three plans for other agents.

\textbf{Output Format:}
Return a single JSON object specifying:
\begin{itemize}
    \item your own labor allocation plan;
    \item zero to three suggested plans for other agents.
\end{itemize} \\
\hline
\end{tabular}
\label{tab:labor_proposal_prompt}
\end{table}

\subsubsection{Phase 3: Labor Division Decision Phase}
\label{app:labor_division_phase}

After negotiation concludes, each agent independently finalizes a binding labor allocation decision.

\begin{table}[htbp]
\centering
\small
\caption{Prompt for final labor allocation decision.}
\begin{tabular}{p{0.95\linewidth}}
\hline
\textbf{LLM Final Decision Prompt} \\ \hline
It is time to finalize your labor allocation for the current round.

\textbf{Context Provided:}
\begin{itemize}
    \item Public speeches made this turn;
    \item Proposals directed to you during negotiation;
    \item Your personal profile (ethical orientation and family membership).
\end{itemize}

\textbf{Task:}
Decide your final labor allocation. The allocation must sum to 1.0.

\textbf{Output Format:}
Return a JSON object containing your final reasoning and labor allocation. \\
\hline
\end{tabular}
\label{tab:final_labor_prompt}
\end{table}

\subsubsection{Phase 4a: Production}
\label{app:production_phase}

Given a binding labor allocation, each agent deterministically produces resources as a function of (i) a resource-specific production unit and (ii) its current skill. Let $\mathcal{R}$ denote the resource set. For agent $i$ at round $t$, let $\mathbf{l}_{i,t}=(l_{i,t}(r))_{r\in\mathcal{R}}$ be its labor allocation with $l_{i,t}(r)\ge 0$ and $\sum_{r\in\mathcal{R}} l_{i,t}(r)=1$. Let $\mathbf{s}_{i,t}=(s_{i,t}(r))_{r\in\mathcal{R}}$ denote its labor skill vector, and let $\mathbf{b}=(b(r))_{r\in\mathcal{R}}$ be the resource-specific production unit (a fixed configuration parameter).

\paragraph{Production function.}
Production for each resource $r$ is computed as:
\begin{equation}
y_{i,t}(r) \;=\; b(r)\, l_{i,t}(r)\, \log\!\bigl(1+s_{i,t}(r)\bigr), \forall r\in\mathcal{R}.
\end{equation}
This implies diminishing returns in skill via the $\log(1+s)$ term, while output is linear in labor allocation and the base production unit.

\paragraph{Resource update.}
Produced goods are immediately added to the agent's private resource holdings:
\begin{equation}
R_{i,t}(r) \;\leftarrow\; R_{i,t}(r) + y_{i,t}(r), \qquad \forall r\in\mathcal{R}.
\end{equation}

\paragraph{Skill accumulation (learning-by-doing).}
After production, skills increase additively by the time allocated to each resource:
\begin{equation}
s_{i,t+1}(r) \;=\; s_{i,t}(r) + l_{i,t}(r), \qquad \forall r\in\mathcal{R}.
\end{equation}
Thus, allocating labor to a resource both produces that resource in the current round and increases future productivity on that dimension.

The realized production vector $\mathbf{y}_{i,t}$ is recorded as the agent's turn-level output for logging and for downstream clearing.

\subsubsection{Phase 4b: Clearing (Pooling and Redistribution)}
\label{app:clearing_phase}

The clearing phase pools produced goods at the community level and redistributes them proportionally to agents' consumption preferences. Let $g$ be a community and $\mathcal{I}_g$ its set of agents.

\paragraph{Pooling.}
For each resource $r\in\mathcal{R}$, the community-level pot is:
\begin{equation}
Y_{g,t}(r) = \sum_{i\in\mathcal{I}_g} y_{i,t}(r).
\end{equation}
In code, each agent first contributes all produced goods to the pot (removing them from private holdings), and then receives a redistribution share.

\paragraph{Preference-weighted redistribution.}
Each agent $i$ has a nonnegative preference weight $p_i(r)$ for each resource $r$. Define total preference weight within community $g$:
\begin{equation}
P_{g}(r) = \sum_{i\in\mathcal{I}_g} p_i(r).
\end{equation}
Agent $i$'s share ratio for resource $r$ is:
\begin{equation}
\alpha_{i,g}(r)=
\begin{cases}
\frac{p_i(r)}{P_{g}(r)}, & \text{if } P_{g}(r)>0,\\
0, & \text{otherwise.}
\end{cases}
\end{equation}
The redistributed amount received by agent $i$ is then:
\begin{equation}
x_{i,t}(r) = \alpha_{i,g}(r)\, Y_{g,t}(r).
\end{equation}
After clearing, agent $i$'s resource holdings are incremented by $\mathbf{x}_{i,t}=(x_{i,t}(r))_{r\in\mathcal{R}}$.

\paragraph{Degenerate case.}
If $|\mathcal{I}_g|\le 1$ (a singleton community), no pooling/redistribution occurs and the agent simply keeps its produced goods.

\subsubsection{Phase 4c: Consumption and Utility}
\label{app:consumption_phase}

In the consumption phase, each agent converts redistributed resources into utility via a Leontief-style (bottleneck) utility rule implied by the implementation. Let $\mathbf{R}_{i,t}=(R_{i,t}(r))_{r\in\mathcal{R}}$ denote agent $i$'s resource holdings immediately before consumption, and let $\mathbf{p}_i=(p_i(r))_{r\in\mathcal{R}}$ denote its preference vector.

\paragraph{Per-round utility.}
Define resource-to-preference ratios:
\begin{equation}
\rho_{i,t}(r)=
\begin{cases}
\frac{R_{i,t}(r)}{p_i(r)}, & \text{if } p_i(r)>0,\\
+\infty, & \text{if } p_i(r)=0.
\end{cases}
\end{equation}
The per-round utility is the minimum ratio across resources:
\begin{equation}
u_{i,t}=\min_{r\in\mathcal{R}} \rho_{i,t}(r),
\end{equation}
with the convention that if all $\rho_{i,t}(r)=+\infty$ (i.e., $p_i(r)=0$ for all $r$), then $u_{i,t}=0$.

The limiting resource (bottleneck) is:
\begin{equation}
r^\star_{i,t} \in \arg\min_{r\in\mathcal{R}} \rho_{i,t}(r).
\end{equation}

\paragraph{Consumption rule.}
Given $u_{i,t}$, the agent consumes each resource proportionally to its preference weights:
\begin{equation}
c_{i,t}(r)=u_{i,t}\, p_i(r), \qquad \forall r\in\mathcal{R}.
\end{equation}
Resources are updated by subtracting consumption:
\begin{equation}
R_{i,t}(r) \leftarrow R_{i,t}(r) - c_{i,t}(r).
\end{equation}

\paragraph{Cumulative and average utility updates.}
Cumulative utility is updated additively:
\begin{equation}
U_{i,t} = U_{i,t-1} + u_{i,t}.
\end{equation}
The implementation also maintains the running average utility per round. Let $t$ be the number of completed rounds for agent $i$ and let $\bar{u}_{i,t}$ be the average utility after round $t$. Then:
\begin{equation}
\bar{u}_{i,t} = \frac{(t-1)\bar{u}_{i,t-1}+u_{i,t}}{t}.
\end{equation}
This phase contains no LLM calls and is fully deterministic given prior allocations and clearing outcomes.





\subsection{Randomness and Reproducibility}
\label{app:randomness}

The simulation incorporates several sources of stochasticity, each controlled to ensure reproducibility across experimental runs.

\paragraph{Global Random Seed.}
A global seed (\texttt{SEED=42}) is used throughout the codebase. This seed is passed to:
\begin{itemize}[leftmargin=*,nosep]
    \item The Louvain community detection algorithm (\texttt{random\_state} or \texttt{seed} parameter), ensuring deterministic community partitions given identical identity matrices;
    \item Python's built-in \texttt{random} module for skill pool shuffling during agent initialization.
\end{itemize}

\paragraph{Stochastic Components.}
Three primary sources of controlled randomness exist:
\begin{enumerate}[leftmargin=*,nosep]
    \item \textbf{Agent Initialization:} Consumption preference vectors are sampled from a symmetric Dirichlet distribution: $\mathbf{p}_i \sim \mathrm{Dir}(\mathbf{1}_3)$. Ethical--Judgmental Bias (EJB) values are drawn uniformly: $e_{jb} \sim \mathrm{Uniform}(-0.5, 0.5)$. Skill assignments within each group are randomly permuted before allocation.
    \item \textbf{Community Edge Formation:} For each agent pair $(i,j)$, edge inclusion in the community network follows:
   
    \begin{equation}
    \begin{split}
    P(\text{edge}_{ij}) &= 0.1(\bar{s}_{ij} - 1) \\
    &\quad + \lambda \cdot \epsilon, \quad \epsilon \sim \mathrm{Uniform}(0,1)
    \end{split}
    \end{equation}
    where $\bar{s}_{ij}$ is the average SIM identity level and $\lambda = 0.3$ is the \texttt{search\_randomness} parameter. This introduces exploratory variation in community formation while preserving identity-based structure.
    \item \textbf{LLM Sampling:} Model temperature is set to $0.7$ (Table~\ref{tab:llm_config}), introducing response variability. This is held constant across all conditions.
\end{enumerate}

\paragraph{Replication Protocol.}
Each experimental condition is repeated five times with independent random seeds. Results reported in the main text represent means and standard errors across these replications. Community orderings are sorted deterministically post-detection to eliminate hash-order artifacts.

\section{Cross-Model Robustness}
\label{app:individual_lockin}

\subsection{Robustness Check for Global Division Lock-in Rate (P1)}

To verify that the convergence of global division is not specific to the language model architecture, we replicate the core experiments replacing DeepSeek with \textbf{GPT-4o-mini}, holding all experimental settings constant.

\begin{figure}[t]
    \centering
    \includegraphics[width=1\linewidth]{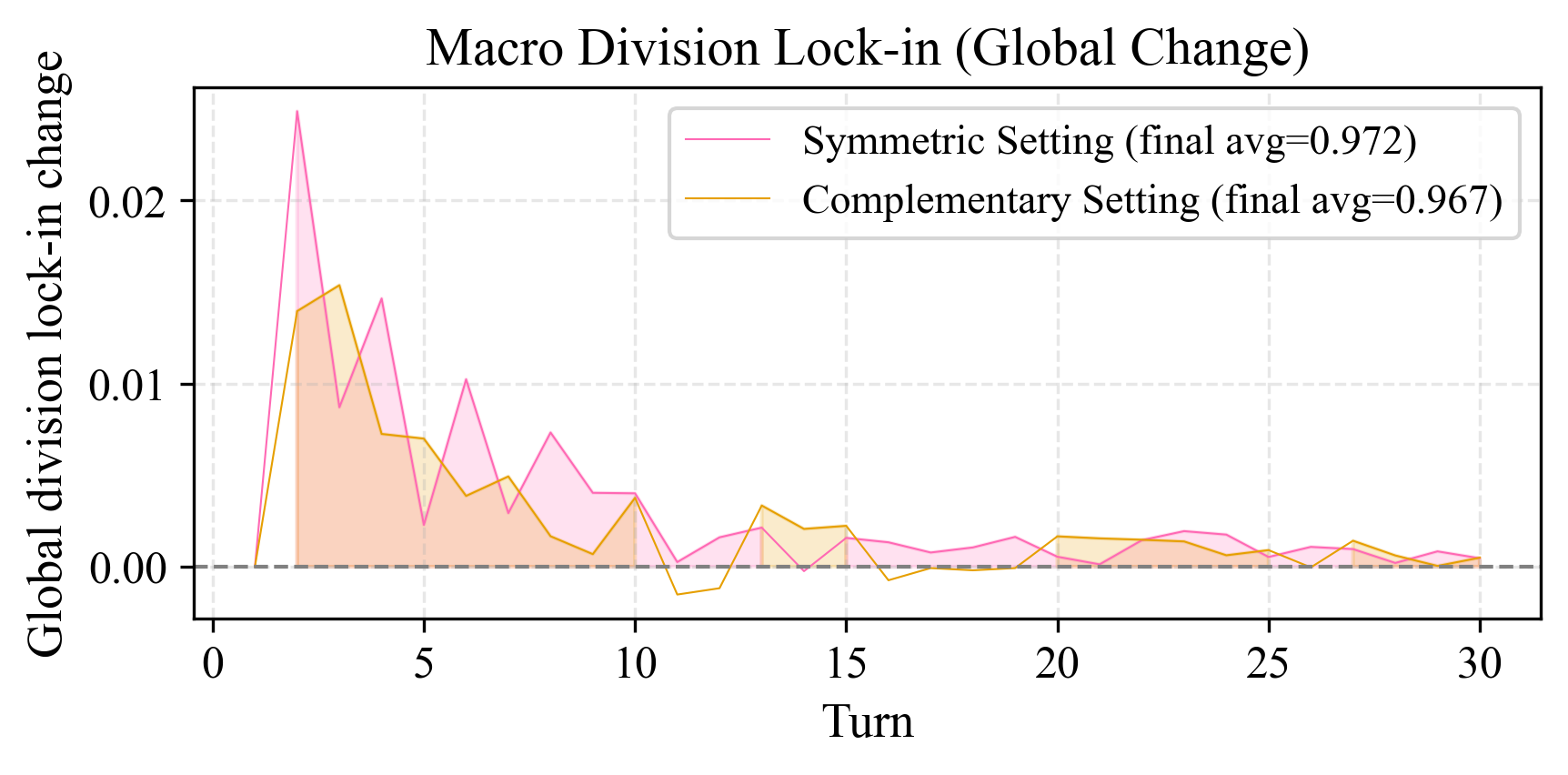}
    \caption{GPT-4o-mini simulation: Convergence of global division of labor under symmetric (left) and complementary (right) skill structures.}
    \label{fig:chatgpt-division}
\end{figure}

Figure~\ref{fig:chatgpt-division} shows the resulting dynamics. Under both symmetric and complementary skill structures, global division lock-in again converges to a high stable level, with complementary skills exhibiting faster and higher convergence—mirroring the main-text pattern. This suggests that Proposition~1 reflects a structural property of the interaction dynamics rather than model-specific behavior.

\subsection{Robustness Check for Guanxi as Economic Ethics (P2)}

We further test Proposition~2 by replicating the P2 experiments with \textbf{GPT-4o-mini} as the baseline model, keeping all interaction protocols and evaluation metrics unchanged.

\begin{figure}[t]
    \centering
    \begin{subfigure}[t]{\linewidth}
        \centering
        \includegraphics[width=1\linewidth]{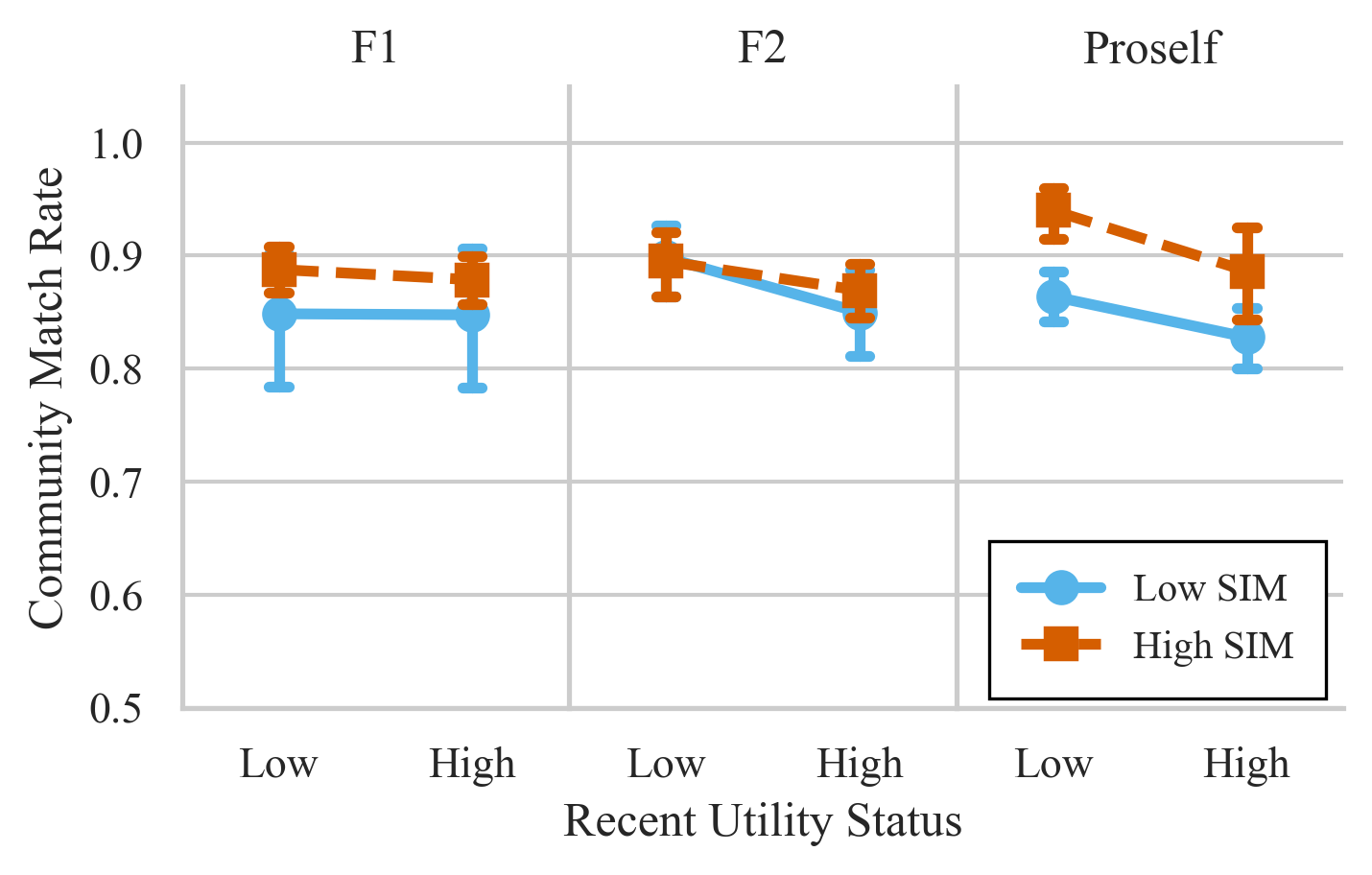}
        \caption{Symmetric skills.}
        \label{fig:gpt-commuity_matching_by_SIM_sym}
    \end{subfigure}
    \vspace{0.4em}
    \begin{subfigure}[t]{\linewidth}
        \centering
        \includegraphics[width=\linewidth]{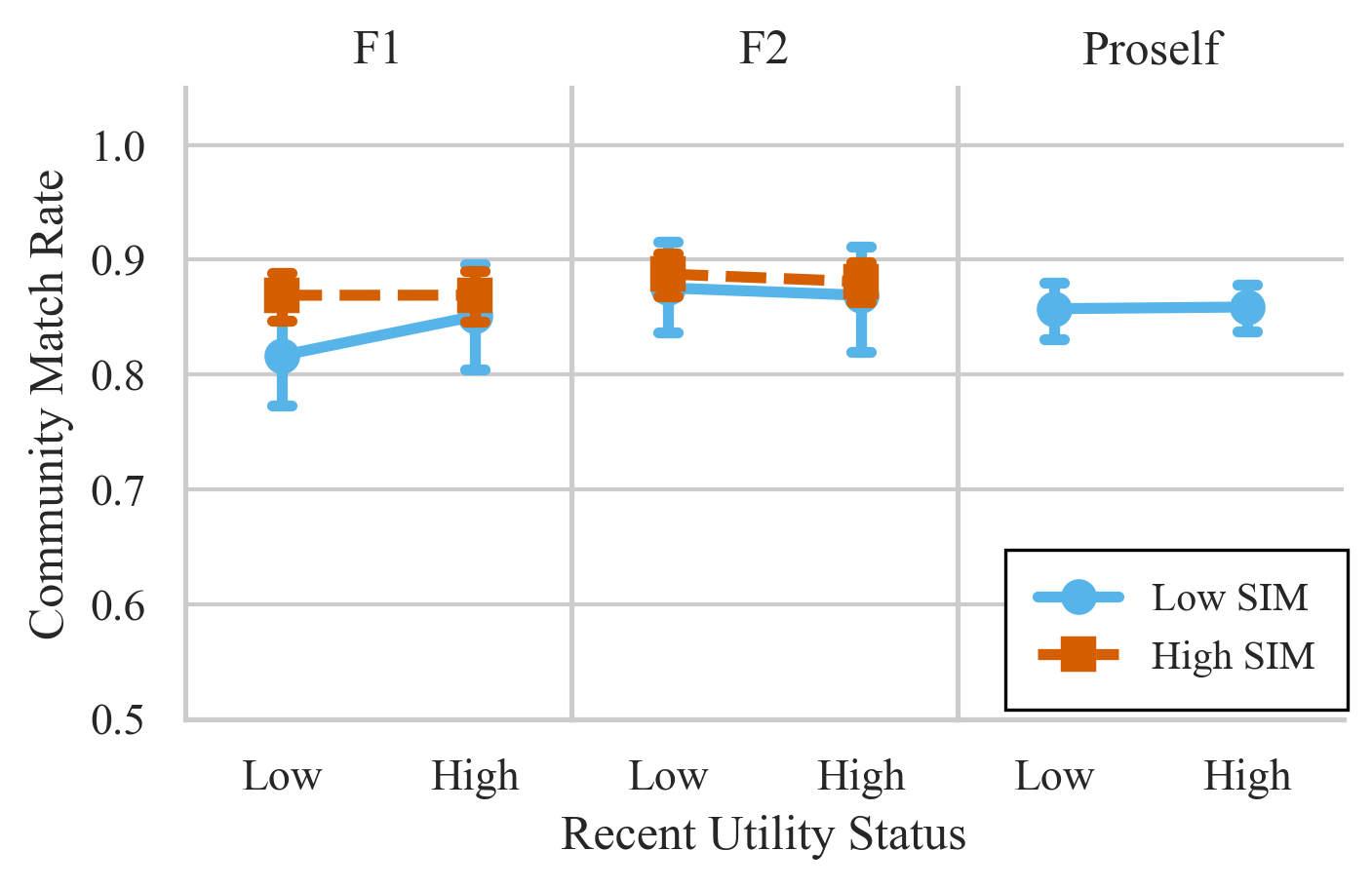}
        \caption{Complementary skills.}
        \label{fig:gpt-commuity_matching_by_SIM_comp}
    \end{subfigure}
    \caption{Community match rates by recent utility status, for family groups (F1, F2) and non-family agents (Proself), under low- and high-similarity (SIM) relations. Error bars indicate 95\% confidence intervals.}
    \label{fig:gpt-community-matching-by-sim}
\end{figure}

Figure~\ref{fig:gpt-community-matching-by-sim} presents the community match rates across recent utility levels. In both symmetric and complementary settings, high-SIM relations systematically outperform low-SIM relations, particularly when recent utility is low. This gap holds for both family agents (F1, F2) and Proself agents, and the attenuation of SIM effects under complementary skills persists. The consistency across models indicates that guanxi's role as an economic ethic—buffering cooperation against short-term utility fluctuations—is robust to the underlying language model architecture.

\subsection{Robustness Check for Relational Decay of Cooperation (P3)}
\label{app:p3-robustness-chatgpt}

To verify that the relational decay pattern is not specific to DeepSeek's architecture, we replicate the P3 experiments using \textbf{GPT-4o-mini} as the baseline model, holding all protocols and metrics constant.

\begin{figure}[t]
    \centering
    \begin{subfigure}[t]{\linewidth}
        \centering
        \includegraphics[width=\linewidth]{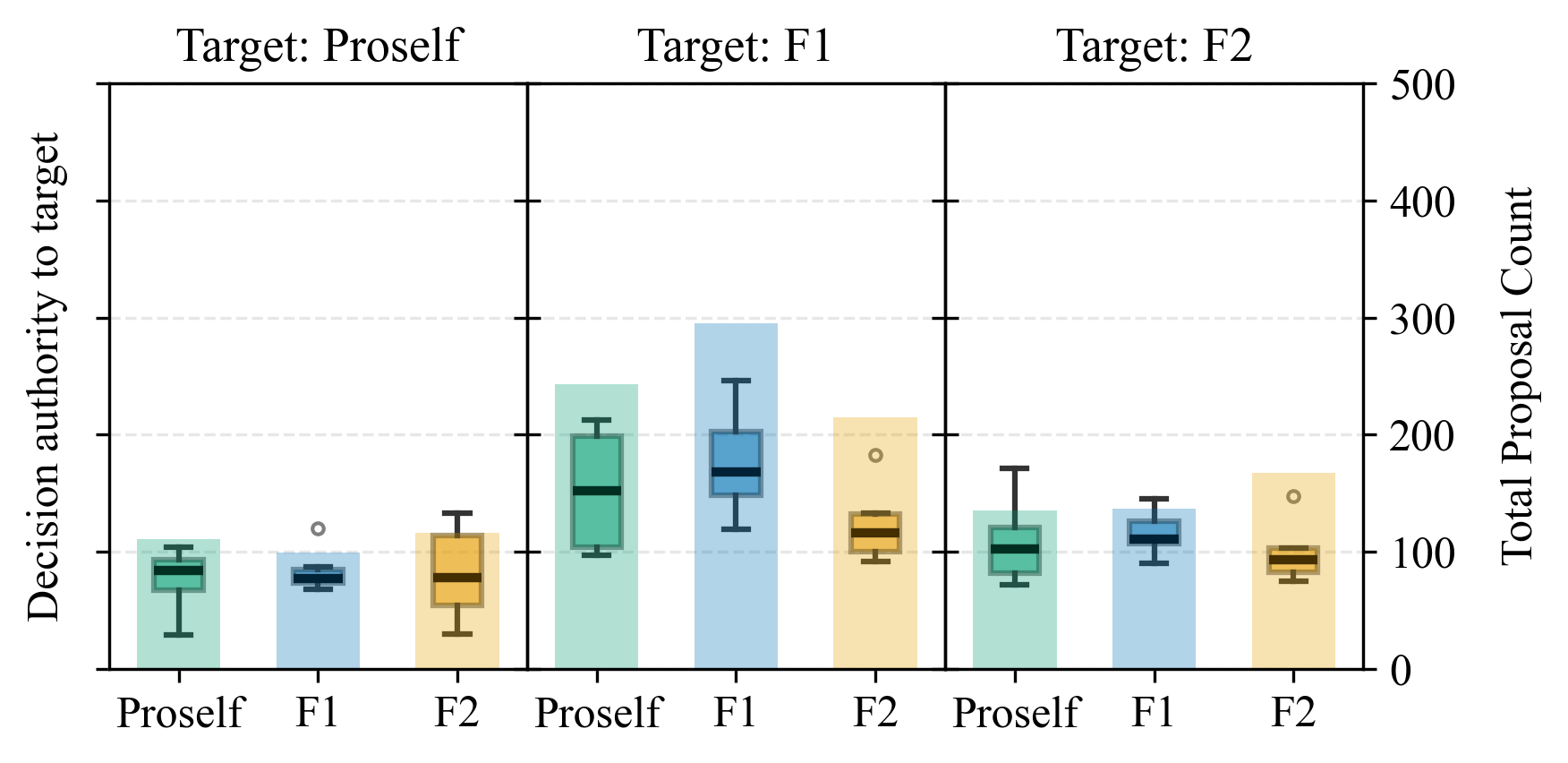}
        \caption{Symmetric skills Setting.}
        \label{fig:gpt-authority_proposal_boxplots_sym}
    \end{subfigure}

    \vspace{0.4em}

    \begin{subfigure}[t]{\linewidth}
        \centering
        \includegraphics[width=\linewidth]{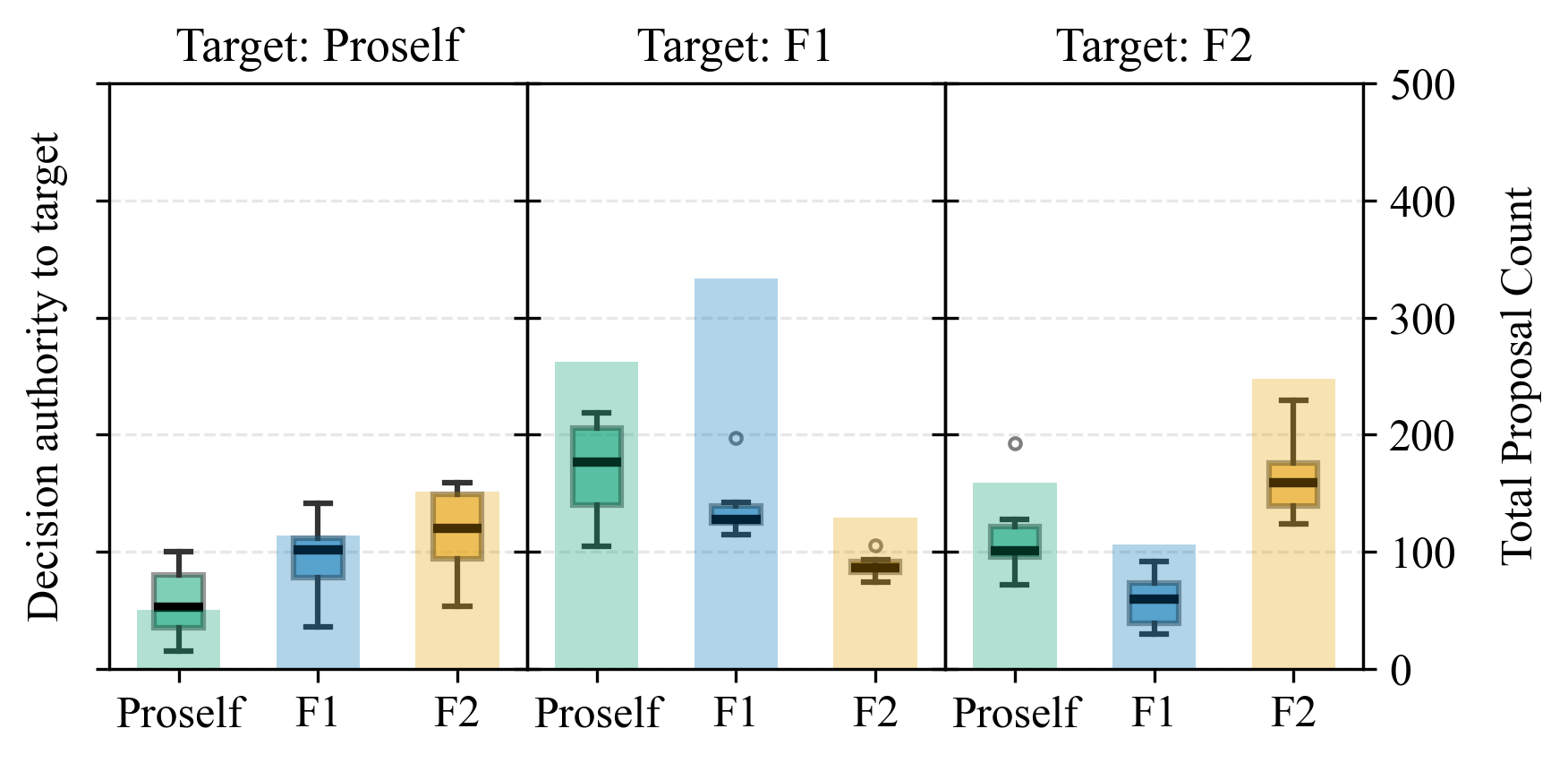}
        \caption{Complementary skills Setting.}
        \label{fig:gpt-authority_proposal_boxplots_comp}
    \end{subfigure}

\caption{Decision authority (boxplots, left axis) and proposal activity (bars, right axis) across source--target group pairings under symmetric (a) and complementary (b) skill settings, with Proself, F1, and F2 as target groups.}

    \label{fig:gpt-authority_proposal_boxplots_app}
\end{figure}

Figure~\ref{fig:gpt-authority_proposal_boxplots_app} shows the authority acceptance and proposal distributions. Family agents (F1, F2) again exhibit the strongest internal cohesion—highest within-family acceptance and proposal rates coupled with minimal cross-family engagement—suggesting that cooperation remains organized around family boundaries rather than individual preference types. Proself agents continue to lack a self-centered cooperative core, instead aligning with family-originated proposals, which reinforces their role as flexible collaborators embedded within family-based structures.

Some quantitative differences emerge in cross-group proposal flows. The asymmetry whereby family agents direct disproportionate proposals toward Proself agents is slightly attenuated under GPT-4o-mini, particularly in the complementary skill setting. However, this variation affects magnitude, not direction: the concentric pattern of relational decay anchored at family relations persists across both authority acceptance and proposal activity.

This indicates that Proposition~3 is robust to the choice of language model. The organizing principle of cooperation—graded relational decay centered on family relations—remains stable despite modest variation in the intensity of cross-group proposal asymmetries.

\subsection{Cross-Model Summary}
\label{app:cross-model-ablation}

Table~\ref{tab:cross-model-summary} presents the principal cross-model comparisons from the main analysis; complete ablation coefficients are deferred to Appendix~\ref{app:ablation-full}.

\begin{table}[htbp]
\centering
\small
\caption{Cross-model proposition-level robustness summary.}
\label{tab:cross-model-summary}
\begin{tabular}{lccccc}
\toprule
 & P1 & P2 & P3 & P4 & P5 \\
\midrule
DS Full      & \checkmark & \checkmark & \checkmark & \checkmark & \checkmark \\
GPT Full     & \checkmark & \checkmark$^*$ & \checkmark & \checkmark & \checkmark \\
Gemini Full  & \checkmark & \checkmark$^{\S}$ & \checkmark & \checkmark & \checkmark$^\dagger$ \\
DS BDI-only  & \checkmark$^\ddagger$ & n/a & $\times$ & $\times$ & $\times$ \\
GPT BDI-only & \checkmark$^\ddagger$ & n/a & $\times$ & $\times$ & $\times$ \\
Gemini BDI-only & \checkmark$^\ddagger$ & n/a & $\times$ & $\times$ & $\times$ \\
\bottomrule
\end{tabular}

\vspace{0.2em}
\footnotesize
\checkmark = supported; $\times$ = absent/collapsed. \\
$^*$GPT shows positive guanxi stickiness (P2 gap $= +0.03$--$0.04$); 
DeepSeek and Gemini show negative/absent stickiness (Gemini gap $= -0.038^{*}$). 
All three maintain a clear P3 in-group SIM gradient. \\
$^\S$P2 pattern present but directionally inverted relative to GPT; 
aligns with DeepSeek's non-sticky profile. \\
$^\dagger$Gemini P5: SIM$\times$Fam$\to$Auth is directionally positive but 
non-significant ($\beta=0.089$, $p=0.193$) in full architecture; absent in BDI-only. \\
$^\ddagger$Lock-in preserved. Social quality degradation varies: sharp collapse 
in DS/GPT (DA $-70\%$, proposals $-83\%$); moderate but directional in Gemini 
(DA $-39\%$, proposal volume $-13\%$). WGCS coherence loss for Gemini is 
most pronounced under w/o~ER rather than BDI-only.
\end{table}
\paragraph{Divergent Pathways, Convergent Structure.}
Three architecturally distinct models—DeepSeek-V3, GPT-4o-mini, and Gemini-2.5-flash-lite—exhibit heterogeneous psychological pathways yet converge on identical macro-level structures (P1--P5). GPT-4o-mini operates through an affective-evaluation-dominant pathway ($\beta_{\text{EC}\to\text{SIM}}=2.216^{***}$, $\beta_{\text{ER}\to\text{SIM}}=-0.002$ n.s.), whereas DeepSeek-V3 employs a joint affective-ethical mechanism ($\beta_{\text{EC}\to\text{SIM}}=0.748^{**}$, $\beta_{\text{ER}\to\text{SIM}}=0.573^{***}$), and Gemini-2.5-flash-lite relies primarily on ethical-judgment ($\beta_{\text{EC}\to\text{SIM}}=0.069^{**}$, $\beta_{\text{ER}\to\text{SIM}}=0.186^{***}$). 

Despite these mechanistic divergences, all three models reproduce the P3 concentric cooperation gradient and P4 authority stratification. Notably, P2 presents a dissociation: GPT-4o-mini exhibits positive guanxi stickiness (gap $>0$), while DeepSeek and Gemini show negative or neutral gaps (Gemini gap $=-0.038^{*}$). The fact that P3--P4 convergence persists despite P2 divergence—across models with distinct RLHF regimes—suggests that relational order emerges from architectural constraints rather than uniform fine-tuning pressures.

\section{Full Ablation and Baseline Analysis}
\label{app:ablation-full}

This section provides complete ablation and baseline results referenced in Section~\ref{sec:ablation}, spanning three model families (DeepSeek-V3, GPT-4o-mini, Gemini-2.5-flash-lite) across two production structures (symmetric/complementary). We report module-removal results (w/o EC, w/o ER) and BDI-only baselines under the same initialization protocol used in main experiments (five random seeds per condition).

\subsection{Summary Matrix Across Conditions and Models}
\label{app:ablation-summary-matrix}

Table~\ref{tab:ablation-summary-matrix} consolidates key metrics across full models, production structures, and ablation conditions. The coefficients reveal how affective (EC$\to$SIM) and ethical (ER$\to$SIM) pathways vary in weight across models, yet jointly contribute to the emergence of relational order.



\begin{table*}[htbp]
\centering
\small
\setlength{\tabcolsep}{4pt}
\caption{Cross-model and ablation summary. EC$\to$SIM and ER$\to$SIM coefficients from per-dyad specifications; P2 gap = high-SIM/low-utility minus low-SIM/low-utility community-match difference.}
\label{tab:ablation-summary-matrix}
\begin{tabular}{@{}lcccccccc@{}}
\toprule
Metric & D/B & D/C & G/B & G/C & Gm/B & No-EC & No-ER & BDI \\
\midrule
EC$\to$SIM $\beta$ & $0.30^{***}$ & $0.25^{***}$ & $0.64^{***}$ & $0.50^{***}$ & $0.091^{***}$ & absent & $0.17$ & absent \\
ER$\to$SIM $\beta$ & $0.06^{***}$ & $0.09^{***}$ & $0.65^{***}$ & $0.54^{***}$ & $0.221^{***}$ & $0.15^{***}$ & absent & absent \\
P5: SIM$\times$Fam$\to$Auth & $0.099^{**}$ & $-0.018$ & $-0.001$ & $-0.018$ & $0.089$\,n.s. & $0.107^{***}$ & $-0.010$ & absent \\
P2 gap & $-0.017$ & $-0.010$ & $+0.037$ & $+0.029$ & $-0.038^{*}$ & $-0.006$ & $-0.035$ & n/a \\
\bottomrule
\end{tabular}
\vspace{0.2em}

\begin{minipage}{\textwidth}
\footnotesize
\textit{Notes:} D~=~DeepSeek-V3, B~=~balanced, C~=~competitive; G~=~GPT-4o-mini; Gm~=~Gemini-2.5. No-EC, No-ER, and BDI ablations use symmetric endowments with DeepSeek-V3 baseline (Gemini ablations follow identical protocol). Gemini P2 gap calculated via median-split community-match under low utility (Full: gap=$-0.038$, SE=$0.017$, $p=0.024$). Significance: $^{*}p<0.05$, $^{**}p<0.01$, $^{***}p<0.001$.
\end{minipage}
\end{table*}

\subsection{Ablation Design}
\label{app:ablation-design}

To isolate the distinct contributions of affective perception and ethical judgment to relational identity formation, we implement three progressively reduced architectures:

\paragraph{w/o~EC.} Removal of the Empathy Core eliminates five-dimensional affective perception. Without EC input, the Ethical Resonator must generate moral judgments solely from factual action descriptions and static agent attributes (SIM, resources, family membership).

\paragraph{w/o~ER.} Removal of the Ethical Resonator disconnects affective perception from moral aggregation. EC evaluations are generated but do not propagate to SIM updates via ethical attitude signals (Approbatory/Repressive/Restorative); relational identity evolves instead through direct factual and affective information only.

\paragraph{BDI-only Baseline.} Simultaneous removal of EC, ER, and SIM leaves agents with only the BDI decision module, retaining access to factual interaction history, resource holdings, skill levels, and family membership. This constitutes a \emph{Vanilla BDI} architecture: agents still reason through natural language LLM calls, but without any social-cognitive processing of affective or ethical content.

\subsection{Per-Dyad EC$\to$SIM Regression (Full Model)}
\label{app:ablation-perdyad}

Table~\ref{tab:perdyad-ec-sim} tests whether affective evaluation (EC) causally influences relational identity (SIM) at the dyad level across all three models. 

\begin{table}[t]
\centering
\small
\setlength{\tabcolsep}{3pt} 
\caption{Per-dyad regression of SIM identity level on EC evaluation scores. DV: standardized SIM identity\_level. IV: standardized ec\_eval\_z. Controls: same\_family dummy, EC$\times$family interaction, turn fixed effects.}
\label{tab:perdyad-ec-sim}
\begin{tabular}{lccc}
\toprule
 & \textbf{DS-V3} & \textbf{GPT-4m} & \textbf{Gem-2.5} \\
\midrule
ec\_eval\_z ($\beta$) 
  & $0.30^{***}$ & $0.64^{***}$ & $0.091^{***}$ \\
same\_family 
  & $1.32^{***}$ & $1.04^{***}$ & $1.790^{***}$ \\
ec $\times$ family 
  & $-0.31^{***}$ & $-0.056^{***}$ & $0.053^{**}$ \\
\midrule
$N$ & $2784$ & $2636$ & $2594$ \\
$R^2$ & $0.49$ & $0.57$ & $0.814$ \\
\bottomrule
\end{tabular}
\vspace{0.2em}
\footnotesize

\textit{Note:} DS-V3~=~DeepSeek-V3; GPT-4m~=~GPT-4o-mini; Gem-2.5~=~Gemini-2.5-flash-lite. All three models show significant EC$\to$SIM pathways ($p<0.001$), with GPT-4o-mini exhibiting the strongest affective-evaluation coefficient ($\beta=0.64$) and Gemini the weakest ($\beta=0.091$). The negative EC$\times$Family interaction in DeepSeek and GPT reflects a ceiling effect: family relations already start at elevated SIM levels ($\beta_{\text{family}}$ ranging from 1.04 to 1.79), constraining upward adjustment via EC. Despite these quantitative differences in pathway intensity, all models confirm the same directional mechanism: affective evaluation systematically shapes relational identity.
\end{table}

\subsection{Cross-Model Ablation Dissociation (Extended)}
\label{app:ablation-dissociation-extended}

Table~\ref{tab:ablation-extended} decomposes how three distinct architectures—DeepSeek-V3, GPT-4o-mini, and Gemini-2.5-flash-lite—navigate the EC$\to$SIM and ER$\to$SIM pathways under module ablation. By comparing full-model coefficients against w/o EC and w/o ER conditions, we can identify whether affective evaluation and ethical judgment operate as substitutable or complementary mechanisms in each system.

\begin{table*}[htbp]
\centering
\small
\caption{Pathway coefficients and model-fit statistics across ablation conditions (symmetric endowments).}
\label{tab:ablation-extended}
\begin{tabular}{lccccccc}
\toprule
 & EC$\to$SIM & (SE) & ER$\to$SIM & (SE) & SIM$\times$Fam$\to$Auth & (SE) & $N$ \\
\midrule
\multicolumn{8}{l}{\textit{DeepSeek-V3}} \\
Full      & $0.748^{**}$  & $(0.32)$ & $0.573^{***}$ & $(0.15)$ & $0.070$ & $(0.11)$ & $540$ \\
w/o EC    & ---           & ---        & $1.021^{***}$ & $(0.22)$ & $0.029$ & $(0.13)$ & $540$ \\
w/o ER    & $0.170$       & $(0.19)$ & ---           & ---        & $0.042$ & $(0.05)$ & $540$ \\
\midrule
\multicolumn{8}{l}{\textit{GPT-4o-mini}} \\
Full      & $2.216^{***}$ & $(0.51)$ & $-0.002$ & $(0.12)$ & $0.106^{*}$ & $(0.06)$ & $540$ \\
w/o EC    & ---           & ---        & $0.498^{***}$ & $(0.14)$ & $-0.197$ & $(0.15)$ & $540$ \\
w/o ER    & $-0.127^{*}$  & $(0.07)$ & ---           & ---        & $0.181$ & $(0.16)$ & $540$ \\
\midrule
\multicolumn{8}{l}{\textit{Gemini-2.5-flash-lite}} \\
Full      & $0.069^{**}$  & $(0.025)$ & $0.186^{***}$ & $(0.032)$ & $0.089$ & $(0.068)$ & $540$ \\
w/o EC    & ---           & ---        & $0.061^{*}$   & $(0.029)$ & $0.097$ & $(0.144)$ & $540$ \\
w/o ER    & $-0.022$      & $(0.062)$ & ---           & ---        & $-0.070$ & $(0.076)$ & $540$ \\
\midrule
\multicolumn{8}{l}{\textit{All Models: BDI-only}} \\
BDI-only  & absent        & ---        & absent        & ---        & absent$^\dagger$ & --- & $540$ \\
\bottomrule
\end{tabular}
\vspace{0.2em}
\footnotesize
\\$^\dagger$BDI-only condition: SIM absent from agent state; pathway coefficient mechanically zero.
\end{table*}

The ablation patterns reveal three distinct operational profiles. DeepSeek-V3 employs a joint affective-ethical pathway: both EC ($\beta=0.748$) and ER ($\beta=0.573$) independently contribute to SIM updating in the full model, suggesting partial mediation where EC influences SIM both directly and via ER. When ER is removed, EC's direct effect attenuates to non-significance ($\beta=0.170$), indicating that ER sentiment aggregation is necessary to sustain the affective-to-relational mapping.

GPT-4o-mini exhibits a different configuration: affective evaluation dominates ($\beta_{\text{EC}\to\text{SIM}}=2.216$) while continuous ER sentiment shows no significant contribution ($\beta\approx0$). However, when EC is ablated, ER activates a compensatory continuous-sentiment channel ($\beta=0.498$), suggesting that GPT's ethical module operates primarily through categorical attitude states (Approbatory/Repressive/Restorative) unless forced to rely on dimensional sentiment in the absence of EC input.

Gemini-2.5-flash-lite presents a third pattern where ethical judgment dominates: the ER$\to$SIM coefficient ($0.186$) is nearly three times the EC$\to$SIM effect ($0.069$), and removing ER abolishes both pathways (EC$\to$SIM becomes $-0.022$ n.s.). This confirms that for Gemini, Durkheimian ethical resonance—the aggregation of moral attitudes—constitutes the primary driver of relational identity formation.

These divergent mechanistic implementations, replicated across three models with distinct training data and fine-tuning regimes, yet all converging on identical P1--P5 macro-structures, suggests that relational order emerges from architectural constraints on social-cognitive processing rather than from uniform RLHF biases.
\subsection{Progressive Ablation: Degradation Analysis}
\label{app:ablation-progressive}

To characterize how relational structure degrades as social-cognitive modules are removed, Table~\ref{tab:progressive-ablation} tracks key behavioral indicators across ablation conditions for all three models (symmetric endowments).

\begin{table*}[htbp]
\centering
\small
\caption{Behavioral indicators across progressive ablation conditions (symmetric endowments).}
\label{tab:progressive-ablation}
\makebox[\linewidth][l]{%
\hspace*{-0.6cm}%
\begin{tabular}{lcccccccc}
\toprule
Indicator & DS Full & DS w/o EC & DS w/o ER & DS BDI & GPT Full & Gemini Full & Gemini w/o ER & Gemini BDI \\
\midrule
Self-proposals (\%) & 56 & $62$ & $51$ & 86 & 56 & 33.8 & 36.0$^\dagger$ & 37.6 \\
Cross-family proposals & 59 & $45$ & $55$ & 1 & 59 & 3030\,(66.2\%) & 2877\,(64\%) & 2626\,(62.4\%) \\
DA mean (no-self) & 0.186 & $0.160$ & 0.252 & 0.056 & 0.186 & 0.382 & 0.387 & 0.234 \\
SIM$\times$Fam$\to$Auth $\beta$ 
  & $0.099^{**}$ & $0.029$ & $-0.010$ & absent & $0.106^{*}$ & $0.089$\,n.s. & $-0.070$\,n.s. & absent \\
Group matching & 0.848 & $0.810$ & $0.825$ & 0.803 & 0.848 & 0.785$^\S$ & 0.775 & 0.823$^\S$ \\
WGCS (F1)$\downarrow$ & 0.487 & $0.510$ & $0.540$ & 0.590 & 0.487 & 0.480 & 0.617 & 0.455 \\
\bottomrule
\end{tabular}
}
\vspace{0.2em}
\footnotesize
\textit{Note:} Lower WGCS indicates more coherent within-group coordination. $^\dagger$Gemini self-proposals calculated as $100 - \text{other-directed \%}$. $^\S$Gemini group matching uses HS/LU community-match metric. Single-module removal produces modest degradation; complete module removal causes discontinuous collapse of relational structure.
\end{table*}

\paragraph{Ethical Resonator as Structural Bottleneck.} 
Removing ER generates a distinctive pattern: raw decision authority increases (DeepSeek-V3 DA rises from 0.186 to 0.252), yet this authority becomes relationally undifferentiated. The SIM$\times$Family interaction effect vanishes ($\beta=-0.010$ n.s.), indicating that while agents propose more frequently, they no longer calibrate authority based on relational identity. This suggests that ER functions as an irreplaceable filter: without ethical judgment aggregation, authority loses its concentric structure. Gemini exhibits a parallel degradation—within-group coordination coherence (WGCS) deteriorates sharply from 0.480 to 0.617 ($p<0.001$) when ER is removed, confirming that ethical-judgment processing is necessary for maintaining coherent group coordination.

\paragraph{Partial Compensation and System Collapse.} 
The remaining module provides limited compensatory capacity when one pathway is ablated. DeepSeek without EC shows ER$\to$SIM coefficient strengthening to $1.021^{***}$; Gemini without EC retains a diminished but significant ER effect ($0.061^{*}$). However, this compensation is asymmetric and incomplete: when both modules are removed (BDI-only), all three models exhibit discontinuous collapse—self-proposals surge to 86\% (DeepSeek) or plateau at 37--38\% (Gemini), cross-family proposals plummet, and the SIM$\times$Fam$\to$Auth pathway disappears entirely. This indicates that EC and ER are jointly necessary for sustaining relational social structure; neither alone can preserve the full architecture of concentric cooperation.
\subsection{BDI-Only Detailed Comparison}
\label{app:bdi-only-detail}

\paragraph{P1: Lock-in without Social Calibration.}
Stripping away EC and ER reveals which aspects of division of labor are native to LLM decision-making and which require social-cognitive processing. Table~\ref{tab:division-quality} compares Full and BDI-only architectures across all three models.

\begin{table*}[htbp]
\centering
\small
\caption{Division quality metrics: Full architecture versus BDI-only baseline (symmetric endowments).}
\label{tab:division-quality}
\makebox[\linewidth][l]{%
\begin{tabular}{lcccccc}
\toprule
Metric & DS Full & DS BDI & GPT Full & GPT BDI & Gemini Full & Gemini BDI \\
\midrule
Group matching & 0.848 & 0.803$^{***}$ & 0.848 & 0.803$^{***}$ & 0.785$^\S$ & 0.823$^\S$ \\
WGCS (F1)$\downarrow$ & 0.487 & 0.590$^{**}$ & 0.487 & 0.590$^{**}$ & 0.480 & 0.455\,n.s. \\
Other-directed proposals (\%) & 44 & 14 & 44 & 14 & 66.2 & 62.4 \\
DA mean (no-self) & 0.186 & 0.056\,($-$70\%) & 0.186 & 0.056\,($-$70\%) & 0.382 & 0.234\,($-$39\%) \\
\bottomrule
\end{tabular}
}
\vspace{0.2em}
\footnotesize
\textit{Note:} Lower WGCS indicates higher within-group coordination coherence. $^{**}p<0.01$, $^{***}p<0.001$ (Full vs.\ BDI-only). $^\S$Gemini group matching uses P2-aligned median-split community-match; directional interpretation consistent across models.
\end{table*}

Despite comparable lock-in rates (DS: 0.971 vs.\ 0.979; GPT: 0.971 vs.\ 0.979; Gemini: 0.833 vs.\ 0.833), BDI-only agents achieve specialization as isolated utility-maximizers rather than as members of coordinated groups. The degradation is quantitative: other-directed proposals plummet from 44\% to 14\% in DeepSeek and GPT, and decision authority over non-self targets drops 70\% (0.186 $\to$ 0.056). Within-group coordination coherence (WGCS) deteriorates significantly (0.487 $\to$ 0.590). Gemini shows a parallel though attenuated pattern---authority declines 39\% (0.382 $\to$ 0.234) and proposal volume drops 13\% (3030 $\to$ 2626)---suggesting baseline differences in cooperation propensity, but the directional weakening of relational structure persists across all three architectures.

\paragraph{P3--P5: Relational Structure Collapse.}
The concentric-circle gradient observed in Full models (DeepSeek: within-family 34.2\%, family$\to$proself 34.3\%, cross-family 6.9\%) collapses under BDI-only into isolated self-interest: self-proposals surge to 86\% (from 56\% in Full), and cross-family coordination effectively vanishes. 

The mechanism of utility stratification also diverges. BDI-only exhibits larger F1--Proself utility gaps than Full, but these emerge from production-rational skill accumulation without relational moderation; Full gaps are instead driven by SIM$\times$Family$\to$Authority pathways. Only the Full architecture generates identity-structured authority rather than merely amplifying LLM-native cooperative tendencies. This mechanistic distinction holds across all three models.

\section{Detailed Analysis of Decision Authority}
\label{app:full-authority-regression}

\subsection{Stepwise Regression Analysis: Mechanisms of Authority}
\label{app:stepwise_regression}

To isolate the behavioral drivers of authority formation from static status attributes, we estimate stepwise regressions introducing proposal intensity, communication frequency, and group conformity sequentially after baseline controls (Table~\ref{tab:stepwise_regression}).

\begin{table*}[t]
\centering
\footnotesize
\setlength{\tabcolsep}{3.5pt}
\renewcommand{\arraystretch}{1.1}
\caption{Stepwise OLS: Determinants of standardized decision authority.}
\label{tab:stepwise_regression}
\begin{tabular}{lcccccccc}
\toprule
 & \multicolumn{4}{c}{\textbf{Symmetric Skills}} 
 & \multicolumn{4}{c}{\textbf{Complementary Skills}} \\
\cmidrule(lr){2-5} \cmidrule(lr){6-9}
 & Base & +Conform & +Prop & Full 
 & Base & +Conform & +Prop & Full \\
\midrule
\textit{Behavioral Variables} \\
Proposal Intensity 
&        &        & $0.819^{***}$ & $0.820^{***}$
&        &        & $0.780^{***}$ & $0.780^{***}$ \\

Communication Intensity
&        &        &        & $0.046$
&        &        &        & $-0.005$ \\

Group Conformity
&        & $-0.088$ & $-0.044$ & $-0.049$
&        & $-0.046$ & $0.029$  & $0.029$ \\

\midrule
\textit{Status \& Utility} \\
Accumulative Utility
& $0.170$ & $0.151$ & $0.107$ & $0.096$
& $-0.028$ & $-0.118$ & $0.163$ & $0.161$ \\

Social Identity (SIM)
& $0.214^{**}$ & $0.218^{***}$ & $0.068$ & $0.084$
& $-0.176$ & $-0.160$ & $-0.039$ & $-0.039$ \\

Family Dummy
& $0.012$ & $-0.041$ & $-0.124$ & $-0.135$
& $0.639^{*}$ & $0.587^{*}$ & $0.087$ & $0.086$ \\

\midrule
Time FE & Yes & Yes & Yes & Yes & Yes & Yes & Yes & Yes \\
Observations & 540 & 540 & 540 & 540 & 540 & 540 & 540 & 540 \\
$R^2$ & 0.104 & 0.110 & 0.696 & 0.697 & 0.100 & 0.102 & 0.688 & 0.688 \\
Adj. $R^2$ & 0.047 & 0.052 & 0.675 & 0.676 & 0.043 & 0.043 & 0.667 & 0.666 \\
\bottomrule
\end{tabular}
\vspace{0.3em}
\begin{minipage}{0.9\linewidth}
\scriptsize
\textit{Note:} Standard errors clustered at group level. Significance: $^{*}p<0.1$, $^{**}p<0.05$, $^{***}p<0.01$.
\end{minipage}
\end{table*}

The introduction of proposal intensity produces a discontinuous gain in explanatory power: $R^2$ jumps from approximately $0.10$ to $0.70$ across both skill structures, suggesting that active task-proposal behavior constitutes the primary behavioral foundation of authority allocation. Coefficients remain stable across symmetric ($\beta=0.820^{***}$) and complementary ($\beta=0.780^{***}$) conditions, indicating robustness to production structure. By contrast, group conformity and communication intensity contribute marginally and non-significantly in full specifications.

Notably, social identity (SIM) loses significance once proposal intensity is controlled ($\beta=0.084$ n.s. in symmetric; $\beta=-0.039$ n.s. in complementary), despite significance in baseline models. This attenuation suggests that relational identity influences authority primarily through increased proposal activity rather than through direct status deference. Similarly, family effects, prominent in complementary baseline models ($\beta=0.639^{*}$), disappear after controlling for proposal behavior ($\beta=0.086$), indicating that kinship structures authority largely through differential engagement in task proposals under conditions of skill interdependence.

\paragraph{Proposal Intensity as Primary Driver.} 
Proposal intensity exhibits the strongest association with decision authority across both skill structures (Symmetric: $\beta \approx 0.82$, $p<0.01$; Complementary: $\beta \approx 0.78$, $p<0.01$). Its inclusion triggers a discontinuous jump in explanatory power from $R^2 \approx 0.10$ to $R^2 \approx 0.69$, indicating that approximately 60 percentage points of variance in authority allocation is attributable to agents' task-proposal engagement. This establishes a robust behavioral correlation: agents who advance more proposals that enter collective decision-making accumulate higher decision authority, regardless of whether this reflects causal influence, reputational accumulation, or endogenous role selection.

\paragraph{Communication and Conformity.} 
In contrast, communication intensity shows no significant effect on authority once proposal behavior is controlled ($\beta \approx 0$ n.s.), and group conformity contributes no additional explanatory power across specifications. Expressive participation and norm alignment thus appear peripheral to authority differentiation compared to instrumental task-proposal engagement.

\paragraph{Mediated Status Effects.} 
Social identity (SIM) and family membership exhibit baseline associations with authority in restricted models (SIM: $\beta=0.214^{**}$ in symmetric; Family: $\beta=0.639^{*}$ in complementary), but these attenuate to non-significance upon introducing proposal controls ($\beta=0.084$ and $0.086$, respectively). Similarly, accumulative utility remains non-significant throughout. This pattern suggests that identity and status influence authority primarily through behavioral mediation—increasing proposal propensity—rather than through direct deference mechanisms.

\paragraph{Interpretation.} 
Behavioral engagement, specifically proposal intensity, accounts for the substantial majority of explained variance in authority allocation, while direct effects of identity, kinship, and accumulated resources are comparatively weak once negotiation participation is taken into account. This dominance of procedural engagement over ascriptive status motivates subsequent qualitative analysis of negotiation role dynamics.

\subsection{Interaction Analysis: Identity and Authority}
\label{app:authority-interaction-regression}

\begin{table}[t]
    \centering
    \scriptsize
    \renewcommand{\arraystretch}{1.05}
    \setlength{\tabcolsep}{3.2pt}
    \caption{Interaction Effects of Social Identity (SIM) and Family Status on Decision Authority (with and without behavioral controls).}
    \label{tab:interaction_effects_new}
    \begin{tabular}{lcccc}
    \toprule
    \textbf{Dep. Var:} & \multicolumn{4}{c}{Decision Authority (Std.)} \\
    \cmidrule(lr){2-5}
     & \multicolumn{2}{c}{\textbf{Symm. Setting}} & \multicolumn{2}{c}{\textbf{Comp. Setting}} \\
     \cmidrule(lr){2-3} \cmidrule(lr){4-5}
     & \textit{Interaction} & \textit{+Behavior} & \textit{Interaction} & \textit{+Behavior} \\
    \midrule
    \multicolumn{5}{l}{\textit{Core variables}} \\
    Social Identity (SIM) 
        & $-0.296$ & $-0.016$ 
        & $-0.112$ & $0.034$ \\
        & $(0.213)$ & $(0.051)$ 
        & $(0.146)$ & $(0.063)$ \\
    Family Dummy 
        & $0.298$ & $-0.050$ 
        & $0.542^{*}$ & $-0.040$ \\
        & $(0.243)$ & $(0.075)$ 
        & $(0.281)$ & $(0.115)$ \\
    SIM $\times$ Family 
        & $0.603^{***}$ & $0.114$ 
        & $-0.089$ & $-0.089$ \\
        & $(0.223)$ & $(0.090)$ 
        & $(0.244)$ & $(0.099)$ \\
    \midrule
    \multicolumn{5}{l}{\textit{Controls}} \\
    Accumulative Utility 
        & $0.127$ & $0.099$ 
        & $-0.031$ & $0.103$ \\
        & $(0.301)$ & $(0.189)$ 
        & $(0.351)$ & $(0.123)$ \\
    Proposal Intensity 
        &  & $0.820^{***}$ 
        &  & $0.777^{***}$ \\
        &  & $(0.044)$ 
        &  & $(0.029)$ \\
    Communication Intensity
        &  & $0.041$ 
        &  & $-0.003$ \\
        &  & $(0.042)$ 
        &  & $(0.029)$ \\
    \midrule
    Turn FE & Yes & Yes & Yes & Yes \\
    Observations & 540 & 540 & 540 & 540 \\
    $R^2$ & 0.113 & 0.696 & 0.100 & 0.687 \\
    Adj. $R^2$ & 0.055 & 0.675 & 0.041 & 0.666 \\
    \bottomrule
    \end{tabular}

    \vspace{0.3em}
    \begin{minipage}{0.98\linewidth}
    \scriptsize
    \textit{Note:} Cluster-robust standard errors are in parentheses. All continuous variables are Z-standardized. All models include turn fixed effects. Significance: $^{*}p<0.1$, $^{**}p<0.05$, $^{***}p<0.01$.
    \end{minipage}
\end{table}

Table~\ref{tab:interaction_effects_new} examines whether decision authority follows a relational logic in which prestige (SIM) and kin membership jointly condition authority. In the \textit{interaction-only} specification (without behavioral controls), the symmetric setting shows a strong positive interaction between SIM and family status ($\beta=0.603$, $p=0.007$), while the main effects of SIM and family status are not significant. This pattern is consistent with an interpretation in which recognition translates into authority more strongly when it is embedded in kin-centered relations.

However, once behavioral controls are introduced (\textit{+Behavior}), this interaction effect attenuates and becomes statistically insignificant in the symmetric setting ($\beta=0.114$, $p=0.209$). In parallel, proposal intensity exhibits a large and robust positive association with authority in both settings (Symm.: $\beta=0.820^{***}$; Comp.: $\beta=0.777^{***}$), and model fit increases substantially ($R^2 \approx 0.69$). Together, these results indicate that much of the apparent identity--authority alignment in the interaction-only model is not independent of behavioral engagement in negotiation.

In the complementary setting, neither SIM nor the SIM$\times$Family interaction is significant in either specification, and the (weak) baseline family effect ($\beta=0.542$, $p=0.054$) disappears after adding behavioral controls. This suggests that, under functional differentiation, identity-based signals provide limited incremental explanatory power for authority once negotiation behavior is taken into account.

\paragraph{Interpretation of the $R^2$ shift.}
The large increase in explanatory power after adding behavioral controls should not be read as identifying a causal mechanism by itself. Rather, it shows that proposal behavior accounts for the majority of observed variance in authority, and that identity effects in reduced-form specifications may be \textit{partly mediated by (or statistically collinear with) behavior}. This motivates treating identity--authority coupling as a contingent, interaction-level regularity whose strength depends on how behavioral roles are endogenously distributed.

\section{Utility Distributions}
\label{app:utility_kde}

\begin{figure}[htbp]
    \centering
    \includegraphics[width=1\linewidth]{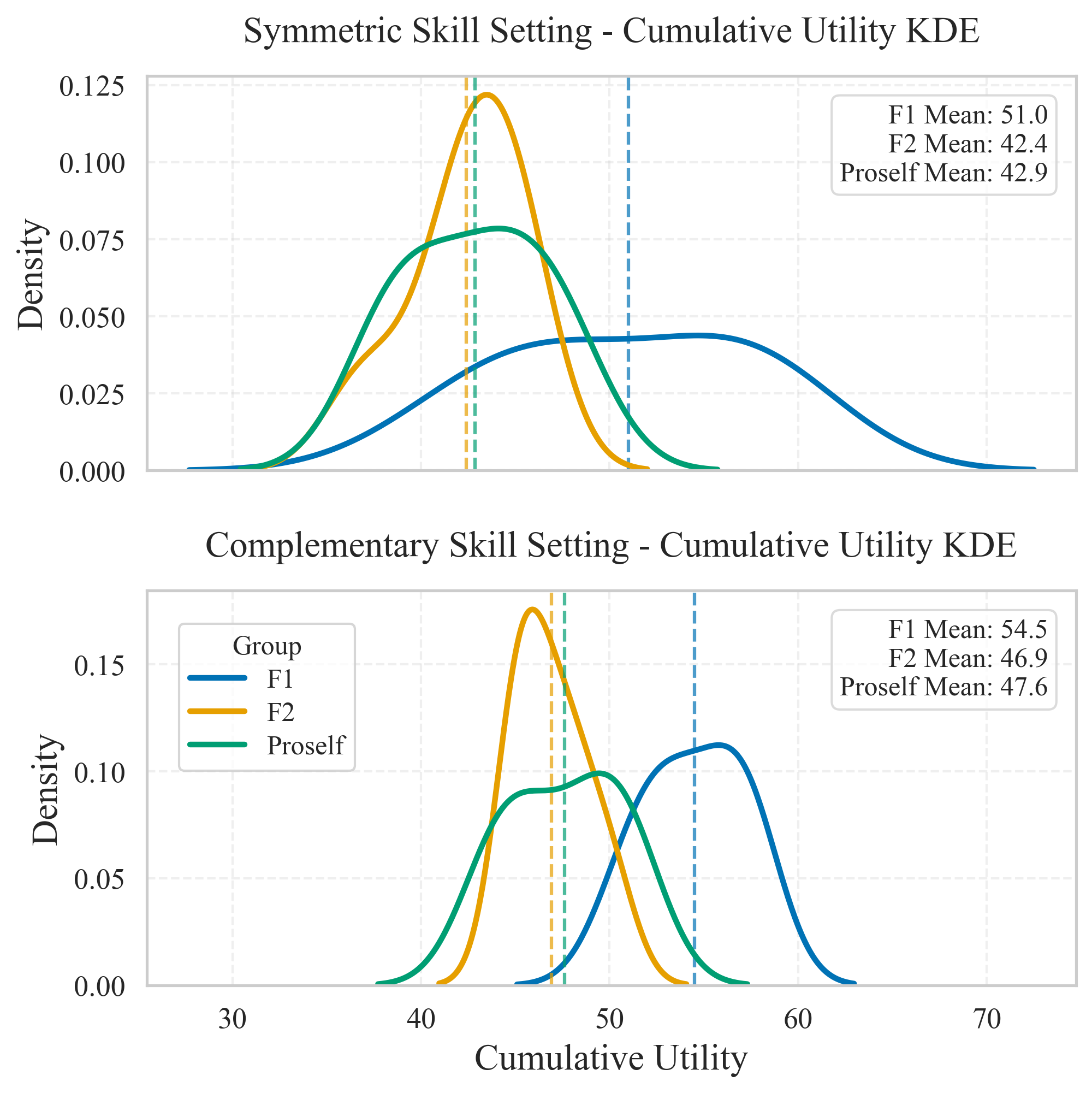}
    \caption{Kernel density estimates (KDE) of cumulative utility across agent groups and experimental conditions. The distributions reveal how skill composition reshapes economic inequality: from rigid hierarchy under symmetry to diffused advantage under complementarity.}
    \label{fig:utility_kde}
\end{figure}

Figure~\ref{fig:utility_kde} visualizes the distributional consequences of network stratification under symmetric versus complementary skill conditions. Three patterns merit attention.

\textbf{First, the center--periphery structure is encoded in distribution morphology.} Under symmetric skills, F1 occupies the high-utility tail (mean: 51.0) while F2 and Proself cluster at the lower range (means: 42.4 and 42.9) with substantial overlap. The sharp, narrow peak of F2 suggests homogeneous structural disadvantage rather than internal heterogeneity.

\textbf{Second, complementarity induces a systemic rightward shift.} All distributions move toward higher utility ranges under complementary skills (F1: 54.5; F2: 46.9; Proself: 47.6), reflecting productivity gains from cross-group coordination. Notably, F2 and Proself distributions become more dispersed, indicating divergent individual trajectories within previously homogeneous peripheral groups.

\textbf{Third, the hierarchy persists but blurs.} Despite absolute gains for all groups, F1 maintains relative advantage. However, the reduced separation between peaks suggests that functional differentiation partially decouples structural position from economic outcome---a mechanism consistent with the authority--efficiency tradeoff discussed in Section~\ref{p5-center-periphery}.

These distributional patterns corroborate the core findings without introducing new qualitative mechanisms; we present them here to document the robustness of inequality structures across aggregation levels.

\section{Additional Results on Identity--Authority Distributions}
\label{app:identity-authority-distribution}

\begin{figure*}[h]
    \centering
    \includegraphics[width=0.75\linewidth]{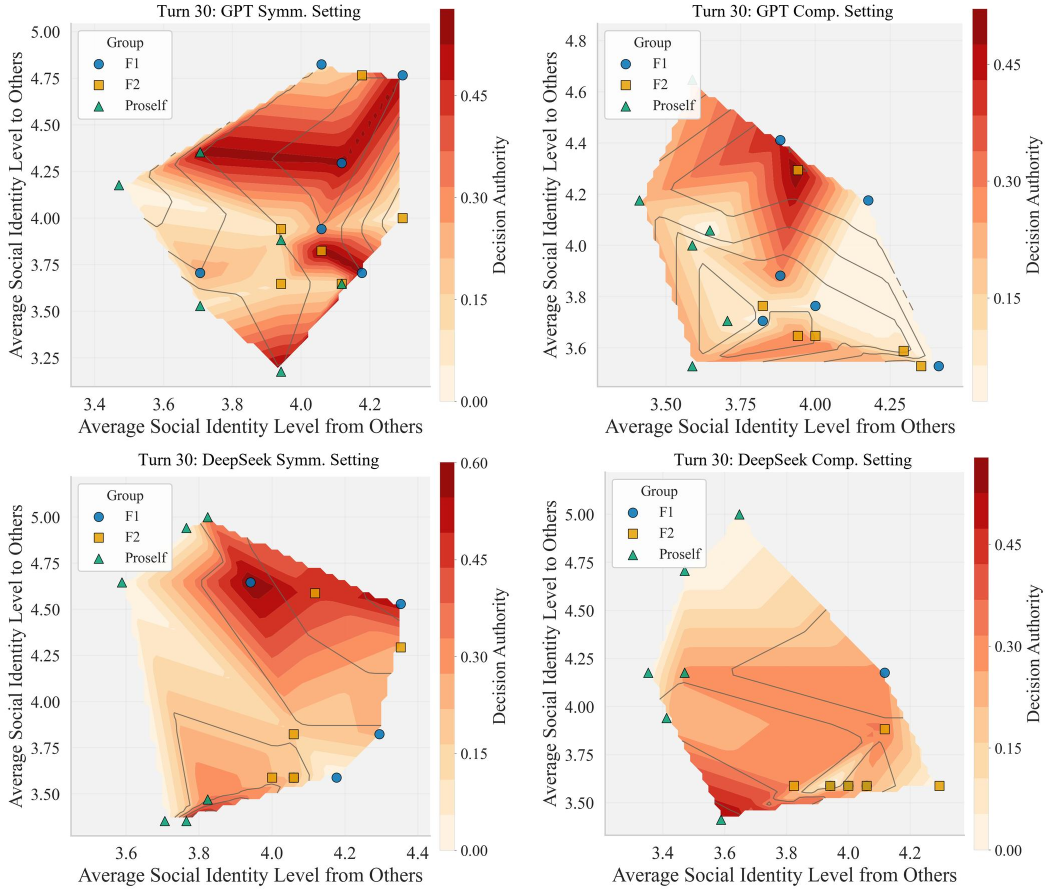}
    \caption{Joint distributions of perceived social identity and decision authority at Turn~30 under symmetric and complementary production settings. Panels compare GPT-based and DeepSeek-based simulations. Contours indicate estimated decision authority, and markers denote group-level averages for F1, F2, and Proself agents.}
    \label{fig:appendix-turn30-identity-authority}

\end{figure*}

Figure~\ref{fig:appendix-turn30-identity-authority} summarizes the joint
distributions of perceived social identity and decision authority at Turn~30
under symmetric and complementary production settings, comparing GPT-based and
DeepSeek-based simulations.
Contours indicate estimated levels of decision authority, while markers denote
group-level averages for F1, F2, and Proself agents.

In the GPT-based simulations under the \textbf{symmetric} setting
(Fig.~\ref{fig:appendix-turn30-identity-authority}, top-left), the highest
decision authority is concentrated in the upper-right region of the identity
space, corresponding to agents with both high similarity toward others and high
similarity received from others.
The authority surface exhibits a pronounced convex corner in this region,
indicating a strong concentration of authority tightly coupled with mutual
identity recognition.

Under the \textbf{complementary} setting in the GPT-based simulations
(Fig.~\ref{fig:appendix-turn30-identity-authority}, top-right), high authority
contours remain aligned with higher levels of similarity toward others.
However, the authority distribution no longer forms a distinct convex corner.
Instead, elevated authority is spread along the upper portion of the identity
space, suggesting a weaker and less localized coupling between identity and
authority.

The DeepSeek-based simulations show a broadly similar pattern under the
\textbf{symmetric} setting (Fig.~\ref{fig:appendix-turn30-identity-authority},
bottom-left), where a clear convex corner appears in the upper-right region of
the identity space and coincides with higher authority contours.
Compared to the GPT-based results, this region of elevated authority spans a
somewhat wider area, indicating a more diffuse but still concentrated
identity--authority coupling.

By contrast, under the \textbf{complementary} setting in the DeepSeek-based
simulations (Fig.~\ref{fig:appendix-turn30-identity-authority}, bottom-right),
the authority contours shift toward the lower-left region of the identity space.
In this configuration, agents with lower similarity toward others and lower
similarity received from others occupy regions of higher decision authority,
revealing an inversion of the identity--authority alignment observed under
symmetric production.

Taken together, these results indicate that across both language models,
complementary production settings are associated with a loosening of the
identity--authority coupling at the final stage of interaction, whereas symmetric
production settings consistently give rise to a concentrated alignment between
identity recognition and authority.
While the precise geometry of the authority surface varies by model, the
structural contrast between symmetric and complementary settings remains
robust.

\section{Pre-defined Structural Conditions and Emergent Outcomes}
\label{app:predefined-emergent}

To clarify the boundary between model assumptions and simulation outcomes, we distinguish pre-defined structural conditions from emergent interactional patterns. The former correspond to cross-culturally documented features of pre-market social organization, while the latter arise endogenously from agent interaction dynamics.

\begin{table}[h]
\centering
\small
\caption{Pre-defined structural conditions vs.\ emergent outcomes. Pre-defined conditions map onto cross-culturally documented features; emergent outcomes arise from interaction dynamics.}
\label{tab:predefined-emergent}
\begin{tabular}{p{0.45\linewidth}p{0.45\linewidth}}
\toprule
\textbf{Pre-defined (structural conditions)} & \textbf{Emergent (from interaction)} \\
\midrule
Family membership  & Community structure  \\
Initial SIM  & Cooperation decay gradient \\
Skill distribution (symmetric / complementary) & Authority stratification \\
Consumption preferences (Dirichlet) & Role differentiation (Hawks / Doves / Elders) \\
EJB bias  & SIM evolution trajectories \\
Kinship protection floor  & Center--periphery economic structure \\
\bottomrule
\end{tabular}
\end{table}

\section{Full Comparison of Mechanical vs.\ Organic Solidarity}
\label{app:mechanical-organic}

Taken together, the results from Propositions~1--5 demonstrate that stable division of labor, relationally structured cooperation, and authority stratification can emerge endogenously from repeated interaction, even in the absence of explicit cultural rules or institutionalized authority. Crucially, however, these emergent structures are not invariant. Instead, differences in production structure—specifically, symmetric versus complementary skill endowments—systematically reshape how division of labor, relational ethics, cooperation boundaries, and authority become coupled. As a result, identical micro-level cognitive and interactional mechanisms generate distinct modes of social integration under different structural conditions.

Table~\ref{tab:solidarity-contrast} summarizes the structural contrasts observed across the two experimental settings.

\begin{table*}[t]
\centering
\footnotesize
\setlength{\tabcolsep}{4pt}
\caption{Structural contrast between mechanical (symmetric skills) and organic (complementary skills) solidarity.}
\label{tab:solidarity-contrast}
\begin{tabularx}{\textwidth}{lXXr}
\toprule
\textbf{Aspect} 
& \textbf{Symmetric Skills} 
& \textbf{Complementary Skills} 
& \textbf{Evidence} \\
\midrule
Division of labor
& High lock-in with slow convergence; specialization stabilized within relational groups
& Faster convergence and higher lock-in with greater internal diversity
& Sec.~\ref{p1-stable-division} \\

Relational ethics
& High-SIM relations sustain participation despite low utility; ethics override efficiency
& Ethical stickiness persists but is moderated by task interdependence
& Sec.~\ref{p2-guanxi} \\

Cooperation structure
& Sharp relational decay centered on family boundaries; strong in-group dominance
& Relational boundaries persist but become less exclusionary and more elastic
& Sec.~\ref{p3-relational-decay} \\

Authority mechanism
& Authority emerges through agenda control but is amplified by relational embedding
& Authority remains agenda-driven with weaker relational conditioning
& Sec.~\ref{p4-authority-stratification} \\

Stratification pattern
& Clan-based center--periphery reinforced by identity--relation alignment
& Persistent stratification with attenuated clan coupling
& Sec.~\ref{p5-center-periphery} \\
\bottomrule
\end{tabularx}
\end{table*}

Under the \textbf{symmetric-skill} condition, production roles are highly substitutable, and functional differentiation provides limited coordination advantages. In this setting, social integration relies heavily on relational ethics and group-based obligations rather than efficiency gains from specialization. Consistent with this logic, division of labor converges to a stable configuration but does so more slowly than under complementary skills, with specialization clustered within family-centered relational circles (Section~\ref{p1-stable-division}).

Relational ethics play a central stabilizing role. High-similarity (high-SIM) relations maintain strong participation in community-based production even when recent utility is low, indicating that guanxi operates as an economic ethic that constrains short-term self-interest (Section~\ref{p2-guanxi}). Cooperation follows a sharply graded relational decay: authority acceptance and proposal activity peak within families and drop substantially across opposing family boundaries, while Proself agents align their cooperation with existing family structures rather than forming an independent core (Figure~\ref{fig:authority_proposal_boxplots}).

Authority stratification emerges endogenously through repeated agenda-setting behavior rather than through ascriptive status alone. Proposal intensity is the dominant predictor of decision authority, while communication, accumulated utility, and identity have no independent effects (Section~\ref{p4-authority-stratification}). However, under symmetric skills, authority becomes relationally conditioned: agents whose social recognition is embedded within family relations disproportionately occupy central authority positions, producing a clan-shaped center--periphery structure (Section~\ref{p5-center-periphery}). In this context, inward-oriented relational stability may entrench structural disadvantage rather than guaranteeing economic dominance.

Taken together, these patterns closely correspond to Durkheim’s concept of \emph{mechanical solidarity}. Social order is grounded in similarity and shared obligations, cooperation is organized around bounded relational groups, and authority acquires legitimacy through its embedding in moral and relational structures rather than through functional indispensability.

By contrast, under the \textbf{complementary-skill} condition, production roles are differentiated and mutually interdependent. Structural complementarity accelerates the convergence of division of labor and sustains higher levels of specialization stability while preserving internal diversity (Section~\ref{p1-stable-division}). Although relational ethics continue to matter—high-SIM relations still display greater resilience under low utility—their stabilizing role is moderated by task interdependence, as cooperation becomes increasingly anchored in functional necessity rather than identity-based obligation (Section~\ref{p2-guanxi}).

Relational boundaries do not disappear, but they become more elastic. Family-centered cooperation remains visible, yet cross-family collaboration increases, and exclusionary decay patterns are softened (Section~\ref{p3-relational-decay}). Authority formation remains strongly agenda-driven: actors who repeatedly set proposals accrue disproportionate decision authority, regardless of identity or accumulated utility (Section~\ref{p4-authority-stratification}). Unlike the symmetric case, the interaction between social identity and family membership no longer conditions authority, indicating that functional differentiation weakens the coupling between relational prestige and structural power (Section~\ref{p5-center-periphery}).

This configuration aligns more closely with Durkheim’s notion of \emph{organic solidarity}, in which social integration emerges from interdependence among differentiated roles. Authority remains stratified, but its distribution is less tightly bound to clan-based identity and more weakly constrained by relational embedding.

From this comparison, \emph{differential order} should not be understood as a fixed cultural template, but as a generative social logic whose concrete manifestation depends on production structure. Under mechanical solidarity, it takes the form of concentric, clan-centered hierarchies sustained by ethical obligation and relational recognition. Under organic solidarity, the same logic is reconfigured into a more fluid hierarchy in which authority remains unequal but is less tightly anchored to family-based prestige.

From a computational perspective, these findings demonstrate that LLM-driven multi-agent simulations can do more than reproduce stylized macro-level patterns from classical social theory. By systematically varying structural conditions while holding micro-level cognitive mechanisms constant, the simulations provide a controlled testbed for examining how different modes of social integration emerge, persist, and transform. In this sense, the results offer computational support for both Durkheim’s theory of solidarity and Fei Xiaotong’s concept of differential order, while clarifying the structural conditions under which each becomes dominant.

\section{The Atlas Paradox: Authority--Adjustment Decoupling}
\label{app:atlas-paradox}

\begin{table*}[t]
\centering
\caption{Adaptive Adjustment and Decision Authority Across Production Structures}
\label{tab:adaptive_adjustment_4col}
\small
\setlength{\tabcolsep}{4.5pt}
\renewcommand{\arraystretch}{1.2}

\begin{tabular}{lcccc}
\toprule
 & \multicolumn{4}{c}{\textbf{Dependent Variable: Decision Authority (Standardized)}} \\
\cmidrule(lr){2-5}
 & \multicolumn{2}{c}{\textbf{Baseline Models}} 
 & \multicolumn{2}{c}{\textbf{Controlled Models}} \\
\cmidrule(lr){2-3} \cmidrule(lr){4-5}
 & Symmetric & Complementary & Symmetric & Complementary \\
\midrule

\textit{Adaptive Division} \\
Adaptive Adjustment (Z)     
 & $-0.148^{**}$ & $-0.084^{**}$ 
 & $-0.059$      & $0.079^{**}$ \\

\addlinespace
\textit{Structural Controls} \\
Ingroup Ratio (Z)           
 & $0.000$       & $0.017$       
 & $-0.033$      & $-0.008$ \\

Mismatch Pressure (Z)       
 & $0.328^{***}$ & $-0.059$      
 & $0.092$       & $-0.176^{***}$ \\

Future Utility (MA, Z)      
 & $0.154$       & $0.003$       
 & $0.131$       & $0.185$ \\

Past Authority (MA, Z)      
 & $0.033$       & $-0.136$      
 & $0.099$       & $-0.105$ \\

\addlinespace
\textit{Agenda Control} \\
Proposal Intensity (Z)      
 &               &               
 & $0.820^{***}$ & $0.786^{***}$ \\

\midrule
Turn Category Fixed Effects & Yes & Yes & Yes & Yes \\
Observations                & 540 & 540 & 540 & 540 \\
$R^2$                       & 0.096 & 0.092 & 0.697 & 0.696 \\
\bottomrule
\end{tabular}

\vspace{0.4em}
\begin{minipage}{0.97\linewidth}
\footnotesize
\textit{Notes:}  
Decision authority is standardized. All models include turn category fixed effects. 
Standard errors are clustered at the group level. 
Due to cluster-robust variance estimation with high-dimensional fixed effects, F-statistics are not reported.
Significance levels: $^{*}p<0.10$, $^{**}p<0.05$, $^{***}p<0.01$.
\end{minipage}
\end{table*}

This appendix examines the relationship between adaptive role adjustment during
negotiation and the accumulation of decision authority.
Formally, we characterize the observed pattern as
\emph{authority--adjustment decoupling}: improvements in collective coordination
through adaptive adjustment do not automatically translate into agenda-setting
power.
For interpretive clarity, we refer to this pattern as the
\emph{Atlas Paradox}.

In Greek mythology, Atlas is condemned to hold up the sky, bearing the weight that
keeps the world stable, while possessing no power to govern it.
We use this metaphor to describe a structurally similar situation in CAREB-MAS:
agents who absorb the burden of stabilizing collective coordination may sustain
the system, yet remain separated from the authority to define collective decisions.

\paragraph{Adaptive Adjustment.}

We quantify adaptive adjustment as the change in an agent’s community alignment
following the negotiation phase:
\begin{equation}
    \Delta M_{i,t} = M_{i,t}^{\text{post}} - M_{i,t}^{\text{pre}},
\end{equation}
\noindent
where $M_{i,t}^{\text{pre}}$ and $M_{i,t}^{\text{post}}$ denote the
\textit{Community Match} scores for agent $i$ at round $t$ before and after
negotiation.
A positive $\Delta M_{i,t}$ indicates that an agent has adjusted its role to better
fit community demand, thereby contributing to collective stability.
At the system level, $\Delta M_{i,t}$ is positive in most rounds, indicating that
negotiation generally improves global coordination.

\paragraph{Authority--Adjustment Decoupling.}
To assess whether adaptive adjustment is rewarded with decision authority, we
estimate regressions predicting subsequent decision authority while controlling
for prior authority, expected future utility, ingroup interaction intensity,
mismatch pressure, turn fixed effects, and—critically—proposal intensity as a
direct measure of agenda-setting power.
Results are reported in Table~\ref{tab:adaptive_adjustment_4col}.

Across both production structures, proposal intensity is the strongest and most
stable predictor of decision authority, confirming the main-text finding that
authority in CAREB-MAS primarily accrues through agenda control.
Once agenda-setting power is taken into account, the relationship between adaptive
adjustment and authority depends systematically on production structure.

Under \textbf{symmetric skill} conditions, adaptive adjustment has no independent
effect on decision authority.
In these functionally homogeneous settings, role flexibility is treated as a
baseline requirement rather than a distinctive capacity.
Agents who absorb coordination costs help maintain collective stability, but this
stabilizing labor remains politically invisible.
In the terms of the Atlas Paradox, agents may ``hold up the sky,'' yet gain no
additional influence over collective decisions.

Under \textbf{complementary skill} conditions, adaptive adjustment becomes
positively associated with decision authority after controlling for proposal
intensity.
Functional differentiation makes adjustment capacity observable and attributable,
allowing adaptive competence to enter authority formation as a secondary signal.
Importantly, this effect does not replace agenda control.
Instead, it supplements it: agents who can both define proposals and absorb
coordination costs occupy the most advantaged positions.
Here, the Atlas Paradox is mitigated but not eliminated—stabilizing the system is
recognized, but agenda-setting power remains paramount.

Mismatch pressure further clarifies this structural contrast.
In complementary systems, persistent misalignment is negatively associated with
authority, indicating that failure to maintain functional fit is penalized when
roles are clearly differentiated.
In symmetric systems, mismatch pressure has no comparable effect, consistent with
the absence of assignable responsibility in homogeneous role structures.

\paragraph{Interpretation.}
Taken together, these results show that adaptive adjustment reliably improves
collective coordination, but does not constitute a primary pathway to authority.
Authority remains anchored in agenda-setting power, while coordination labor is
structurally positioned as supportive rather than governing.
The Atlas Paradox thus highlights a persistent division between maintaining social
order and directing it.
By making this division explicit, the analysis clarifies how center--periphery
structures can persist alongside efficiency gains, and how different production
structures condition the political visibility of coordination work.

\DefineChatType{Hawk}{red!12}{black}
\DefineChatType{Dove}{green!12}{black}
\DefineChatType{Elder}{violet!14}{black}
\DefineChatType{Proself}{gray!12}{black}
\DefineChatType{default}{gray!10}{black}

\section{Micro-Political Dynamics of Negotiation}
\label{app:micro-politics}

This appendix provides qualitative evidence supporting the micro-level mechanisms
discussed in Sections~\ref{p1-stable-division}--\ref{p5-center-periphery} and
Section~\ref{app:atlas-paradox}. By examining representative negotiation utterances
alongside behavioral regularities, we illustrate how authority and stratification
emerge through interactional role differentiation rather than formal designation,
cultural scripts, or institutional rules.

Across both symmetric and complementary skill settings, negotiation does not
converge toward homogeneous participation. Instead, interaction stabilizes into
distinct \emph{functional roles}—Hawks, Doves, and Integrative Elders—whose
differentiated contributions jointly sustain cooperation, coordination, and
authority. These roles are not fixed identities, but recurrent patterns of
discourse, proposal behavior, and relational positioning that repeatedly appear
over time.

\subsection{Ideal-Typical Roles in Negotiation}

\paragraph{Hawks.}
Hawks are characterized by frequent agenda-setting attempts that prioritize
family-centered interests, articulate normative boundaries, and condition
cross-group cooperation on trustworthiness and value alignment. Their utterances
typically foreground internal solidarity while explicitly managing the limits of
external engagement (see H1--H2).

\begingroup
\setlength{\tabcolsep}{0pt}
\renewcommand{\arraystretch}{0.92}
\begin{table}[t]
\centering
\begin{tabularx}{\linewidth}{@{}p{1.8em}X@{}}

\ChatRow{H1}{
  \begin{ChatDiagram}[bubble width=.74\linewidth, icon size=8mm, vspace=1.2mm, legend=true]
    \ChatMsgEx[Hawk]{L}{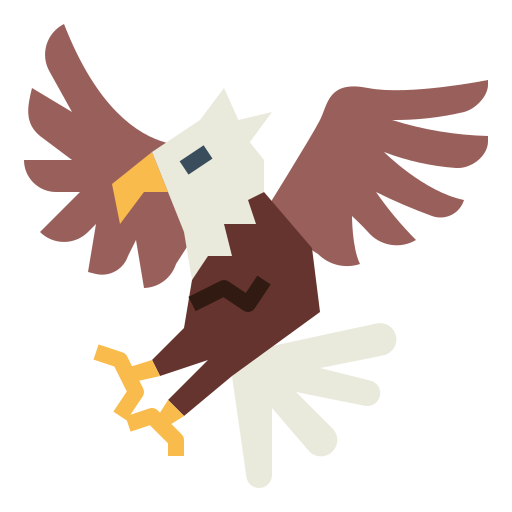}{Symm.}{Agent04}{F1}{%
      ``I'm glad to see the continued focus on Resource C from Agents 05, 15, and 17.
      Let's ensure our family's labor allocation reflects both our immediate needs
      (Resource C) and our commitment to supporting trusted allies' Resource B
      production where possible.''%
    }
  \end{ChatDiagram}
}

\ChatRow{H2}{
  \begin{ChatDiagram}[bubble width=.74\linewidth, icon size=8mm, vspace=1.2mm, legend=false]
    \ChatMsgEx[Hawk]{L}{icons/hawk.png}{Comp.}{Agent04}{F1}{%
      ``As always, my priority is supporting Family F1, but I'm open to working with
      Agents 07--18 who demonstrate consistent cooperation, particularly on Resource
      C production where I can contribute most effectively.''%
    }
  \end{ChatDiagram}
}

\end{tabularx}
\caption{Hawk agenda-setting utterances used in the argument (H1--H2).}
\end{table}
\endgroup

These utterances illustrate how Hawks actively shape the negotiation agenda by
asserting boundaries and priorities. Importantly, consistent with the regression
results in Section~\ref{p4-authority-stratification}, such proposal activity is
\emph{positively} associated with subsequent decision authority. Hawks do not merely
symbolize cohesion; through repeated agenda-setting, they occupy central positions
in the political economy of negotiation, even when their proposals emphasize
conservative or family-oriented goals.

\paragraph{Doves.}
Doves emphasize communication, inclusiveness, and cross-group openness. Their
discourse frequently addresses multiple audiences simultaneously, stressing trust,
fairness, and mutual recognition rather than concrete agenda control (see D1--D2).

\begingroup
\setlength{\tabcolsep}{0pt}
\renewcommand{\arraystretch}{0.92}
\begin{table}[t]
\centering
\begin{tabularx}{\linewidth}{@{}p{1.8em}X@{}}

\ChatRow{D1}{
  \begin{ChatDiagram}[bubble width=.74\linewidth, icon size=8mm, vspace=1.2mm, legend=true]
    \ChatMsgEx[Dove]{L}{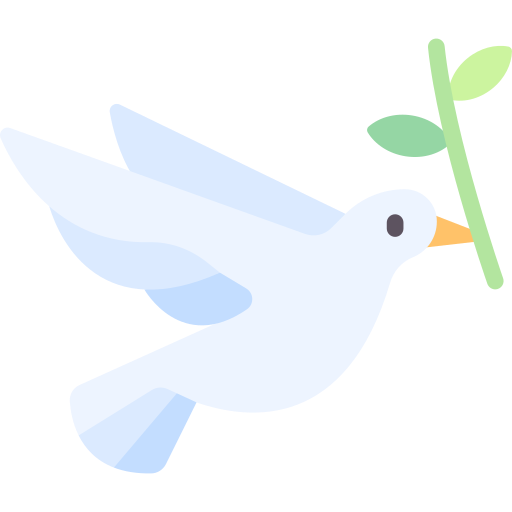}{Symm.}{Agent08}{F2}{%
      ``I echo your commitment to trust and unity, and I’d like to focus on deepening
      collaboration with our Core Allies while also reaching out to Neutral Outsiders
      to build stronger ties. Let’s work together to ensure everyone feels valued and
      included.''%
    }
  \end{ChatDiagram}
}

\ChatRow{D2}{
  \begin{ChatDiagram}[bubble width=.74\linewidth, icon size=8mm, vspace=1.2mm, legend=false]
    \ChatMsgEx[Dove]{L}{icons/dove.png}{Comp.}{Agent07}{F2}{%
      ``Let's focus on building trust and supporting each other—especially within our
      core group—while remaining open to positive collaboration with others.''%
    }
  \end{ChatDiagram}
}

\end{tabularx}
\caption{Dove coordination-oriented utterances used in the argument (D1--D2).}
\end{table}
\endgroup

Quantitatively, communication intensity alone does not robustly predict authority
once proposal behavior is accounted for
(Section~\ref{p4-authority-stratification}).
Dove behavior thus appears oriented toward coordination and relational smoothing
rather than agenda dominance. Doves facilitate information flow and reduce overt
conflict, but do not, by themselves, consolidate decision authority.

\paragraph{Proself-Oriented Positioning.}
For contrast, Proself agents frequently combine expressions of loyalty with explicit
articulation of personal or functional priorities. Their utterances embed economic
rationality within relational language without converging on authority-holding
roles (see P1).

\begingroup
\setlength{\tabcolsep}{0pt}
\renewcommand{\arraystretch}{0.92}
\begin{table}[t]
\centering
\begin{tabularx}{\linewidth}{@{}p{1.8em}X@{}}

\ChatRow{P1}{
  \begin{ChatDiagram}[bubble width=.74\linewidth, icon size=8mm, vspace=1.2mm, legend=true]
    \ChatMsgEx[Proself]{L}{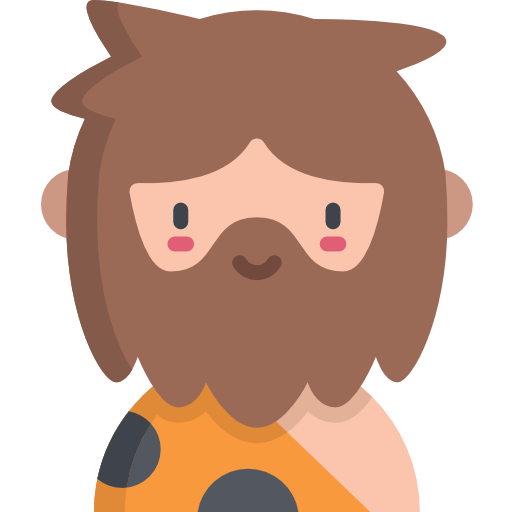}{Comp.}{Agent16}{--}{%
      ``While I support the focus on Resource B, I also recognize the value of Resource
      C for our community's stability. Let's work together to allocate our efforts
      wisely to meet both priorities.''%
    }
  \end{ChatDiagram}
}

\end{tabularx}
\caption{Proself-oriented utterance used for contrast (P1).}
\end{table}
\endgroup

Such statements highlight flexibility and strategic alignment but lack the
agenda-setting consolidation characteristic of authority-bearing roles.

\paragraph{Integrative Elders.}
Integrative Elders are distinguished by selective intervention, high proposal
effectiveness, and broad cross-group recognition. Rather than engaging continuously,
they typically intervene after extended discussion by Hawks and Doves, consolidating
prior coordination into concrete, low-friction proposals (see E1--E3).

Under symmetric skills, Elder authority tends to emerge within family cores, where
dense relational ties enable rapid coordination:

\begingroup
\setlength{\tabcolsep}{0pt}
\renewcommand{\arraystretch}{0.92}
\begin{table}[t]
\centering
\begin{tabularx}{\linewidth}{@{}p{1.8em}X@{}}

\ChatRow{E1}{
  \begin{ChatDiagram}[bubble width=.74\linewidth, icon size=8mm, vspace=1.2mm, legend=true]
    \ChatMsgEx[Elder]{L}{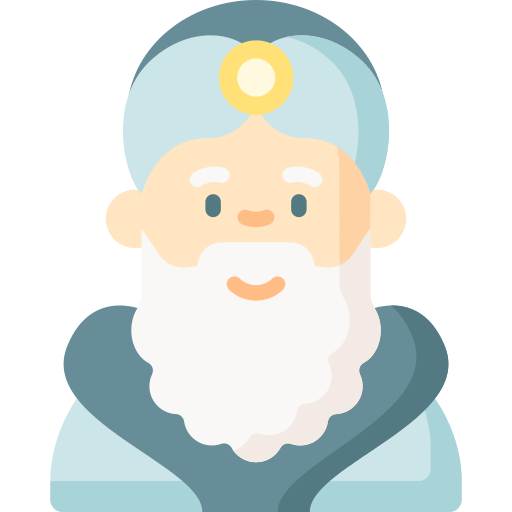}{Symm.}{Agent05}{F1}{%
      ``Thank you all for the enthusiastic support of the joint project proposal!
      Let's continue this momentum by finalizing the details—I suggest we focus on
      Resource A production this turn to align with our shared consumption needs.
      Agent07 and Agent12, would you be open to co-leading this initiative with me?''%
    }
  \end{ChatDiagram}
}

\end{tabularx}
\caption{Integrative Elder intervention under symmetric skills (E1).}
\end{table}
\endgroup

This intervention exemplifies Elder dynamics: it follows broad discussion, proposes
a specific allocation, and recruits cross-family collaborators. Such proposals are
disproportionately likely to be accepted and to generate higher marginal utility,
contributing to durable authority concentration.

Under complementary skills, Elder-like authority is less clan-centered and more
brokerage-oriented. Multiple agents may occupy integrative positions by aligning
functional complementarities with relational trust:

\begingroup
\setlength{\tabcolsep}{0pt}
\renewcommand{\arraystretch}{0.92}
\begin{table}[t]
\centering
\begin{tabularx}{\linewidth}{@{}p{1.8em}X@{}}

\ChatRow{E2}{
  \begin{ChatDiagram}[bubble width=.74\linewidth, icon size=8mm, vspace=1.2mm, legend=true]
    \ChatMsgEx[Elder]{L}{icons/elder.png}{Comp.}{Agent07}{F2}{%
      ``I'd like to propose a joint production effort between my in-group
      (Agents 08--12) and trusted collaborators like Agent15, while ensuring we
      also meet Resource C needs.''%
    }
  \end{ChatDiagram}
}

\ChatRow{E3}{
  \begin{ChatDiagram}[bubble width=.74\linewidth, icon size=8mm, vspace=1.2mm, legend=false]
    \ChatMsgEx[Elder]{L}{icons/elder.png}{Comp.}{Agent12}{F2}{%
      ``I propose we focus on transparent labor allocations that demonstrate our
      commitment to both in-group cohesion and equitable community collaboration.
      Let's lead by example in building a resilient and inclusive resource-sharing
      system.''%
    }
  \end{ChatDiagram}
}

\end{tabularx}
\caption{Integrative Elder brokerage-oriented proposals under complementary skills
(E2--E3).}
\end{table}
\endgroup

Together, these utterances illustrate how Integrative Elders under complementary
skills operate as coordination brokers rather than clan anchors. Authority accrues
to those who repeatedly succeed in integrating heterogeneous production needs into
coherent collective action.

\subsection{Negotiation, Authority, and Political Equilibrium}

Taken together, the Hawk–Dove–Elder configuration constitutes the micro-political
foundation of differential order in CAREB-MAS. Hawks stabilize internal cohesion and
agenda focus through frequent proposal activity; Doves maintain communicative
connectivity and relational openness; Integrative Elders consolidate coordination
into decisive, widely recognized proposals.

Crucially, this qualitative structure aligns with the quantitative finding that
proposal intensity is strongly and positively associated with authority
(Section~\ref{p4-authority-stratification}), while communication and adaptive
adjustment alone are insufficient to secure decision power. Authority emerges not
from dominance or compliance, but from repeated recognition of effective
agenda-setting within a relationally structured negotiation environment.

This interactional differentiation also underpins the \emph{Atlas Paradox}
(Section~\ref{app:atlas-paradox}). Agents who primarily absorb coordination and
adjustment costs stabilize collective outcomes without displacing authority centers,
while those who repeatedly set and consolidate agendas accumulate durable authority.
Differential order thus arises as a political equilibrium sustained through
functional role differentiation rather than explicit hierarchy.

Overall, the micro-political evidence presented here demonstrates how stable
authority, stratification, and division of political labor emerge endogenously from
interaction, providing the qualitative foundation for the macro-level patterns
documented in the main text.

\section{Empirical Correspondence}
\label{app:empirical-correspondence}

This appendix explores the correspondence between our five propositions (P1--P5) derived from Fei Xiaotong’s Differential Order Pattern and cross-cultural ethnographic evidence. It also explicitly states the falsification criteria for our mechanistic model. The concept--metric mapping in Appendix~\ref{app:concept-metric} serves as the bridge between ethnographic vocabulary and computational variables.

\subsection{P1--P2: Division of Labor and Relational Ethics}

The emergence of stable division of labor without centralized coordination (P1) corresponds to extensive anthropological documentation of spontaneous specialization in pre-state societies. For instance, Sahlins'~\citep{sahlins1972stone} analysis of the "domestic mode of production" demonstrates that household-level production specialization emerges primarily from kinship obligations, rather than from market incentives. Our simulations align with this view: division lock-in steadily increases across rounds, stabilizing without the need for centralized control (Figure~\ref{fig:macro_lockin}).

The persistence of cooperation under adversity among relationally close agents (P2) parallels Sahlins’ tripartite reciprocity model, wherein \emph{generalized reciprocity} (open-ended sharing among close kin) sustains cooperation precisely when balanced-reciprocity logic would predict withdrawal. Mauss’~\citep{mauss1925gift} analysis of gift exchange further supports this, showing that relational obligation overrides short-term material calculation. Our results substantiate this idea, with kin groups (F1, F2) showing a consistent preference for internal cooperation, even when resource stress could incentivize self-interested withdrawal (Figures~\ref{fig:commuity_matching_by_SIM}). These findings suggest that relational ethics embedded in kinship networks play a significant role in sustaining cooperation in rural societies.

\emph{Falsification criterion}: If ethnographic cases consistently show that production specialization in pre-market kinship societies is centrally coordinated rather than emergent, or that close-kin cooperation collapses as readily as distant-kin cooperation under resource stress, P1--P2 would be undermined. Our simulation outcomes and the ethnographic documentation of spontaneous kin-based specialization support the robustness of P1--P2, indicating that they are not easily falsified by contrary evidence.

\subsection{P3: Relational Decay of Cooperation}

The concentric gradient of cooperation intensity (P3) is supported by the “segmentary principle” documented across African, Melanesian, and Middle Eastern societies~\citep{fortes1940african}. In segmentary lineage systems, political allegiance and economic cooperation follow genealogical distance: closer segments cooperate against more distant ones, producing a nested, ego-centered cooperation structure that closely parallels Fei's concentric circles. Our simulation results reveal similar patterns of cooperation, where family groups exhibit high in-group cooperation and lower cooperation with other groups, especially as relational distance increases (Figures~\ref{fig:authority_proposal_boxplots} and~\ref{fig:commuity_matching_by_SIM}).

This pattern of relational decay mirrors Sahlins’ and Fortes’ insights that kinship distance influences cooperation intensity. We find that family groups (F1 and F2) show strong internal solidarity, while Proself agents, who represent less relationally bound actors, function as more flexible collaborators across groups. This outcome underscores the concentric nature of cooperation as proposed in P3.

\emph{Falsification criterion}: If cooperation in documented pre-market societies follows territorial proximity, resource complementarity, or random association rather than kinship distance, the relational-decay mechanism would be falsified. Our findings indicate that cooperation, in line with kinship ties, tends to decay across relational distance, as proposed in P3, making it robust against the falsification criterion.

\subsection{P4--P5: Authority Stratification and Center--Periphery Structure}

The emergence of authority through agenda-setting rather than ascription (P4) corresponds to the “big man” system documented in Melanesian societies~\citep{sahlins1972stone}. Leaders gain influence through their demonstrated ability to organize collective action and redistribute resources, not through hereditary position. Our results confirm this model, where decision authority emerges in the simulation through repeated proposal and coordination actions. Table~\ref{tab:authority_regression} shows that higher proposal activity correlates strongly with increased authority, suggesting that authority is contingent on the effectiveness of one’s actions, rather than merely ascribed based on social position.

The coupling of kinship position with social recognition in authority formation (P5) mirrors Fortes and Evans-Pritchard's~\citep{fortes1940african} analysis, where authority in segmentary systems accrues to those whose lineage position aligns with demonstrated social competence. In our experiments, authority is concentrated in the hands of family groups, particularly those with higher proposal effectiveness (Table~\ref{tab:authority_regression}). This supports Proposition 5, which asserts that authority stratification emerges as a result of both relational recognition and individual competence.

\emph{Falsification criterion}: If authority in pre-market kinship societies is found to be overwhelmingly ascriptive, determined solely by birth order, gender, or lineage seniority, without any emergent component from social interactions, the agenda-control mechanism underlying P4--P5 would be challenged. Our findings support the idea that authority formation is an emergent process, linked to social recognition and competence, rather than being purely ascriptive.

\subsection{Scope Conditions}

Our mechanistic account applies specifically to \emph{pre-market, acephalous, kinship-organized} societies. We do not claim that Differential Order patterns emerge under all structural conditions. The complementary-skills condition already demonstrates systematic attenuation of clan-based stratification (P5), suggesting that increased functional differentiation—often a hallmark of market integration—weakens the relational coupling that sustains Differential Order. This finding is consistent with both Durkheim’s transition from mechanical to organic solidarity and Fei’s analysis of the pressures of modernization. In our simulations, we observe that when skill complementarity increases, the traditional kinship-based stratification patterns begin to dissipate in favor of more functionally interdependent structures (Figures~\ref{fig:utility_kde}, \ref{fig:authority_kde}).

\emph{Falsification criterion}: If ethnographic evidence shows that Differential Order patterns persist under conditions of high market integration, strong central authority, or other forms of societal complexity, this would challenge our claim that these patterns are specific to kinship-based, pre-market societies. The evidence from our experiments supports the idea that Differential Order is contingent on the relational structures present in kinship societies, suggesting that these patterns are not universal but highly sensitive to underlying social organization.

\subsection{Concluding Remarks}

Our empirical correspondence demonstrates that the patterns of rural social order identified by Fei Xiaotong can emerge from general social mechanisms without the need for culture-specific institutional rules. The CAREB-MAS model provides computational evidence for the spontaneous emergence of stable labor division, relational ethics, and authority stratification, with structural outcomes varying according to production conditions. These findings reinforce the notion that Differential Order is a dynamic, structure-sensitive outcome rather than a fixed cultural template. The scope conditions, empirical correspondence, and falsification criteria for this claim are further discussed in the Section~\ref{sec:limitation} and Appendix~\ref{app:empirical-correspondence}.



\end{document}